\PassOptionsToPackage{unicode}{hyperref}
\PassOptionsToPackage{hyphens}{url}
\PassOptionsToPackage{dvipsnames,svgnames,x11names}{xcolor}
\documentclass[
  12pt]{article}

\usepackage{bm}
\usepackage{dsfont}

\usepackage{multirow}
\usepackage{multicol}

\usepackage{graphicx,amscd,amsmath,amssymb,amsfonts,verbatim, amsthm, dsfont}

\newtheorem{theorem}{Theorem}

\theoremstyle{definition}

\usepackage{algorithm}

\newcommand\numberthis{\addtocounter{equation}{1}\tag{\theequation}}

\usepackage{iftex}
\ifPDFTeX
  \usepackage[T1]{fontenc}
  \usepackage[utf8]{inputenc}
  \usepackage{textcomp} 
\else 
  \usepackage{unicode-math}
  \defaultfontfeatures{Scale=MatchLowercase}
  \defaultfontfeatures[\rmfamily]{Ligatures=TeX,Scale=1}
\fi
\usepackage{lmodern}
\ifPDFTeX\else  
\fi
\IfFileExists{upquote.sty}{\usepackage{upquote}}{}
\IfFileExists{microtype.sty}{
  \usepackage[]{microtype}
  \UseMicrotypeSet[protrusion]{basicmath} 
}{}
\makeatletter
\@ifundefined{KOMAClassName}{
  \IfFileExists{parskip.sty}{%
    \usepackage{parskip}
  }{
    \setlength{\parindent}{0pt}
    \setlength{\parskip}{6pt plus 2pt minus 1pt}}
}{
  \KOMAoptions{parskip=half}}
\makeatother
\usepackage{xcolor}
\makeatletter
\ifx\paragraph\undefined\else
  \let\oldparagraph\paragraph
  \renewcommand{\paragraph}{
    \@ifstar
      \xxxParagraphStar
      \xxxParagraphNoStar
  }
  \newcommand{\xxxParagraphStar}[1]{\oldparagraph*{#1}\mbox{}}
  \newcommand{\xxxParagraphNoStar}[1]{\oldparagraph{#1}\mbox{}}
\fi
\ifx\subparagraph\undefined\else
  \let\oldsubparagraph\subparagraph
  \renewcommand{\subparagraph}{
    \@ifstar
      \xxxSubParagraphStar
      \xxxSubParagraphNoStar
  }
  \newcommand{\xxxSubParagraphStar}[1]{\oldsubparagraph*{#1}\mbox{}}
  \newcommand{\xxxSubParagraphNoStar}[1]{\oldsubparagraph{#1}\mbox{}}
\fi
\makeatother

\usepackage{longtable,booktabs,array}
\usepackage{calc} 
\usepackage{etoolbox}
\makeatletter
\patchcmd\longtable{\par}{\if@noskipsec\mbox{}\fi\par}{}{}
\makeatother
\IfFileExists{footnotehyper.sty}{\usepackage{footnotehyper}}{\usepackage{footnote}}
\makesavenoteenv{longtable}
\usepackage{graphicx}
\makeatletter
\def\maxwidth{\ifdim\Gin@nat@width>\linewidth\linewidth\else\Gin@nat@width\fi}
\def\maxheight{\ifdim\Gin@nat@height>\textheight\textheight\else\Gin@nat@height\fi}
\makeatother
\setkeys{Gin}{width=\maxwidth,height=\maxheight,keepaspectratio}
\makeatletter
\def\fps@figure{htbp}
\makeatother

\makeatletter
\@ifpackageloaded{caption}{}{\usepackage{caption}}
\AtBeginDocument{%
\ifdefined\contentsname
  \renewcommand*\contentsname{Table of contents}
\else
  \newcommand\contentsname{Table of contents}
\fi
\ifdefined\listfigurename
  \renewcommand*\listfigurename{List of Figures}
\else
  \newcommand\listfigurename{List of Figures}
\fi
\ifdefined\listtablename
  \renewcommand*\listtablename{List of Tables}
\else
  \newcommand\listtablename{List of Tables}
\fi
\ifdefined\figurename
  \renewcommand*\figurename{Figure}
\else
  \newcommand\figurename{Figure}
\fi
\ifdefined\tablename
  \renewcommand*\tablename{Table}
\else
  \newcommand\tablename{Table}
\fi
}
\@ifpackageloaded{float}{}{\usepackage{float}}
\floatstyle{ruled}
\@ifundefined{c@chapter}{\newfloat{codelisting}{h}{lop}}{\newfloat{codelisting}{h}{lop}[chapter]}
\floatname{codelisting}{Listing}

\makeatother
\makeatletter
\@ifpackageloaded{caption}{}{\usepackage{caption}}
\@ifpackageloaded{subcaption}{}{\usepackage{subcaption}}
\makeatother

\ifLuaTeX
  \usepackage{selnolig}  
\fi
\usepackage[]{natbib}
\usepackage{bookmark}

\IfFileExists{xurl.sty}{\usepackage{xurl}}{} 
\hypersetup{
  pdftitle={Title},
  pdfauthor={Author 1; Author 2},
  pdfkeywords={3 to 6 keywords, that do not appear in the title},
  colorlinks=true,
  linkcolor={blue},
  filecolor={Maroon},
  citecolor={Blue},
  urlcolor={Blue},
  pdfcreator={LaTeX via pandoc}}

\newcommand{\anon}{1}

\begin{document}

\def\spacingset#1{\renewcommand{\baselinestretch}%
{#1}\small\normalsize} \spacingset{1}


\if1\anon
{
  \title{\bf Semi- and non-parametric approaches to  individualized treatment regimes in the presence of causal mediation}
 \author{Misha Dolmatov$^{1}$, Erica E. M. Moodie$^{1}$, David A. Stephens$^{2}$,\\ Dipankar Bandyopadhyay$^{3,}$\thanks{
    The authors gratefully acknowledge the United Network for Organ Sharing for providing the motivating OPTN data, and the context for the work. The work is partially funded by grants R01DE031134 and R21DE031879 from the United States National Institutes of Health.} \hspace{.3cm}\\[4pt] 
    Department of Epidemiology, Biostatistics and Occupational Health$^1$, \\ McGill University, Montreal, QC, Canada\\[4pt]
    Department of Mathematics and Statistics$^2$, \\ McGill University, Montreal, QC, Canada\\[4pt]
    Department of Biostatistics$^3$, School of Public Health, \\ Virginia Commonwealth University, Richmond, VA, USA
    }
  \maketitle
} \fi

\if0\anon
{
  \bigskip
  \bigskip
  \bigskip
  \begin{center}
    {\LARGE\bf Semi- and non-parametric approaches to  individualized treatment regimes in the presence of causal mediation}
\end{center}
  \medskip
} \fi

\bigskip
\begin{abstract}
Individualized treatment rules (ITRs) map an individual patient's characteristics to their recommended treatment value. Typically, the optimal ITR is defined as the rule which maximizes a mean counterfactual outcome; the resulting ITR maximizes the effect of treatment along all causal pathways to the outcome, including indirect pathways through mediating variables. Although maximizing the total effect is often sufficient, explicitly incorporating causal mediation in an ITR analysis has several potential benefits such as enhanced interpretability, and additional flexibility in targeting specific causal pathways. For this purpose, we introduce novel Bayesian semiparametric and nonparametric estimators for conditional mediation effects in the presence of multiple mediators and show how they can be used to estimate optimal ITRs. We demonstrate the proposed methodology via an application to optimal kidney allocation with hepatitis C positive donors.
\end{abstract}

\noindent%
\vfill

\newpage
\spacingset{1.8} 


\section{Introduction}

The goal of precision medicine is to improve patient outcomes by tailoring treatment to individual-level characteristics. Tailoring can be achieved by way of individualized treatment rules (ITRs), which map a patient's covariate information to their optimal treatment, typically defined as the treatment which maximizes a mean counterfactual outcome. This definition of the optimal ITR targets the effect of treatment along all directed causal pathways to the chosen outcome variable; in particular, the standard optimal ITR incorporates indirect effects of the exposure through mediating variables. Several methods for estimating (total effect) ITRs have been proposed  including Q-learning \citep{murphy2003}, G-estimation \citep{Robins2004}, and dynamic weighted ordinary least squares \citep[dWOLS;][]{WallacedWOLS}.
In cases where only the total effect of treatment is of interest, these methods are sufficient. However, explicitly integrating causal mediation into an ITR analysis has several potential benefits.

First, decomposing conditional total treatment effects into direct, indirect, and potentially even individual mediator effects can produce ITRs which are more interpretable: for any fixed patient profile, optimality of the chosen treatment can be attributed to its direct effect on the outcome, or its indirect effects through selected mediators. Second, it may be the case that maximizing the effect of treatment across all possible causal pathways is undesirable \citep{nabi2018estimation}. By targeting specific mediating pathways, it is possible to formulate ITRs that more accurately reflect the research question at hand.
Lastly, there may be specific external constraints or costs that must be taken into account and which depend on mediators of the treatment effect. For instance, treatment allocation may cause side-effects that have an additional financial cost. Ideally, the optimal ITR should maximize the study outcome while also limiting costs by reducing incidence of side-effects.      

Identification of pure (natural) direct and indirect effects \citep{robins1992direct} with multiple mediators often requires stringent assumptions on the data-generating mechanism. Interventional mediator effects have been proposed as a more conceptually sound alternative to traditional pure mediation effects \citep{vansteelandt2014intervention}. These effects, which target a hypothetical population where mediator values are drawn from a specific interventional distribution, can be point-identified in the presence of post-treatment mediator-outcome confounding, do not require cross-world independence assumptions, and can be mapped to a target trial \citep{chen2025targettrial}. In the presence of multiple mediators, the contribution of each individual mediator to the overall indirect effect is also of interest. However, in the observational data setting, the causal ordering of mediators is often unknown, making estimation of individual mediator effects challenging. A notable recent development in multiple mediation is the introduction of the so-called exit effects, which quantify mediator-specific contributions to the indirect effect in an ordering-agnostic way \citep{xia2022exit}. 

Most existing mediation methods target marginal mediation effects. In the case of identifying the optimal ITR, it is their conditional equivalents that are most relevant. Methodology for estimating conditional mediation effects is relatively sparse in comparison \citep{rubinstein2023hetmed}. 

In this paper, we develop novel Bayesian semiparametric and nonparametric estimators of conditional mediation effects, and showcase how they can be used to identify optimal ITRs.   
The proposed semiparametric estimator is a doubly robust G-estimator derived from a partially linear model that allows for misspecification of a portion of the outcome model under the assumption of correct specification of a ``blip'' model, which captures the effect of the exposure and mediators on the outcome. Partially linear models are popular tools for the estimation of mediation effects with multiple mediators as they often allow for straightforward  estimation and statistical inference.
However, existing partially linear mediation approaches are either singly-robust \citep{vanderweele2014multmed}, do not allow for exposure-mediator and mediator-mediator interactions, \citep{hines2021partlinear},  or rely on parametric modelling of mediators to simplify estimation of indirect effects \citep{cai2022highmed, yang2025dmlmed}. In cases where correct specification of a blip model is unlikely, we also provide a quantization-based nonparametric estimator that makes no parametric assumptions regarding the outcome model. Existing nonparametric estimators of conditional mediation effects only consider discrete mediators, discretizing continuous mediators when present \citep{benkeser2021npmed, rubinstein2023hetmed}. Optimal vector quantization offers a more principled approach of approximating continuous mediator distributions by discrete measures \citep{graf2000foundquant,pages2015introduction}.  

We study the frequentist properties of these estimators and provide explicit convergence rates. Finally, we demonstrate how the Bayesian bootstrap \citep{rubin1981bayesboot} can be employed to obtain valid non-asymptotic uncertainty quantification for conditional mediation effects and the resulting optimal ITR. 

The paper is organized as follows. In section \ref{sec:defs}, we define optimal ITRs under the interventional data distribution and show how they can be written in terms of mediation estimands. In section \ref{sec:estimands}, we review the decomposition of total effects into direct, indirect and individual mediator effects, and give the conditions under which they are identified in the observed data. Section \ref{sec:bayesboot} is a short overview of the Bayesian bootstrap. Sections \ref{sec:semi} and \ref{sec:nonpar} introduce the proposed semiparametric and nonparametric estimators along with their theoretical properties. Simulation results for a variety of data generating mechanisms are provided in section \ref{sec:sims}. In section \ref{sec:application}, we showcase an application to the problem of optimal kidney allocation with Hepatitis C positive donors using data from the Organ Procurement and Transplantation Network (OPTN). Finally, section \ref{sec:discussion} concludes with a discussion. Additional theoretical results, proofs, and simulation details are relegated to appendices A-C. 

\section{Definitions and interventional framework}\label{sec:defs}



We consider a single stage decision problem where a treatment  $A \in \mathcal{A} \subseteq \mathbb{R}$ is to be allocated based on a vector of pre-treatment covariates $\bm{X} \in \mathcal{X}$ in order to maximize an end of study outcome $Y \in \mathbb{R}$. Without loss of generality, we assume that larger values of $Y$ are more desirable. Additionally, we measure the vector of mediators $\bm{M} = (M_1, \dots, M_\kappa) \in \mathcal{M}$. The exposure $A$, the outcome $Y$, and mediators $\{M_k\}_{k=1}^\kappa$ are assumed to be either continuous or binary. The observed data are the collection of random variates $\{O_i\}_{i=1}^n \sim \mathbb{P}^\mathcal{O}$, where $O_i = (\bm{X}_i, A_i, \bm{M}_i, Y_i)$. For a given $a \in \mathcal{A}$, we let $\bm{M}(a)$ denote the vector of mediators if, possibly contrary to the fact, treatment value $A = a$ had been assigned. Similarly, for $a, a' \in \mathcal{A}$, $Y(a', \bm{M}(a))$ denotes the potential outcome that results from treatment $a'$ and mediator vector $\bm{M}(a)$. An ITR is a function $d : \mathcal{X} \to \mathcal{A}$ that maps a subject's covariate information to the set of possible treatments; assigning treatment according to an ITR $d$ yields the potential outcome $Y(d, \bm{M}(d))$. For causal identification and estimation of these potential outcomes, we require counterfactual consistency,  the stable unit treatment value assumption (SUTVA) \citep{rubin1980}, and positivity of both treatment and mediator densities. Additional estimand-specific assumptions are introduced in subsequent sections.

Natural mediation effects are not identified when there are confounders $\bm{L}$ of the mediator-outcome relationship that are affected by treatment \citep{avin2005identifiability}. For instance, they are not identified under the following generic observational data generating distribution:
\begin{align*}
     f_\mathcal{O}(o) &= f_{\bm{X}}(\bm{x}) f_{A|\bm{X}}(a|\bm{x}) f_{\bm{L}|A,\bm{X}} (\bm{l}|a, \bm{x}) f_{\bm{M}| \bm{X}, A, \bm{L}}(\bm{m}| \bm{x}, a, \bm{l}) f_{Y| \bm{X}, A, \bm{L}, \bm{M}}(y| \bm{x}, a, \bm{l}, \bm{m}).
\end{align*}
We instead consider the estimation of mediation effects under the so-called \textit{interventional} distribution given by 
\begin{align}\label{eq:dgm}
     f_\mathcal{E}(o) &= f_{\bm{X}}(\bm{x}) f_{A|\bm{X}}(a|\bm{x}) f_{\bm{L}|A,\bm{X}} (\bm{l}|a, \bm{x}) f_{\bm{M}| \bm{X}, A}(\bm{m}| \bm{x}, a) f_{Y| \bm{X}, A, \bm{L}, \bm{M}}(y| \bm{x}, a, \bm{l}, \bm{m}).
\end{align}
This choice of data generating process corresponds to a hypothetical intervention that sets the value of the mediator $\bm{M}$ by sampling from the conditional distribution $ f_{\bm{M}| \bm{X}, A}$, eliminating all dependence on post-treatment confounders $\bm{L}$. The mediation effects associated with this particular distribution are termed 
interventional effects \citep{vansteelandt2014intervention}. 

Alternative interventional data generating distributions have also been proposed.  One may consider (\ref{eq:dgm}) with the conditional joint distribution $f_{\bm{M}|\bm{X}, A}$ replaced with $\prod_{k=1}^\kappa f_{M_k | A, \bm{X}}$, corresponding to a hypothetical world where the components of $\bm{M}$ are generated independently conditional on covariates and exposure. The margins of the mediator vector can then be modelled separately, considerably simplying estimation and inference for mediation effects. ~\cite{vansteelandt2017interventional} utilized this approach in order to estimate interventional effects in the presence of multiple mediators. This interventional distribution is most relevant when the dependence between mediators is expected to be weak such that intervening on any particular mediator would have limited follow-on effects on the remaining mediators \citep{chen2025targettrial}. For the remainder of the paper, we focus on the more general data generating distribution in (\ref{eq:dgm}). 

 Expectations indexed by $\mathcal{E}$ are taken with respect to the interventional (experimental) distribution in (\ref{eq:dgm}). The optimal ITR under the interventional distribution is defined as
\begin{align*}
    d_\mathcal{E}^{\text{opt}}(\bm{X}) &= \arg \max_{d} \mathbb{E}_\mathcal{E}[Y(d, \bm{M}(d))] - \mathbb{E}_\mathcal{E}[ C(d, \bm{M}(d), \bm{X})],
\end{align*} 
where $C(a, \bm{m}, \bm{x}) \in \mathbb{R}$ is a known function that represents some additional cost (e.g., financial) incurred upon allocating treatment $a$ and observing mediator value $\bm{m}$ at covariate level $\bm{x}$. The cost function $C$ can also be used to incorporate resource constraints or side-effects into decision-making \citep{vander2024costside}. Costs are specified on the outcome scale. 

Now, let $a_0$ indicate some chosen reference value of the treatment variable. Following standard arguments, the optimal ITR can equivalently be written as
\begin{align*}
    d_\mathcal{E}^{\text{opt}}(\bm{X}) &= \arg \max_a \mathbb{E}_\mathcal{E}[Y(a, \bm{M}(a))| \bm{X}] - \mathbb{E}_\mathcal{E}[C(a, \bm{M}(a), \bm{X})]\\
     &= \arg \max_a \mathbb{E}_\mathcal{E}[Y(a, \bm{M}(a))| \bm{X}] -  \mathbb{E}_\mathcal{E}[Y(a_0, \bm{M}(a_0))| \bm{X}] - \mathbb{E}_\mathcal{E}[C(a, \bm{M}(a), \bm{X})]\\
     &= \arg \max_a \tau_\mathcal{E}(a, \bm{X}) - \mathbb{E}_\mathcal{E}[C(a, \bm{M}(a), \bm{X})],
\end{align*}
where $\tau_\mathcal{E}(a, \bm{X})$ is the interventional conditional \textit{total} effect. The conditional total effect captures the causal effect of shifting the exposure from $a_0$ to $a$ along all directed pathways towards the outcome variable with covariates set to $\bm{X}$. 

In certain cases, only some specific causal pathways are of interest. For instance, \cite{nabi2018estimation} consider an ITR which maximizes the effect of antiretroviral therapy on patients' CD4 counts while fixing adherence behaviour to its counterfactual value under a reference treatment. Without loss of generality, suppose we are interested in optimizing the effect of the treatment through mediators $\bm{M}$ while fixing the distribution of $M_1$ to its value under treatment value $a_0$. Let $\bm{M}_{-k} = (M_1, \dots, M_{k-1}, M_{k+1}, \dots, M_\kappa)$ denote the subvector of mediators with the $k$th mediator removed. 
This leads to the following \textit{path-specific} ITR:
\begin{align*}
    d_{\mathcal{E}, -1}^{\text{opt}}(\bm{X}) &= \arg \max_a \mathbb{E}_\mathcal{E}[Y(a, \bm{M}_{-1}(a), M_1(a_0))| \bm{X}] - \mathbb{E}_\mathcal{E}[C(a, \bm{M}_{-1}(a), M_1(a_0), \bm{X})]\\
     &= \arg \max_a \tau_{\mathcal{E}, -1}(a, \bm{X}) - \mathbb{E}_\mathcal{E}[C(a, \bm{M}_{-1}(a), M_1(a_0), \bm{X})],
\end{align*}
where $\tau_{\mathcal{E}, -1}(a, \bm{X}) =\mathbb{E}_\mathcal{E}[Y(a, \bm{M}_{-1}(a), M_1(a_0))| \bm{X}] -\mathbb{E}_\mathcal{E}[Y(a_0, \bm{M}_{-1}(a_0), M_1(a_0))| \bm{X}]$ is the conditional total effect with $M_1$ fixed at its counterfactual value under exposure level $a_0$ and covariates set to $\bm{X}$. This total effect captures the causal effect of the exposure along all directed pathways towards the outcome, excluding pathways through $M_1$.

\section{Estimands and identifying assumptions}\label{sec:estimands}

\subsection{Direct and indirect effects}
Without loss of generality, we assume that $a_0 =0$. The  total effect $\tau_\mathcal{E}$ can be further decomposed into direct and indirect effects:
\begin{align*}
    \tau_\mathcal{E}(a,\bm{x}) &=\mathbb{E}_\mathcal{E}[Y(a, \bm{M}(a)) | \bm{X} = \bm{x}] -\mathbb{E}_\mathcal{E}[Y(0, \bm{M}(0)) | \bm{X} = \bm{x}] =   \zeta_\mathcal{E}(a,\bm{x}) +  \delta_\mathcal{E}(a,\bm{x}),\\
    \zeta_\mathcal{E}(a,\bm{x}) &=\mathbb{E}_\mathcal{E}[Y(a, \bm{M}(a))| \bm{X} = \bm{x}] -\mathbb{E}_\mathcal{E}[Y(0, \bm{M}(a)) | \bm{X} = \bm{x}],\\
    \delta_\mathcal{E}(a,\bm{x}) &=\mathbb{E}_\mathcal{E}[Y(0, \bm{M}(a))| \bm{X} = \bm{x}] -\mathbb{E}_\mathcal{E}[Y(0, \bm{M}(0)) | \bm{X} = \bm{x}].
\end{align*}
The conditional \textit{direct} effect $\zeta(a, \bm{x})$ represents the effect of shifting the exposure from $0$ to $a$ on the outcome with the mediator distribution being held to its counterfactual value under treatment value $a$ and covariates set to $\bm{x}$. Conversely, the \textit{indirect} effect $\delta(a, \bm{x})$ reflects the effect of the exposure through the mediator vector with the exposure held constant at  level $0$ and covariates fixed at $\bm{x}$.

For conditional direct and indirect effects to be identified, the data generating mechanism must satisfy the following:

(A1)  Sequential ignorability \citep{imai2010}: for all $a, a' \in \mathcal{A}$ and all $\bm{m} \in \mathcal{M}$,
        $\{Y(a', \bm{m}), \bm{M}(a)\} \perp A | \bm{X}, \text{ and }
        Y(a', \bm{m}) \perp \bm{M}(a) | A=a, \bm{X}.$

Sequential ignorability requires that the set of baseline covariates $\bm{X}$ is rich enough to adjust for confounding in the exposure-mediator, exposure-outcome and mediator-outcome relationships across all mediators. This assumption holds for the interventional distribution $f_\mathcal{E}$ in (\ref{eq:dgm}) by design. 
Under sequential ignorability, direct and indirect effects are identified as $ \zeta_\mathcal{E}(a,\bm{x}) =Q(a, a, \bm{x}) - Q(0, a, \bm{x})$ and $ \delta_\mathcal{E}(a,\bm{x}) = Q(0, a, \bm{x}) -Q(0, 0, \bm{x})$ , where
\begin{align*}
        Q(a, a', \bm{x}) &= \int\mathbb{E}_\mathcal{O}[Y| \bm{X}=\bm{x}, A= a, \bm{M} = \bm{m}] d\mathbb{P}^\mathcal{O}_{\bm{M}|\bm{X} = \bm{x}, A = a'}(\bm{m}).
\end{align*}
\subsection{Individual mediator effects}
The indirect effect measures the effect of the exposure through the mediator vector. To recover each mediator's contribution to the overall indirect effect, we can further decompose it into the sum of \textit{individual mediator indirect} effects $\delta^k_\mathcal{E}$ and an interaction effect $\delta^{\text{INT}}$:
\begin{align*}
 \delta_\mathcal{E}^k(a,\bm{x}) &=\mathbb{E}_\mathcal{E}[Y(0, \bm{M}_{-k}(0), M_k(a))| \bm{X} = \bm{x}] -\mathbb{E}_\mathcal{E}[Y(0, \bm{M}_{-k}(0), M_k(0)) | \bm{X} = \bm{x}],\\
 \delta_\mathcal{E}^{\text{INT}}(a, \bm{x}) &=   \delta_\mathcal{E}(a,\bm{x}) - \sum_{k=1}^\kappa \delta_\mathcal{E}^k(a,\bm{x}).
\end{align*}
The individual mediator indirect effect for mediator $k$, $\delta^k_\mathcal{E}$, is the effect of the exposure through $M_k$ with the rest of the mediator vector set to its value under the reference treatment. The interaction effect $\delta^{\text{INT}}$ is the difference between the overall indirect effect and the sum of the individual mediator indirect effects, and reflects possible effect modification of individual mediator effects by other mediators. These effects are modified versions of the \textit{exit} effects introduced in \cite{xia2022exit}.

When the causal ordering of the mediators is unknown, individual mediator indirect effects are generally non-identified. If they are of interest, the interventional data-generating distribution must also satisfy the following:
\begin{enumerate}
    \item[(A2)] $\mathbb{E}_\mathcal{E}[Y(0, \bm{m}_{-k}, M_k(a))-Y(0, \bm{m}_{-k}, M_k(0)) |\bm{M}_{-k}(0) = \bm{m}_{-k}, \bm{X}] \\    =\mathbb{E}_\mathcal{E}[Y(0,  \bm{m}_{-k}, M_k(a))-Y(0, \bm{m}_{-k}, M_k(0) ) |\bm{X}], \quad \forall k = 1, \dots, \kappa, \quad \forall \bm{m}_{-k}.$
\end{enumerate}
This assumption requires that, conditional on covariates, the indirect effect through any individual mediator does not depend on the counterfactual value of the rest of the mediator vector under the reference treatment. In particular, we do not require mediators or their counterfactual values to be conditionally independent given covariates. 

Under (A2), individual mediator effects  are identified as $\delta^k_\mathcal{E}(a,\bm{x}) =Q_k(0, 0, a,\bm{x}) -Q_k(0, 0, 0,\bm{x}) $ \citep{xia2022exit}, where $ Q_k(a,a', a'', \bm{x}) $ equals
\begin{align*}
    \int \Bigl(\int \mathbb{E}_\mathcal{O}[Y| \bm{X}=\bm{x}, A= a,M_k = m_k, \bm{M}_{-k} =\bm{m}_{-k}] d\mathbb{P}^\mathcal{O}_{\bm{M}_{-k}| \bm{X} = x, A = a'}(\bm{m}_{-k})\Bigr)d\mathbb{P}^\mathcal{O}_{M_k| \bm{X} = x, A = a''}(m_k).
\end{align*}

\subsection{Path-specific effects}
 
The mediation effects introduced in the previous subsections can trivially be generalized to the path-specific case. We define the conditional path-specific total, direct, indirect and individual mediator effects as:
\begin{align*}
    \tau_{\mathcal{E}, -1}(\bm{x}) &=\mathbb{E}_\mathcal{E}[Y(a, \bm{M}_{-1}(a), M_1(0))| \bm{X} = \bm{x}] -\mathbb{E}_\mathcal{E}[Y(0, \bm{M}_{-1}(0), M_1(0)) | \bm{X} = \bm{x}],\\
    \zeta_{\mathcal{E}, -1}(\bm{x}) &=\mathbb{E}_\mathcal{E}[Y(a, \bm{M}_{-1}(a), M_1(0)) | \bm{X} = \bm{x}] - \mathbb{E}_\mathcal{E}[Y(0, \bm{M}_{-1}(a), M_1(0)) | \bm{X} = \bm{x}],\\
    \delta_{\mathcal{E}, -1}(\bm{x}) &=\mathbb{E}_\mathcal{E}[Y(0, \bm{M}_{-1}(a), M_1(0))| \bm{X} = \bm{x}] -\mathbb{E}_\mathcal{E}[Y(0, \bm{M}_{-1}(0), M_1(0)) | \bm{X} = \bm{x}].
\end{align*}
The path-specific interaction effect is defined analogously as $ \delta^{\text{INT}}_{\mathcal{E}, -1}(a, \bm{x}) =   \delta_{\mathcal{E},-1}(a,\bm{x}) - \sum_{k=2}^\kappa \delta^k_{\mathcal{E},-1}(a,\bm{x})$. We note that the path-specific individual mediator effects are equal to their value in the non-path-specific case, i.e., $ \delta^k_{\mathcal{E}, -1} = \delta^k_{\mathcal{E}}$ for $k = 2, \dots, \kappa$.
The assumption required to identify them is then the same as in the general case (A2). This simplification only arises because we are fixing $M_1$ to its value under the reference treatment, the same reference value that we use to compute the mediation effects. For a generic path-specific ITR that targets other causal pathways, different assumptions may be required.

Sequential ignorability is not sufficient to identify path-specific direct and indirect effects. For these estimands, we also require the following assumption:
\begin{enumerate}
    \item[(A3)] $\mathbb{E}_\mathcal{E}[Y(a, \bm{M}_{-1}(a), m_1)-Y(0, \bm{M}_{-1}(0), m_1) | M_1(0) = m_1, \bm{X}] \\    =\mathbb{E}_\mathcal{E}[Y(a, \bm{M}_{-1}(a), m_1)-Y(0, \bm{M}_{-1}(0), m_1) |\bm{X}], \quad \forall m_1.$
\end{enumerate}
This condition holds provided that, conditional on covariates,  the total effect of the exposure along paths which exclude $M_1$ does not depend on the counterfactual value of $M_1$ under the reference treatment. Although (A3) appears to be similar to (A2), it is much stronger, and is often equivalent to mean counterfactual independence of $M_1$ and $\bm{M}_{-1}$. As such, these effects are best suited to data where $M_1$ is known to be weakly associated with the rest of the mediator vector. Further details and examples are provided in appendix A. 

Path-specific effects are then identified as $\zeta_{\mathcal{E}, -1}(a,\bm{x}) =Q(a, 0,\bm{x}) -Q(0, 0,\bm{x}) $, and $\delta_{\mathcal{E}, -1}(a,\bm{x}) =Q(0, a,\bm{x})-Q(0, 0,\bm{x}) $, where $ Q_{-1}(a,a', \bm{x})$ equals
\begin{align*}
       \int \Bigl(\int \mathbb{E}_\mathcal{O}[Y| \bm{X}=\bm{x}, A= a,M_1 = m_1, \bm{M}_{-1} = \bm{m}_{-1}] d\mathbb{P}^\mathcal{O}_{\bm{M}_{-1}| \bm{X} = x, A = a'}(\bm{m_{-1}})\Bigr)d\mathbb{P}^\mathcal{O}_{M_1| \bm{X} = x, A = 0}(m_1).
\end{align*}
The identification formula follows by the same argument as for the mediator-specific effects.  Having defined our estimands of interest, we now move on to estimation and inference. 

\section{Semiparametric estimation}\label{sec:semi}
For a generic exposure $A$, which may be a continuous random variable, neither marginal nor conditional mediation effects are pathwise differentiable parameters in the nonparametric model \citep{kennedy2024spreview, diaz2013doseresp}. As a result, standard influence function-based approaches cannot be applied in a straightforward manner to construct estimators that converge at the parametric rate. To efficiently estimate conditional mediation effects, we can either work within a semiparametric submodel of the full nonparametric model, which requires making strong assumptions about the data generating mechanism, or adopt a fully nonparametric approach and be subject to the curse of dimensionality. We explore both of these options below. 

\subsection{Partially linear mediation model}\label{subsec:partlinear}
In the following section, we assume that $Y$ is continuous. We posit the following partially linear outcome model: 
\begin{align*}
   \mathbb{E}_\mathcal{O}[Y|\bm{X}, A, \bm{M}]  &= \mu_0(\bm{X}) + \Gamma(\bm{X}, A, \bm{M}) \\
    &= \mu_0(\bm{X}) + \gamma(\bm{X},A) + \eta(\bm{X},  \bm{M}) + \xi(\bm{X}, A,  \bm{M})\\
    &= \mu_0(\bm{X}) + A \bm{X}_\gamma \bm{\psi}_\gamma +   \sum_{k=1}^\kappa M_k \bm{X}^{(k)}_\eta \bm{\psi}_\eta^{(k)} +  A   \sum_{k=1}^\kappa M_k \bm{X}^{(k)}_\xi\bm{\psi}_\xi^{(k)}. \numberthis \label{eq:spmodel} 
\end{align*}
The function $\mu_0$ is termed the \textit{treatment-mediator-free} component as it only depends on the covariate vector $\bm{X}$. The second component $\Gamma$ is referred to as the \textit{blip} function and characterizes the joint effect of the exposure and the mediators on the outcome. 
The functions $\gamma, \eta$, and $\xi$ parametrize the effect of treatment, the mediator effects, and the mediator-treatment interactions, respectively.  

When the exposure variable $A$ is continuous, correct specification of the partially linear model will often require the inclusion of additional transformations of $A$ (e.g., polynomial, spline terms). This can easily be accommodated by appending the requisite term to the blip, along with its interactions with the mediators.
Interaction terms between mediators and transformations of mediators can also be represented: for example, to produce an interaction between $M_1$ and $M_2$, we can simply define a new variable $M_{\kappa + 1} = M_1 \cdot M_2$ and include it in the same way as other mediators. However, the inclusion of these extra terms requires some minor modifications to the general estimation procedure outlined in section \ref{subsec:gest}. We outline these changes in appendix A, along with an extension to right-censored survival outcomes.  For the rest of this section, we assume that the model in (\ref{eq:spmodel}) holds exactly, with no interactions between mediators or additional transformations.

Under the partially linear model, conditional direct and indirect effects are given by
\begin{align*}
    \zeta_\mathcal{E}(a; \bm{x}) &= a \Bigl( \bm{x}_\gamma \bm{\psi}_\gamma + \sum_{k=1}^\kappa \mathbb{E}_\mathcal{O}[M_k | \bm{X} = \bm{x}, A = a] \bm{x}^{(k)}_\xi\bm{\psi}_\xi^{(k)}\Bigr), \\ 
     \delta_\mathcal{E}(a; \bm{x}) &=  \sum_{k=1}^\kappa (\mathbb{E}_\mathcal{O}[M_k |\bm{X} = \bm{x}, A = a] - \mathbb{E}_\mathcal{O}[M_k| \bm{X} = \bm{x}, A = 0])\bm{x}^{(k)}_\eta \bm{\psi}_\eta^{(k)}.
\end{align*} 
Furthermore, when identified, mediator-specific indirect effects can be expressed as
\begin{align*}
   \delta_{\mathcal{E}}^k(a; \bm{x}) &=  (\mathbb{E}_\mathcal{O}[M_k |\bm{X} = \bm{x}, A = a] - \mathbb{E}_\mathcal{O}[M_k| \bm{X} = \bm{x}, A = 0])\bm{x}^{(k)}_\eta \bm{\psi}_\eta^{(k)}, \quad k = 1, \dots, \kappa
\end{align*}
In particular, when no mediator interactions are included in the model, $  \delta_{\mathcal{E}}^{\text{INT}} = 0$. Similarly, for the path-specific effects, 
\begin{align*}
     \zeta_{\mathcal{E}, -1}(a; \bm{x}) &= a \Bigl( \bm{x}_\gamma \bm{\psi}_\gamma + \sum_{k=2}^\kappa \mathbb{E}_\mathcal{O}[M_k | \bm{X} = \bm{x}, A = a] \bm{x}^{(k)}_\xi\bm{\psi}_\xi^{(k)}\Bigr), \\ 
    \delta_{\mathcal{E}, -1}(a; \bm{x}) &=  \sum_{k=2}^\kappa (\mathbb{E}_\mathcal{O}[M_k |\bm{X} = \bm{x}, A = a] - \mathbb{E}_\mathcal{O}[M_k| \bm{X} = \bm{x}, A = 0])\bm{x}^{(k)}_\eta \bm{\psi}_\eta^{(k)}.
\end{align*}
We note that all mediation effects only depend on the outcome model through the parameter vector $\bm{\psi} = ( \bm{\psi}_\gamma, \bm{\psi}_\eta,  \bm{\psi}_\xi)$. In the next section, we develop a doubly-robust G-estimator of $\bm{\psi}$ that allows for flexible estimation of the nuisance component $\mu_0$.
\subsection{Doubly robust G-estimation}\label{subsec:gest}
First, define the pseudo-outcome $\Tilde{Y} = Y -  \Gamma(\bm{X}, A, \bm{M})$, which is the outcome with the joint effect of the exposure and mediators removed. 
Estimation of $\bm{\psi}$ is based on the following set of unbiased estimating equations:
\begin{align*}
    \bm{U}_1(\bm{\psi}) &= \bm{X}_\gamma (A- \mathbb{E}_\mathcal{O}[A|\bm{X}]) (\Tilde{Y}- \mu_0(\bm{X})),\\
    \bm{U}_2^{(k)}(\bm{\psi}) &= \bm{X}_\eta^{(k)} (M_k - \mathbb{E}_\mathcal{O}[M_k|\bm{X}, A])(\Tilde{Y}- \mu_0(\bm{X})), \; k = 1, \dots, \kappa, \\
    \bm{U}_3^{(k)}(\bm{\psi}) &= \bm{X}_\xi^{(k)} (A- \mathbb{E}_\mathcal{O}[A|\bm{X}]) (M_k - \mathbb{E}_\mathcal{O}[M_k|\bm{X}, A])(\Tilde{Y}- \mu_0(\bm{X})), \; k = 1, \dots, \kappa.
\end{align*}
The joint estimating function is given by $\bm{U}(\bm{\psi}) = (\bm{U}_1(\bm{\psi}), \bm{U}_2(\bm{\psi}), \bm{U}_3(\bm{\psi}))^\top$, where $\bm{U}_j(\bm{\psi})) = (\bm{U}_j^{(1)}(\bm{\psi}), \dots, \bm{U}_j^{(\kappa)}(\bm{\psi}))^\top$ for $j = 2,3$. 

Since the treatment-mediator-free component $\mu_0$ and the conditional means of the exposure and mediators are unknown, they must be estimated from the data. Let $\bm{\alpha}$ be a possibly infinite dimensional parameter which indexes the nuisance models $\mu_0(\bm{X}; \bm{\alpha})$,  $\mathbb{E}_\mathcal{O}[A|\bm{X}; \bm{\alpha}]$, and $\mathbb{E}_\mathcal{O}[M_k| \bm{X}, A; \bm{\alpha}]$ for $k = 1, \dots, \kappa$. We denote by $\hat{\bm{\alpha}}_n$ the estimate of $\bm{\alpha}$ obtained from the data. The proposed G-estimator $\hat{\bm{\psi}}_n$ solves the vector-valued equation $ \mathbb{P}_n \bm{U}(\bm{\psi}, \hat{\bm{\alpha}}_n) = \bm{0}$,
where $\mathbb{P}_n(.)$ denotes the empirical expectation. The frequentist properties of $\hat{\bm{\psi}}_n$ are described by theorems \ref{thm:estimation} and \ref{thm:asymnorm}, with proofs provided in appendix A.

\begin{theorem}\label{thm:estimation}
    Assuming the semiparametric model in equation (\ref{eq:spmodel}) holds, the G-estimator $\hat{\bm{\psi}}_n$ is consistent for the true value of the parameter $\bm{\psi}_0$ provided either of the following sets of nuisance models is correctly specified:
    \begin{enumerate}
        \item[(i)] The treatment-mediator-free component $\mu_0(\bm{X};\bm{\alpha})$  
        \item[(ii)] The treatment model $\mathbb{E}_\mathcal{O}[A|\bm{X};\bm{\alpha}]$ and the mediator models $\{\mathbb{E}_\mathcal{O}[M_k|\bm{X}, A;\bm{\alpha}]\}_{k=1}^\kappa$.
    \end{enumerate}
\end{theorem} 

\begin{theorem}\label{thm:asymnorm}
Suppose that the estimators $(\hat{\bm{\psi}}_n, \hat{\bm{\alpha}}_n)$ are consistent for $(\bm{\psi}_0, \bm{\alpha}_0)$ and the nuisance parameter estimators $\hat{\bm{\alpha}}_n$ converge at a sufficiently fast rate, i.e., for all $\bm{\psi} \in \Psi$,
\begin{align*}
        \|A - \mathbb{E}_\mathcal{O}[A| \bm{X}; \hat{\bm{\alpha}}_n]\|_2\cdot  \|\Tilde{Y} - \mu_0(\bm{X}; \hat{\bm{\alpha}}_n)\|_2 &= o_p(n^{-1/2}),\\
        \|M_k - \mathbb{E}_\mathcal{O}[M_k| \bm{X},A; \hat{\bm{\alpha}}_n]\|_2\cdot  \|\Tilde{Y} - \mu_0(\bm{X}; \hat{\bm{\alpha}}_n)\|_2 &= o_p(n^{-1/2}), \; k = 1, \dots, \kappa, \\
        \|A - \mathbb{E}_\mathcal{O}[A| \bm{X}; \hat{\bm{\alpha}}_n]\|_3 \cdot \|M_k - \mathbb{E}_\mathcal{O}[M_k| \bm{X},A; \hat{\bm{\alpha}}_n]\|_3\cdot  \|\Tilde{Y} - \mu_0(\bm{X}; \hat{\alpha}_n)\|_3 &= o_p(n^{-1/2}), \; k = 1, \dots, \kappa,
    \end{align*}
where $\|\cdot\|_q$ denotes the $L^q(\mathbb{P}_\mathcal{O})$ norm. Then, under suitable regularity conditions,
    \begin{align*}
        \sqrt{n}(\hat{\bm{\psi}}_n - \bm{\psi}_0) = - V_{\bm{\psi}_0, \bm{\alpha}_0}^{-1} \frac{1}{\sqrt{n}}\sum_{i=1}^n  U_i(\bm{\psi}_0; \bm{\alpha}_0) + o_p(1)
    \end{align*}
    for some invertible matrix $V_{\bm{\psi}_0, \bm{\alpha}_0}$.
\end{theorem}
Regularity conditions are omitted from the statements of the theorems for readability; the full set of assumptions is given in appendix A. Notably, we assume the function class defined by $U(\bm{\psi}, \bm{\alpha})$ is Donsker, which limits the complexity of the unknown nuisance models. Alternatively, cross-fitting allows one to completely bypass the need for complexity restrictions at the cost of an increased computational burden \citep{kennedy2024spreview}. 

The G-estimator $\hat{\bm{\psi}}_n$ is doubly robust; it is consistent for $\bm{\psi}$ provided either the treatment-mediator-free model or the treatment and mediator models are correctly specified. Additionally, $\hat{\bm{\psi}}_n$ exhibits ``rate multiply robustness'', i.e., it is root-$n$ consistent as long as the products of the $L^q$ norms of the nuisance model residuals converge at the parametric rate. This enables the use of flexible nuisance estimators such as random forests or neural nets, which tend to converge at rates slower than root-$n$.

As shown in section \ref{subsec:partlinear}, all mediation effects of interest can be inferred directly from the blip parameters $\bm{\psi}$ and the conditional mediator means $\mathbb{E}_\mathcal{O}[M_k| \bm{X}, A]$. Theorem \ref{thm:semi_rates} summarizes the asymptotic properties of mediation effect estimates based on the G-estimator $\hat{\bm{\psi}}_n$. Its proof can be found in appendix A. 

\begin{theorem}\label{thm:semi_rates}
Suppose the assumptions of Theorem 2 hold. Let $(\hat{\bm{\psi}}_n, \hat{\bm{\alpha}}_n)$ be estimators of the blip parameters and nuisance parameters for the partially linear model. Suppose further that the cost function has the form $  C(a, \bm{x}, \bm{m}) = f(a, \bm{x}) + \sum_{k=1}^\kappa c_k M_k$. Then, for $a,  \in \mathcal{A}$, $\bm{x} \in \mathcal{X}$,
\begin{align*}
    |\hat{\zeta}_\mathcal{E}(a,\bm{x}) - \zeta_\mathcal{E}(a,\bm{x})| = O_p(b_n), \quad 
    |\hat{\delta}_\mathcal{E}(a, \bm{x}) - \delta_\mathcal{E}(a,\bm{x})| &= O_p(b_n),\\
    |\mathbb{E}_\mathcal{O}[C(a, \bm{x}, \bm{M})| \bm{X} = \bm{x}, A = a; \hat{\bm{\alpha}}_n] -\mathbb{E}_\mathcal{O}[C(a, \bm{x}, \bm{M})|\bm{X} = \bm{x}, A = a]| &= O_p(b_n),
\end{align*}
where $ b_n = \max(n^{-1/2}, r_n)$, and
\begin{align*}
   \left| \mathbb{E}[M_k|A = a, \bm{X} = \bm{x}] - \mathbb{E}[M_k|A = a, \bm{X} = \bm{x}; \hat{\bm{\alpha}}_n]\right| = O_p(r_n), \quad \forall k = 1, \dots, \kappa.
\end{align*}
The same result holds for individual mediator and path-specific effects.
\end{theorem}
Theorem \ref{thm:semi_rates} makes clear that despite being able to estimate the blip parameters $\bm{\psi}$ at the parametric rate, the overall convergence rates for the mediation effect estimators are determined by the convergence rates of the mediator mean models. 

\subsection{Bayesian estimation and bootstrap inference}\label{sec:bayesboot}

Robust statistical inference for causal effects can often be achieved under the frequentist paradigm by using standard tools from semiparametric theory. However, full uncertainty quantification for the optimal ITR and functions thereof is complex due to the non-smoothness of the maximization operation required to define the optimal treatment \citep{moodie2010except}. Procedures such as the m-out-of-n bootstrap have been proposed to address these limitations, but these approaches may require additional modelling steps or the use of heuristics \citep{bibhas2013nonreg}.
 
Instead, we opt for a semiparametric Bayesian approach based on the Bayesian bootstrap \citep{rubin1981bayesboot}.
The Bayesian bootstrap allows us to obtain valid finite-sample statistical inference for any estimand $\theta(\mathbb{P}_\mathcal{O}) = \arg \max_u\mathbb{E}_\mathcal{O}[f(O, u)]$ that can be written as an optimizer of an expectation with respect to the data generating distribution $\mathbb{P}_\mathcal{O}$. In particular, this includes any parameter defined as the solution of an estimating equation or as the minimizer of an expected loss. Draws from the posterior of $\theta$  can be obtained by drawing sets of weights $\bm{\omega}^{(1)}, \dots, \bm{\omega}^{(B)} \sim \text{Dirichlet}(1, \dots, 1)$, and computing
\begin{align*}
   \theta^{(b)} = \arg \max_u\sum_{i=1}^n \omega_i^{(b)} f(o_i, u), \quad b = 1, \dots, B.
\end{align*}
The set $\{\theta^{(1)}, \dots, \theta^{(B)}\}$ constitutes a valid set of draws from the posterior $p_n(\theta)$ under the Bayesian bootstrap prior. Although computationally intensive, the Bayesian bootstrap allows for straightforward uncertainty quantification for complex estimands such as the optimal ITR.

\subsection{Bayesian inference for the partially linear model}
The Bayesian bootstrap can be employed to yield valid finite-sample statistical inference for conditional mediation effects under the partially linear model. Under the Bayesian bootstrap prior, sampling from the posterior distribution of the blip parameters $\bm{\psi}$ is straightforward: for every set of drawn Dirichlet weights, fit the nuisance models using the sample weights, and solve the resulting weighted estimating equation. Mediation effects can then be computed for each set of sampled blip parameters and fitted nuisance models. The full algorithm is provided in appendix A. 
\section{Nonparametric estimation}\label{sec:nonpar}
The mediation effects of interest can all be written in terms of integrals of the conditional mean outcome with respect to the conditional distribution of the mediator vector. 
Specifying a parametric form for the blip function in the partially linear model allows for integrals with respect to the mediator vector to be evaluated explicitly by computing moments of mediators and transformations of mediators. Alternatively, one can  estimate the conditional distribution of the mediator vector directly and compute these expectations numerically. However, estimating conditional joint distributions nonparametrically is generally difficult, especially when $\bm{M}$ has continuous components. In this section, we consider approximating the conditional mediator distribution by a discrete distribution, allowing for direct computation of mediation effects. In the statistical learning literature, this task is referred to as \textit{optimal vector quantization} \citep{graf2000foundquant,pages2015introduction}. 

\subsection{Optimal vector quantization}
For the following section, we assume the outcome $Y$ is continuous or binary. We also assume that $\bm{M}$ is absolutely continuous with respect to the Lebesgue measure on $\mathbb{R}^d$. We extend our methodology to the case where $\bm{M}$ has both discrete and continuous components in section \ref{subsec:mixed}. Let $\bm{Z} \in L^2(\mathbb{R}^d, \mathbb{P})$ be an arbitrary absolutely continuous random vector and let $\|\cdot\|$ denote the Euclidean norm on $\mathbb{R}^d$. The goal of optimal quantization is to approximate the distribution of $\bm{Z}$ by a distribution supported on a finite set of points $G = (\bm{g}_1, \dots, \bm{g}_N)$. The set $G$ is known as the \textit{quantizer} or the \textit{quantization grid}, and its cardinality $N$ is the \textit{quantization level}. Associated to any given quantization grid $G$ is the nearest-neighbour projection
\begin{align*}
    \pi_G(\bm{z}) = \sum_{i=1}^N \bm{g}_i \mathds{1}_{S_i(G)}(\bm{z}), \quad S_i(G) = \{\bm{z} \in \mathbb{R}^d: \| \bm{z} - \bm{g}_i\| = \min_{1\le j \le N} \|\bm{z} - \bm{g}_j\| \},
\end{align*}
which maps a point $\bm{z}\in \mathbb{R}^d$ to its closest quantization grid point. The vector $\hat{\bm{Z}} = \pi_G(\bm{Z})$ is the \textit{quantization} of $\bm{Z}$ by $G$, and represents a discrete approximation of $\bm{Z}$ by points in $G$.

Of special interest is conditional quantization, where the objective is to quantize the conditional measure $\mathbb{P}_{\bm{Z}| \bm{V} = \bm{v}}$ for some random vector $\bm{V}$. For any given $\bm{v}$ and quantization level $N$, the $L^2(\mathbb{R}^d)$ optimal conditional quantization grid for $\bm{Z}$ is defined as: 
\begin{align}
    G_{\text{opt}}(\bm{v}) = \arg \min_{G  = (\bm{g}_1, \dots, \bm{g}_N)} \mathbb{E}\Bigl( \min_{1\le j \le N}\| \bm{Z} - \bm{g}_j \|^2 \; \Bigl| \bm{V} = \bm{v} \Bigr).
\end{align}
The optimal quantization grid minimizes the conditional expected distortion between $\bm{Z}$ and an arbitrary quantization of $\bm{Z}$ by a grid $G$.

Given observed data $\{(\bm{Z}_i, \bm{V}_i)\}_{i=1}^n$ and a vector $\bm{v}$, an estimate of the optimal conditional quantization grid $\hat{G}_{\text{opt}}(\bm{v})$ can be computed via algorithm \ref{alg:quant} \citep{loubes2017condquant}.
\begin{algorithm}
    \caption{Conditional quantization}\label{alg:quant}
    \textbf{Input}: Data $(\bm{Z}_1, \bm{V}_1), \dots, (\bm{Z}_n, \bm{V}_n)$, input vector $\bm{v}$, number of nearest neighbours $s$ and number of quantization points $N \le s$.
    \begin{enumerate}
        \item Initialize grid with $\bm{g}^{(0)} = (\bm{g}_1^{(0)}, \dots, \bm{g}_N^{(0)})$.
        \item For $t > 0$, alternate between the following steps until convergence:
        \begin{enumerate}
        \item \textit{Assignment step}: Obtain cluster assignments $I_j^{(t)} = \{1\le i \le n : \|\bm{Z}_i - \bm{g}_j\| = \min_{1\le l\le N}\| \bm{Z}_i - \bm{g}_l\|\}$ for all $j = 1, \dots, N$. 
        \item \textit{Update step}: Update centroids via the weighted mean 
        \begin{align*}
            \bm{g}_j^{(t+1)} = \frac{\sum_{i \in I_j^{(t)}} W_{n,i}(x)\bm{Z}_i }{\sum_{i \in I_j^{(t)}} W_{n,i}(x) }, \quad j = 1, \dots, N,
        \end{align*}
        where $ W_{n,i}(x) = \frac{1}{s} \mathds{1}(\bm{V}_i \text{ is among the }s \text{ nearest neighbours of } \bm{v}).$
    \end{enumerate}
    \end{enumerate}
\end{algorithm}
We note that algorithm \ref{alg:quant} is equivalent to performing $N$-means clustering using only the $s$ nearest neighbours of $\bm{v}$ in the observed data. The measure of the conditional quantization of $\bm{Z}$ by $\hat{G}_{\text{opt}}(\bm{v})$ is then given by 
\begin{align*}
    \hat{\mathbb{P}}_{\bm{Z}| \bm{V}= {\bm{v}}} = \sum_{j=1}^N p_j \mathds{1}_{\{\bm{g}_j\}}, \quad p_j &= \frac{|\{1\le i\le n: \|\bm{Z}_i - \bm{g}_j\| = \min_{1\le l\le N} \|\bm{Z}_i - \bm{g}_l\|\}|}{n}.
\end{align*}
The probability of each quantization point $g_j$ under this measure is simply the proportion of sample points $\{\bm{Z}_i\}_{i=1}^n$ that are closer to $g_j$ than any other point in $\hat{G}_{\text{opt}}$.  

Optimal quantization allows us to compute a discrete approximation of the conditional distribution of the mediator vector $\bm{M}$ given covariates $\bm{X}$ and exposure $\bm{A}$.  Mediation effects can then be computed directly by integrating with respect to this discrete measure. Let $\hat{\mu}_n(\bm{X}, A, \bm{M})$ be some estimate of the  outcome model $\mathbb{E}_\mathcal{O}[Y| \bm{X}, A, \bm{M}]$. For fixed $\bm{x} \in \mathcal{X}, a \in \mathcal{A}$, let $\hat{\mathbb{P}}_{\bm{M}| \bm{X} = \bm{x}, A = a} = \sum_{j=1}^N p_{j}^a \mathds{1}_{\{\bm{g}_{j}^a\}}$ denote the measure of the estimated conditional quantization of $\bm{M}$. Without loss of generality, we assume that the grid size does not vary with the covariates or exposure. Direct effects, indirect effects, and the expected cost can then be estimated as
\begin{align*}
    \hat{\zeta}_\mathcal{E}(\bm{x}, a) &= \sum_{j= 1}^N p_j^a\hat{\mu}_n(\bm{x}, a, \bm{g}_{j}^a)  - \sum_{j= 1}^N p_j^a\hat{\mu}_n(\bm{x}, 0, \bm{g}_{j}^a),\\
    \hat{\delta}_\mathcal{E}(\bm{x}, a) &= \sum_{j= 1}^N p_j^a\hat{\mu}_n(\bm{x}, 0, \bm{g}_{j}^a)  - \sum_{j= 1}^N p_j^0\hat{\mu}_n(\bm{x}, 0, \bm{g}_{j}^0),\\
    \hat{\mathbb{E}}_\mathcal{O}[C(a, \bm{x}, \bm{M})| \bm{X} = \bm{x}, A = a] &= \sum_{j= 1}^N p_j^a C(\bm{x}, a, \bm{g}_{j}^a). 
\end{align*}
Similarly, for $k \in \{1, \dots, \kappa\}$, let $\hat{\mathbb{P}}_{\bm{M}_{-k}| \bm{X} = \bm{x}, A = a}  = \sum_{j=1}^N p_{j, -k}^a \mathds{1}_{\{\bm{g}_{j, -k}^a\}}$,  $\hat{\mathbb{P}}_{M_{k}| \bm{X} = \bm{x}, A = a}  = \sum_{j=1}^N p_{j, k}^a \mathds{1}_{\{g_{j, k}^a\}}$ denote the measures of the conditional quantizations of $\bm{M}_{-k}$ and $M_k$, respectively. Letting $\hat{\mu}_n(\bm{X}, A, \bm{M}) = \hat{\mu}_n(\bm{X}, A, \bm{M}_{-k}, M_k)$, the $k$th individual mediator effect can be estimated as
\begin{align*}
   \hat{\delta}^k_\mathcal{E}(\bm{x}, a) &= \sum_{j= 1}^N \sum_{l=1}^N p_{j, -k}^0 p_{l,k}^a\hat{\mu}_n(\bm{x}, 0, \bm{g}_{j, -k}^0, g_{l, k}^a)  - \sum_{j= 1}^N \sum_{l=1}^N p_{j, -k}^0 p_{l,k}^0\hat{\mu}_n(\bm{x}, 0, \bm{g}_{j, -k}^0, g_{l, k}^0) .
\end{align*}
Path-specific effects can be estimated in an analogous way. Theorem \ref{thm:quant_rate} summarizes the asymptotic properties of mediator effect estimators based on optimal quantization. 

\begin{theorem}\label{thm:quant_rate}
Suppose $\bm{M} \in L^2(\mathbb{R}^{\kappa}, \mathbb{P}_\mathcal{O})$ is a compactly supported random vector with support $\mathcal{M}$. Let $\hat{\mu}_n(\bm{X}, A, \bm{M}) = \hat{\mu}_{0n}(\bm{X}) + \hat{\Gamma}_n(\bm{X}, A, \bm{M})$ denote an estimate of the outcome model $\mathbb{E}_\mathcal{O}[Y| \bm{X}, A, \bm{M}] = \mu_0(\bm{X}) + \Gamma(\bm{X}, A, \bm{M})$. We assume the following regularity conditions: (1) the estimated blip function $\hat{\Gamma}_n$ is a.s.~Lipschitz continuous in $\bm{m}$ over $\mathcal{M}$ with random positive Lipschitz constant $K_n = O_p(1)$, (2) the cost function $C$ is Lipschitz continuous in $\bm{m}$ over $\mathcal{M}$. Furthermore, we assume that for all $\bm{x}\in \mathcal{X}, a \in \mathcal{A}$,
\begin{align*}
    \mathbb{E}_\mathcal{O}\Bigl[ |\hat{\Gamma}_n(\bm{x}, a, \bm{M}) - \Gamma(\bm{x}, a, \bm{M})|\; \Bigr| \bm{X} = \bm{x}, A = a\Bigr] = O_p(c_n), \quad c_n \to 0.
\end{align*}
Fix a number of nearest neighbours $s$ satisfying $s\asymp n^{\frac{2}{d_X + 3}}$, where $d_X = \text{dim}(\bm{X})$, and a quantization level $N \le s$. For $a \in \mathcal{A}, \bm{x} \in \mathcal{X}$, let $\hat{\mathbb{P}}_{\bm{M}| \bm{X} = \bm{x}, A = a} = \sum_{j=1}^N p_{j}^a \mathds{1}_{\{\bm{g}_{j}^a\}}$ denote the conditional quantization of $\bm{M}$ with hyperparameters $N, s$. Then, 
\begin{align*}
     |\hat{\zeta}_\mathcal{E}(a,\bm{x}) - \zeta_\mathcal{E}(a,\bm{x})| = O_p(a_n), \quad  |\hat{\delta}_\mathcal{E}(a,\bm{x}) - \delta_\mathcal{E}(a,\bm{x})| &= O_p(a_n)\\
    |\hat{\mathbb{E}}_\mathcal{O}[C(a, x, \bm{M})| \bm{X} = \bm{x}, A = a] -\mathbb{E}_\mathcal{O}[C(a, x, \bm{M})|\bm{X} = \bm{x}, A = a]| &= O_p(b_n),
\end{align*}
where $a_n = \max(b_n, c_n)$ and $b_n = \max\left(\left(\sqrt{\frac{\log n}{n}}\right)^{\frac{1}{d_X + 3}}, N^{-1/\kappa}\right)$.
Similar rates can be derived for the individual mediator and path-specific effects. 
\end{theorem}
Some technical assumptions have been omitted from the statement of the theorem for readability. The full set of assumptions, the rates for the other mediation effects, as well as the proof of the theorem can be found in appendix A. Details regarding the choice of hyperparameters $N$ and $s$ are also provided. 

The overall convergence rate of the nonparametric quantization estimator depends on both the $L^1$ convergence rate of the estimated outcome model as well as the convergence rate of the quantized mediator distribution. As is typical for nearest neighbour approaches, estimating the conditional distribution of the mediator vector via  quantization suffers from the curse of dimensionality. In this case, the asymptotic convergence rate degrades rapidly as both the dimensionality of the mediator vector and the dimensionality of the covariates increase. Dimensionality reduction methods such as principal components analysis (PCA) can be used to mitigate the slower convergence rate when the dimension of the covariate vector is large \citep{jolliffe2002pca}. Propensity score methods can also be used to reduce the dimension of the conditioning set; this is described in the following section.

\subsection{Extensions}
\subsubsection{Dimensionality reduction via the propensity score}
Due to the curse of dimensionality, the performance of the nonparametric estimator deteriorates as the dimensionality of the covariate vector increases. This is undesirable, as substantial confounder adjustment may be essential to robust causal inference. The use of a balancing score may be employed to improve performance. For this subsection, we limit ourselves to the case where $A$ is binary.

Suppose that we can partition the covariates into two subvectors $\bm{X} = [\bm{U}, \bm{V}]$, the first of which contains confounders and variables predictive of the outcome, and the second including tailoring variables that modify the direct or indirect effect of treatment on the outcome. Formally, we make two homogeneity assumptions. First, we assume that the outcome model $\mu(\bm{X}, A, \bm{M}) = \mathbb{E}_\mathcal{O}[Y|\bm{X}, A, \bm{M}]$ can be partitioned in the following way: $\mu(\bm{X}, A, \bm{M}) = \mu_0(\bm{X}) + \Gamma(\bm{V}, A, \bm{M})$, where $\mu_0, \Gamma$ are arbitrary. Second, we assume that the conditional direct and indirect effects only vary within levels of the subvector $\bm{V}$, i.e., 
\begin{align*}
    \zeta_\mathcal{E}(a,\bm{x}) &=\mathbb{E}_\mathcal{E}[Y(a, \bm{M}(a)) -Y(0, \bm{M}(a)) | \bm{X} = \bm{x}]
    = \mathbb{E}_\mathcal{E}[Y(0, \bm{M}(a)) -Y(0, \bm{M}(0)) | \bm{V} = \bm{v}],\\
     \delta_\mathcal{E}(a,\bm{x}) &=\mathbb{E}_\mathcal{E}[Y(a, \bm{M}(a)) -Y(0, \bm{M}(0)) | \bm{X} = \bm{x}]
    = \mathbb{E}_\mathcal{E}[Y(a, \bm{M}(a)) -Y(0, \bm{M}(0)) | \bm{V} = \bm{v}],
\end{align*}
for all $\bm{x}\in \mathcal{X}, a \in \mathcal{A}$. Now, let $\pi(\bm{x}) = \mathbb{P}_\mathcal{O}(A = 1|\bm{X} = \bm{x})$ denote the propensity score. Since the propensity score is a function of $\bm{X}$, we also have that
\begin{align*}
     \zeta_\mathcal{E}(a,\bm{x}) 
    &= \mathbb{E}_\mathcal{E}[Y(a, \bm{M}(a)) - Y(0, \bm{M}(a)) | \bm{V} = \bm{v}, \pi(\bm{X}) = \pi(\bm{x})],\\
    \delta_\mathcal{E}(a,\bm{x}) 
    &= \mathbb{E}_\mathcal{E}[Y(0, \bm{M}(a))-Y(0, \bm{M}(0)) | \bm{V} = \bm{v}, \pi(\bm{X}) = \pi(\bm{x})].
\end{align*}
Under the distribution $f_\mathcal{E}$, $\pi(\bm{X})$ is a balancing score for the relationship between $\bm{M}$ and $A$, i.e., $\bm{M}(a) \perp A | \pi(\bm{X})$. As a result, direct and indirect effects can be identified as $   \zeta_\mathcal{E}(a,\bm{x}) = Q(a, a, \bm{x}) -Q(0, a, \bm{x})$, $ \delta_\mathcal{E}(a,\bm{x}) =Q(0, a, \bm{x}) -Q(0, 0, \bm{x})$, where 
\begin{align*}
        Q(a, a', \bm{x}) &= \int\mathbb{E}_\mathcal{O}[Y| \bm{X}=\bm{x}, A= a, \bm{M} = \bm{m}] d\mathbb{P}^\mathcal{O}_{\bm{M}|\bm{V} = \bm{v}, \pi(\bm{X}) = \pi(\bm{x}), A = a'}(\bm{m}).  
\end{align*}
The individual mediator effects and the path-specific effects can be identified under equivalent homogeneity assumptions. Identification formulas and the relevant proofs can be found in appendix A. We also provide a short argument detailing the difficulty of extending this approach to the continuous exposure setting. Thus, to compute conditional mediation effects, it suffices to quantize $\bm{M}$ conditional on $\bm{V}$, the exposure $A$ and the propensity score $\pi(\bm{X})$; the resulting convergence rate now scales inversely to  $\dim(\bm{V}) + 1$ instead of $\dim(\bm{X})$. Of course, in an observational study, the true propensity score is unknown and must be estimated from the data. Additional uncertainty induced by the estimation of the propensity score is carried forward using the Bayesian bootstrap.  

\subsubsection{Mixed mediator types}\label{subsec:mixed}
Suppose now that the mediator vector can be partitioned into subvectors $\bm{M} = [\bm{M}^c \; \bm{M}^b]$, which contain the absolutely continuous and binary mediators, respectively. Then, the conditional joint distribution of $\bm{M}$ under $f_\mathcal{E}$ can be written as
\begin{align*}
    f^\mathcal{E}_{\bm{M}| \bm{X}, A}(\bm{m}| \bm{x}, a) &= f_{\bm{M}^c| \bm{X}, A, \bm{M}^b}(\bm{m}^c| \bm{x}, a, \bm{m}^b) f_{\bm{M}^b | \bm{X}, A} (\bm{m}^b | \bm{x}, a)\\
    &=  f_{\bm{M}^c| \bm{X}, A, \bm{M}^b}(\bm{m}^c| \bm{x}, a, \bm{m}^b) \prod_{i = 1}^{\dim(\bm{M}^b)} f_{\bm{M}_i^b | \bm{X}, A, \bar{\bm{M}}^b_{i-1}} (m_i^b | \bm{x}, a, \bar{\bm{m}}^b_{i-1}),
\end{align*}
where $\bar{\bm{M}}^b_{k} = (M^b_1, \dots, M^b_k)$. This yields a simple method for sampling from the conditional joint mediator distribution. First, the binary subvector is sampled by succesively sampling from the one dimensional conditionals of each binary mediator. Then, the continuous subvector can be sampled conditional on the sampled value of the binary subvector. 

We adopt the nested sampling approach to generalize the nonparametric estimator to the mixed mediator case. The joint distribution of the binary subvector can be estimated via any regression-based approach; each of the one dimensional conditionals is amenable to standard regression modelling. The continuous subvector $\bm{M}^c$ can then be quantized conditional on the covariate vector $\bm{X}$, the exposure $A$ and the value of the binary subvector $\bm{M}^b$. Estimation of mediation effects can then be carried out via Monte Carlo simulation. See appendix A for details.  

\subsection{Bayesian inference for the quantization estimator} 
The estimated optimal quantizer of the mediator vector is a random discrete measure. As a result, deriving the asymptotic distribution of mediation effect estimators based on conditional quantization is non-trivial, and standard frequentist inference is difficult. On the other hand, full uncertainty quantification via the Bayesian bootstrap is straightforward. Bayesian bootstrap weights can easily be integrated into the estimation of the optimal quantization grid by multiplying the nearest neighbour weights $W_{n,i}$ by the observation weight $\omega_i$. The full algorithm is presented in appendix A. 

\section{Simulations}\label{sec:sims}
We evaluate the performance of the semiparametric and nonparametric estimators through a varied selection of simulation scenarios.

\subsection{Performance metrics and hyperparameters}

We can assess the overall performance of a given regime $d$ by computing its value $V(d) =\mathbb{E}[Y(d, \bm{M}(d))] - \mathbb{E}[C(d, \bm{M}(d))]$, which is the cost-adjusted expected outcome under $d$, and the regime risk $\text{R}(d) = \mathbb{E}[L(d(\bm{X}), d^{\text{opt}}(\bm{X}))] $, which measures the average discrepancy between $d$ and the true optimal regime for a given loss function $L$. 
$L$ is chosen to be the 0-1 loss $L(a, a') = \mathds{1}(a \ne a')$ for binary exposures, and the squared loss $L(a, a') = (a - a')^2$ for continuous exposures. For the purpose of comparison, we define the uniform regime $ d_U \sim \mathcal{U}(\mathcal{A})$ which assigns treatment uniformly over the support $\mathcal{A}$. We can also compute the average difference between the estimated conditional mediation effects and their true values. Letting $\theta(a, \bm{X})$ be an arbitrary conditional mediation estimand (direct, indirect, etc.), we compute the average mean squared error (AMSE) of the mediation effect estimator $\hat{\theta}(a, \bm{X})$ across all values in the support of the exposure $\mathcal{A}$:
\begin{align*}
      \text{AMSE}(\hat{\theta}) &= \frac{1}{|\mathcal{A}|}\sum_{a\in \mathcal{A}}\mathbb{E}\left(\theta(a,\bm{X}) - \hat{\theta}(a, \bm{X})\right)^2.
\end{align*}
When $A$ is continuous, the AMSE can be computed over a grid of pre-specified  values.

In order to assess the frequentist characteristics of the semiparametric and nonparametric approaches, estimation is performed over $500$ replicate training sets ($n_{\text{train}} = 1000$). For each training set, posterior inference is carried out using $B = 100$ Bayesian bootstrap draws. For each set of drawn weights, the optimal ITR is identified and performance metrics are calculated using the empirical measure of a fixed large independent testing set ($n_{\text{test}} = 10000$). Values of the performance metrics are then averaged over the Bayesian bootstrap draws and the replicate training sets to obtain a final performance indicator.

For both estimators, we assume correct specification of the outcome, treatment, and mediator models as well as the censoring model when applicable. Details of model specification are given in appendix B. For the nonparametric estimator, the number of nearest neighbours $s$ is chosen to ensure that the asymptotic rates of Theorem \ref{thm:quant_rate} hold: $s = 3n^{\frac{2}{\text{dim}(\bm{X}) + 3}}$. The number of quantization points is set to $N = s/3$ in order to balance computational efficiency and performance. Additional simulations exploring the double-robustness of the semiparametric estimator can also be found in appendix B. 

\subsection{Results}
To ease interpretation of the results, we give a brief description of each setting and provide the outcome scale (Median [Q1, Q3]) derived from the test set. Setting 1 involves a binary exposure and a binary outcome, with a single continuous mediator. Only the nonparametric estimator is applicable in this setting. In setting 2, the outcome is continuous with a binary exposure and 3 continuous mediators ($7.83$ [$6.47$, $9.32$]); dimensionality reduction via the propensity score is employed for the nonparametric estimator. Setting 3 covers the censored outcome case with a single binary mediator and a binary exposure, as in the applied data example ($0.11$ [$-0.16$, $1.53$]). The provided outcome scale is for the log transformed survival time. Setting 4 demonstrates ITR estimation for a continuous exposure with two mediators, one binary and the other continuous ($2.37$ [$1.88$, $2.96$]). Setting 5 is a modification of Setting 2 which allows for the estimation of the path-specific ITR ($7.87$ [$6.48$, $9.41$]). For each scenario, we estimate direct, indirect, and individual mediator effects (when applicable and identified). For details regarding the data generating distributions, see appendix B. 
 
Table \ref{tab:simres} displays performance metrics of the ITRs estimated using the semiparametric and nonparametric estimators for all simulation settings. Overall, both approaches yield ITRs that achieve values comparable to that of the true optimal regime and have low regime risk. The performance of the two ITR estimators is comparable, with neither dominating the other across all scenarios. As for estimation of mediation effects, the two methods performed adequately in all settings, producing estimators with moderate to low AMSEs.


\section{Mediated ITR analysis of optimal kidney allocation} \label{sec:application}
\subsection{Description of the problem and data}
We apply the proposed methodology to kidney transplant data from the OPTN database. We analyze data from patients enrolled on the kidney transplantation waiting list from January 1, 2001 to December 31, 2022. The exposure of interest is the donor's HCV status, with 0 denoting an HCV-negative donor and 1 denoting a donor who was infected with HCV. For a given transplant recipient, our objective is to determine the optimal donor HCV status in order to maximize overall survival time. Although we expect HCV-negative kidneys to be optimal for a majority of patients, there are individuals (e.g, recipients already infected with HCV) for whom transplants from HCV-positive donors will not lead to significantly worse outcomes. Identifying these patients would allow for a more efficient allocation of HCV-negative kidneys, with these being allocated to patients who benefit most. Moreover, we hypothesize that a portion of the effect of the donor's HCV status is mediated by graft rejection; we conjecture that a mismatch in donor-recipient HCV status leads to increased rates of graft rejection, which in turn leads to reduced survival times. 

For this problem, we adopt the semiparametric partially linear approach and estimate conditional mediation effects with the binary indicator of graft rejection as a mediator. We use donor age in years (scaled by 10), donor type (living or dead), and recipient HCV status as tailoring covariates in the blip function. We include a simple linear term for the mediator, along with all its three-way interactions with the exposure and covariates. For the sake of simplicity, we conduct a complete-case analysis, removing patients with missing event indicators, missing survival times, or missing covariates.  We note that the goal of the analysis is to illustrate our methodology rather than to provide clinical treatment recommendations. Patient characteristics are presented in appendix C. The final dataset contains 311,474 patients, with 15.9\% experiencing graft rejection,  and a censoring rate of 68\%. 

\subsection{Methodology}
The full dataset was split into a training and testing set, both of equal sizes $(n_{\text{train}} = n_{\text{test}} = 155,737)$. Blip parameters for the partially linear mediation model were estimated in both datasets using $B = 500$ Bayesian bootstrap replicates. Details of nuisance model specification are given in appendix C. The model fitted using the training set was then used to estimate optimal treatment values in the testing set. The optimal ITR based on the training set was then compared to the optimal ITR computed using the coefficients estimated in the testing set. We emphasize that the true optimal ITR is not known and must be estimated in the testing set. Despite the absence of a ground truth, this validation scheme can give some indication regarding the variability of the estimation procedure and its stability when it comes to predicting optimal treatment on a holdout dataset.

\subsection{Results}
\begin{figure}[H]
    \centering
    \includegraphics[]{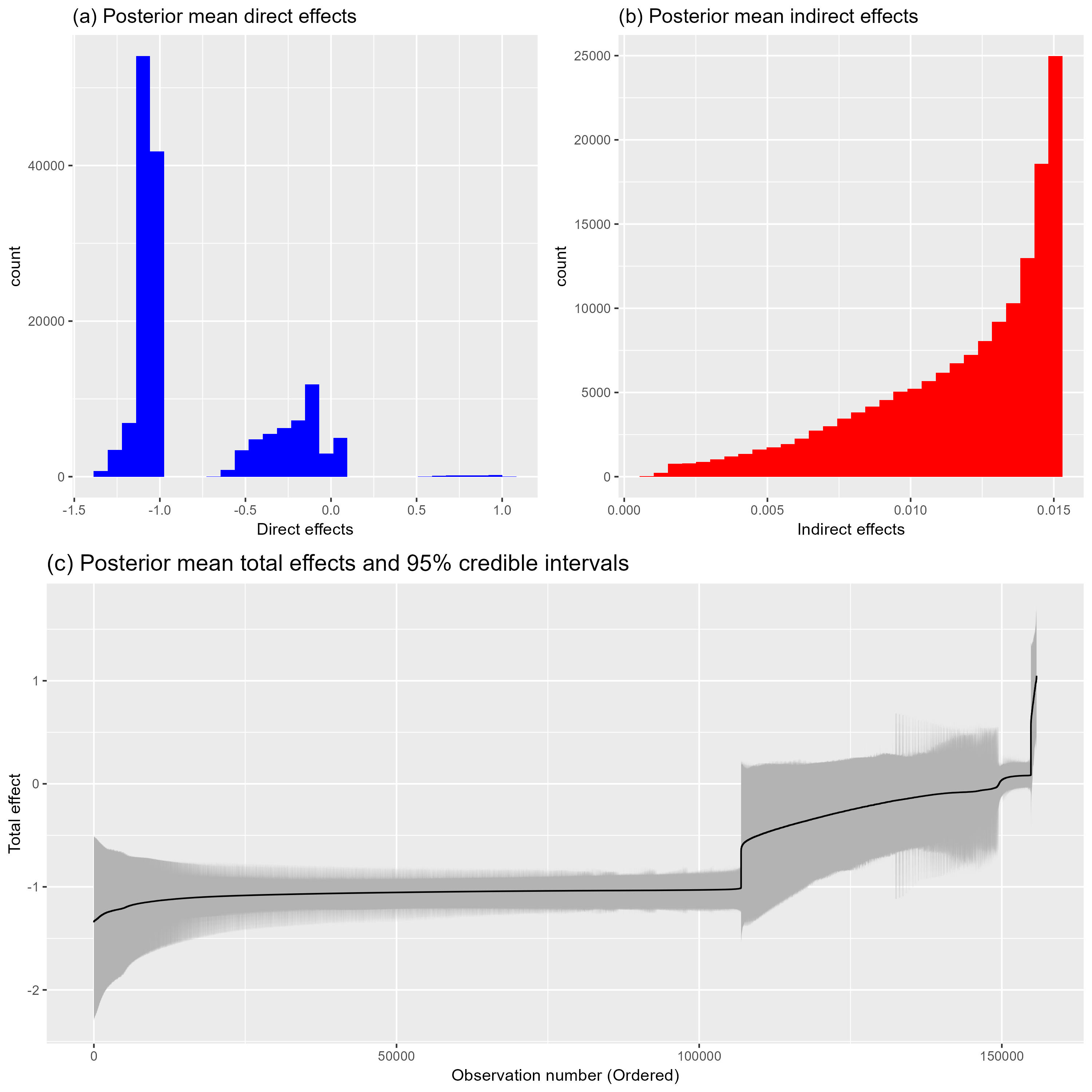}
    \caption{Summary of estimated mediation effects in the testing subsample of $n = 155,737$ subjects from the OPTN data. Distributions of posterior mean direct and indirect effects are presented in panels (a) and (b), respectively. Posterior mean total effects along with their 95\% credible intervals are displayed in panel (c). Note: axes in panels (a)-(c) use different scales.}
    \label{fig:data_analysis}
\end{figure}

Table \ref{tab:data_analysis_estimates} displays posterior means and 95\% credible intervals for blip parameters estimated in the training subsample. Unsurprisingly, receiving a kidney from an HCV-positive donor is associated with a reduction in log survival time; this negative effect is significantly attenuated when the recipient is already HCV-positive. Moreover, graft failure appears to amplify the negative effect, while receiving the kidney from a living donor seemingly lessens it. Donor age had no significant tailoring effect, either statistically or clinically. Figure \ref{fig:data_analysis} presents estimated direct, indirect and total effects for individuals in the testing subsample of the OPTN data. Overall, the indirect effect of donor HCV status on survival through graft rejection was estimated to be negligible relative to its direct effect. For most subjects, the estimated ITRs recommend transplants with HCV-negative donors over transplants with HCV-positive donors, which is consistent with previous literature. However, $31\%$ of patients have estimated total effects with an associated $95$\% credible interval that includes $0$. For these patients, there is no significant evidence of harm in using an HCV-positive donor over an HCV-negative donor as both yield similar outcomes. However, only $3.63\%$ of patients actually received kidneys from an HCV-positive donor. Under the estimated regime, a larger proportion of the study population could have been allocated kidneys from an HCV-positive donor without impacting overall survival. Our findings appear to be robust as the estimated ITR performed well at identifying the optimal treatment on the holdout test set; the posterior mean AMSEs for the direct and indirect effects were $0.070 \;(0.014, 0.21)$ and $1.5\text{x}10^{-4} \;(3.30\text{x}10^{-7}, 5.17\text{x}10^{-4})$, respectively, while the posterior mean risk under the $0$-$1$ loss was $0.13\;    (0.01, 0.28)$.

\section{Discussion}\label{sec:discussion}
In this paper, we developed novel semiparametric and nonparametric estimators for conditional direct, indirect, and individual mediator effects and showcased how they can be used to compute optimal interventional ITRs. We provided asymptotic convergence rates for the two estimators and demonstrated the use of the Bayesian bootstrap to obtain valid finite-sample statistical inference for mediation effects and the resulting optimal treatment allocation. We showcased the performance of our approach in a number of simulations settings, and applied our methodology to the problem of optimal kidney allocation in the presence of mediation by graft rejection. 

Multiple avenues for future research are available. First, one could consider a penalized version of the doubly robust G-estimator for the semiparametric linear model. 
An important weakness of our approach is the reliance on correct specification of the blip model. To make correct specification more plausible, the analyst may be tempted to include a wide selection of candidate tailoring variables and transformations of the exposure and mediator variables. However, when the number of terms in the blip model is large, standard unpenalized estimating equations are not appropriate. This issue can be resolved by considering penalized versions of the doubly-robust estimating equations in section \ref{subsec:gest} \citep{johnson2008penalized}. Bayesian bootstrap-based inference can be carried out in the same way as in the unpenalized case, taking the solution to the penalized estimating equation as the definition of the parameter. Penalization could also be used to perform mediator selection. Second, sharper rates for the conditional quantization estimator could potentially be obtained by considering additional smoothness assumptions. For example, \cite{kennedy2024minimax} derived minimax rates for the conditional average treatment effect (CATE) in terms of the Hölder smoothness of the outcome model, the propensity score model, and the underlying CATE function itself, allowing for near parametric rates of convergence in the high smoothness regime. It would be of interest to derive similar rates for the nonparametric estimator by imposing smoothness conditions on the conditional density of the mediator vector, or directly on the targeted conditional mediation effect.

\begin{center}

{\large\bf SUPPLEMENTARY MATERIAL}

\end{center}

\begin{description}
\item[Appendices A, B, C:] Theoretical results, proofs, and algorithms are provided in appendix A. Appendix B contains simulation details and additional settings. Supplementary information regarding the data analysis is available in appendix C. (pdf)
\end{description}

\bibliography{refs}

\begin{table}[H]
    \centering
    \caption{Performance metrics for nonparametric (Np) and semiparametric (Sp) estimators in simulations settings 1-5. Optimal regimes were estimated over $500$ replicate training sets ($n_{\text{train}} = 1000$) and evaluated on a fixed testing set ($n_{\text{test}} = 10000$). Posterior inference was carried out on each training set using $B = 100$ Bayesian bootstrap draws. Regimes $d^{\text{opt}}, d_U$ are the true optimal regime and the uniform regime, respectively.}
    \begin{tabular}{c|c|cc|cc|cc|cc}
    \hline 
        \multirow{2}{*}{Metrics} & S1 & \multicolumn{2}{c}{S2} & \multicolumn{2}{c}{S3}  & \multicolumn{2}{c}{S4}& \multicolumn{2}{c}{S5}\\
        & Np  & Np & Sp & Np & Sp  & Np& Sp & Np & Sp\\
        \hline
        $\text{AMSE}(\hat{\zeta}_\mathcal{E}; a)$ & 0.01 &0.10 & 0.48 & 0.02 & 0.02 & 0.01 &0.03 &0.08 & 0.38\\
         $\text{AMSE}(\hat{\delta}_\mathcal{E}; a)$ & 0.01 &0.34 & 0.26 & 0.06 & 0.06 & 0.01 & 0.04&0.06 & 0.09\\
         $\text{AMSE}(\hat{\delta}^1_\mathcal{E}; a)$&- &  0.28 & 0.17 &- &- &- & - & - & -\\
         $\text{AMSE}(\hat{\delta}^2_\mathcal{E}; a)$&- &  0.21 & 0.04 &- &- &-& - &0.23 & 0.04\\
         $\text{AMSE}(\hat{\delta}^3_\mathcal{E}; a)$&- &  0.23 & 0.05 &- &- &-& - &0.24 & 0.05\\
         R($\hat{d}^{\text{opt}}$) &0.05  & 0.03 & 0.02 & 0.12 & 0.12 & 0.01 & 0.02&0.03 & 0.08\\
         $V(\hat{d}^{\text{opt}})$ & 0.61 &9.15 & 9.16 &1.03 & 1.03 & 2.43 & 2.42&10.15 & 10.13\\
         \hline 
         $V(d_U)$ & 0.51  & 8.13 & 8.13 & 0.64 & 0.64 & 2.30 & 2.30& 9.10& 9.10 \\
         $V(d^{\text{opt}})$ & 0.62  & 9.17 & 9.17 & 1.04 & 1.04 & 2.45 & 2.45& 10.17 & 10.17 \\
         \hline
    \end{tabular}
    \label{tab:simres}
\end{table}

\begin{table}[H]
    \centering
    \caption{Posterior means and 95\% credible intervals (CIs) for the blip parameters in the partially linear mediation model estimated using a random training subsample of $n = 155,737$ patients from the OPTN data. Credible intervals were computed using $B=500$ Bayesian bootstrap replicates. Donor and recipient variables are prefixed by Don and Rec, respectively.}
    \begin{tabular}{|c|c|c|}
    \hline
     Parameters& 
     Posterior mean & 95\% CI\\
    \hline
    DonHCV & $-0.88$& $(-1.65, -0.16)$ \\
    DonHCV x RecHCV & $1.15$& $(0.91, 1.40)$\\
    DonHCV x DonType & $0.44$& $(-0.50, 1.30)$\\
    DonHCV x DonAge & $-3.33\text{x} 10^{-4}$& $(-1.45\text{x}10^{-3}, 8.82\text{x} 10^{-4})$\\
    GraftFailure & $-0.45$ & $(-0.48, -0.41)$ \\
    GraftFailure x DonHCV & $-0.46$& $(-1.79, 0.71)$ \\
    GraftFailure x DonHCV x RecHCV & $-0.10$& $(-0.53, 0.36)$\\
    GraftFailure x DonHCV x DonType & $0.75$& $(-0.51, 1.91)$\\
    GraftFailure x DonHCV x DonAge & $9.40\text{x} 10^{-4}$& $(-1.00\text{x}10^{-3}, 3.00\text{x} 10^{-3})$\\
    \hline
    \end{tabular}
    \label{tab:data_analysis_estimates}
\end{table}

\end{document}


\title{\bf Supplementary materials: Appendix A}
\date{}
\maketitle

\def\spacingset#1{\renewcommand{\baselinestretch}%
{#1}\small\normalsize} \spacingset{1}
\spacingset{1.8} 

\section{Estimands and identifying assumptions (example)}
We provide a concrete example to illustrate some of the identification conditions required for estimation of conditional mediation effects. Assume the following partially linear model holds:
\begin{align*}
    \mathbb{E}[Y| A, X, M_1, M_2, M_3] &= \mu_0(X) + \gamma A + \gamma_2 AX + \eta_1 M_1 + \eta_2 M_2 + \eta_3 M_3 + \xi_1 AM_1 + \xi_2 AM_2 + \xi_3 AM_3,
\end{align*}
along with the following nonparametric specification for the mediators:
\begin{align*}
    M_1 &= f_1(X, A) + \varepsilon_1\\
    M_2 &= f_2(X, A) + \varepsilon_2\\
    M_3 &= f_3(X, A) + \varepsilon_3,
\end{align*}
where the $f_i$ functions are deterministic and the error vector $\bm{\varepsilon} = (\varepsilon_1, \varepsilon_2, \varepsilon_3)^\top$ has an unspecified dependence structure. 
\subsection{Individual mediator indirect effects}
For the above model, the assumption needed to identify the individual mediator effects
\begin{align*}
     &\mathbb{E}_\mathcal{E}[Y(0, \bm{m}_{-k}, M_k(a)) - Y(0, \bm{m}_{-k}, M_k(0)) |\bm{M}_{-k}(0) = \bm{m}_{-k}, \bm{X}] =\\    &\quad \mathbb{E}_\mathcal{E}[Y(0,  \bm{m}_{-k}, M_k(a)) - Y(0, \bm{m}_{-k}, M_k(0) ) |\bm{X}], \quad \forall k = 1, \dots, \kappa,
\end{align*}
is equivalent to 
\begin{align*}
    \eta_k\mathbb{E}_\mathcal{E}[M_k(a) - M_k(0) | \bm{M}_{-k}(0) = \bm{m}_{-k}, \bm{X}] =     \eta_k\mathbb{E}_\mathcal{E}[M_k(a) - M_k(0) |\bm{X}],\quad k = 1, \dots, 3 .
\end{align*}
This assumption holds for the chosen mediator specification as for all $k = 1, \dots, 3$,
\begin{align*}
    M_k(a) - M_k(0) = f_k(X, a) + \varepsilon_k - (f_k(X, 0) + \varepsilon_k) = f_k(X, a) - f_k(X, 0),
\end{align*}
which does not depend on the error vector $\bm{\varepsilon}$, and is thus independent of the other mediators and their potential outcomes.
\subsection{Path-specific effects}
The assumption required to identify the path-specific direct and indirect effects 
\begin{align*}
    &\mathbb{E}_\mathcal{E}[Y(a, \bm{M}_{-1}(a), m_1) - Y(0, \bm{M}_{-1}(0), m_1) | M_1(0) = m_1, \bm{X}] =\\    &\quad \mathbb{E}_\mathcal{E}[Y(a, \bm{M}_{-1}(a), m_1) - Y(0, \bm{M}_{-1}(0), m_1) |\bm{X}] 
\end{align*}
is equivalent to:
\begin{align*}
    \eta_k \mathbb{E}_\mathcal{E}[M_k(a) - M_k(0)|M_1(0) = m_1, \bm{X}] &=  \eta_k\mathbb{E}_\mathcal{E}[M_k(a) - M_k(0)| \bm{X}],\\
    \xi_k \mathbb{E}_\mathcal{E}[M_k(a)|M_1(0) = m_1, \bm{X}] &=  \xi_k \mathbb{E}_\mathcal{E}[M_k(a)|\bm{X}], \quad k = 2, 3.
\end{align*}
The first condition holds for the chosen mediator specification, as shown above. However, the second condition only holds in full generality (regardless of coefficient values) provided a stronger independence assumption holds, such as
\begin{align*}
    M_k(a) \perp M_1(0) | \bm{X} \iff f_k(X, a) + \varepsilon_k \perp f_1(X, 0) + \varepsilon_1 | X\iff \varepsilon_k \perp \varepsilon_1 | X,
\end{align*}
i.e., if $M_k$ and $M_1$ are conditionally independent given $X$. This is considerably stronger than the assumption required to identify individual mediator indirect effects.

For additional discussion of identifiability conditions for multiple mediation, see for example \cite{taguri2018multmed}, \cite{bellavia2018decomposition}, and \cite{zhou2022ordermed}.

\section{Semiparametric estimator}
\subsection{Mediator-mediator interactions and other transformations}
Without loss of generality, we consider the case where the outcome model contains all first order interactions between mediators. In this case, the outcome model has the form
\begin{align*}\label{eq:out_model}
   \mathbb{E}_\mathcal{O}[Y|\bm{X}, A, \bm{M}]  
    &= \mu_0(\bm{X}) + A \bm{X}_\gamma \bm{\psi}_\gamma +   \sum_{k=1}^\kappa M_k \bm{X}^{(k)}_\eta \bm{\psi}_\eta^{(k)} +\sum_{i\ne j} M_i M_j \bm{X}^{(ij)}_\eta \bm{\psi}_\eta^{(ij)}\\ &+   A   \sum_{k=1}^\kappa M_k \bm{X}^{(k)}_\xi\bm{\psi}_\xi^{(k)} + A\sum_{i\ne j} M_i M_j \bm{X}^{(ij)}_\xi\bm{\psi}_\xi^{(ij)} .
\end{align*}
Under this model, direct, indirect, and mediator-specific mediation effects can be represented in the following manner:
\begin{align*}
     \zeta_\mathcal{E}(a; \bm{x}) &= a \Bigl( \bm{x}_\gamma \bm{\psi}_\gamma + \sum_{k=1}^\kappa \mathbb{E}_\mathcal{O}[M_k | \bm{X} = \bm{x}, A = a] \bm{x}^{(k)}_\xi\bm{\psi}_\xi^{(k)} + \sum_{i\ne j} \mathbb{E}_\mathcal{O}[M_iM_j | \bm{X} = \bm{x}, A = a] \bm{x}^{(ij)}_\xi\bm{\psi}_\xi^{(ij)}\Bigr), \\ 
     \delta_\mathcal{E}(a; \bm{x}) &=  \sum_{k=1}^\kappa (\mathbb{E}_\mathcal{O}[M_k |\bm{X} = \bm{x}, A = a] - \mathbb{E}_\mathcal{O}[M_k| \bm{X} = \bm{x}, A = 0])\bm{x}^{(k)}_\eta \bm{\psi}_\eta^{(k)} \\
     &+ \sum_{i\ne j}\Bigl(\mathbb{E}_\mathcal{O}[M_i M_j |\bm{X} = \bm{x}, A = a] - \mathbb{E}_\mathcal{O}[M_i M_j| \bm{X} = \bm{x}, A = 0]\Bigr)\bm{x}^{(ij)}_\eta \bm{\psi}_\eta^{(ij)},\\
      \delta_{\mathcal{E}}^k(a; \bm{x}) &=  (\mathbb{E}_\mathcal{O}[M_k |\bm{X} = \bm{x}, A = a] - \mathbb{E}_\mathcal{O}[M_k| \bm{X} = \bm{x}, A = 0])\bm{x}^{(k)}_\eta \bm{\psi}_\eta^{(k)}\\
     &+ \sum_{j \ne k}  \mathbb{E}_\mathcal{O}[M_j |\bm{X} = \bm{x}, A = 0] \cdot (\mathbb{E}_\mathcal{O}[M_k |\bm{X} = \bm{x}, A = a] - \mathbb{E}_\mathcal{O}[M_k| \bm{X} = \bm{x}, A = 0])\bm{x}^{(kj)}_\eta \bm{\psi}_\eta^{(kj)}.
\end{align*}
In this case, the interaction effect $\delta^{\text{INT}}$ is no longer null, and can be estimated by subtracting the sum of mediator-specific effects from the total indirect effect. Path-specific effects can be written in an analogous fashion. 
Estimation of this partially linear model can be performed in the same way as in the standard case using the relabeling strategy described in section 5.1.
The case where the model includes transformations of mediators follows in the same way. 

Now, suppose the outcome model includes transformations of the exposure $A$, i.e., there is a finite set of basis functions $f_1, \dots, f_l: \mathcal{A} \to \mathbb{R}$ such that 
\begin{align*}
     \mathbb{E}_\mathcal{O}[Y|\bm{X}, A, \bm{M}]  
    &= \mu_0(\bm{X}) + \sum_{k=1}^\kappa M_k \bm{X}^{(k)}_\eta \bm{\psi}_\eta^{(k)} +  \sum_{i =1 }^lf_i(A) \Bigl(\bm{X}_\gamma^{(i)} \bm{\psi}^{(i)}_\gamma  +    \sum_{k=1}^\kappa M_k \bm{X}^{(ik)}_\xi\bm{\psi}_\xi^{(ik)}\Bigr).
\end{align*}
For this model, mediation effects can be expressed in the following way:
\begin{align*}
      \zeta_\mathcal{E}(a; \bm{x}) &= \sum_{i =1 }^lf_i(a) \Bigl(\bm{x}_\gamma^{(i)} \bm{\psi}^{(i)}_\gamma  +    \sum_{k=1}^\kappa \mathbb{E}_\mathcal{O}[M_k | \bm{X} = \bm{x}, A = a]  \bm{x}^{(ik)}_\xi\bm{\psi}_\xi^{(ik)}\Bigr)\\ &-  \sum_{i =1 }^lf_i(0) \Bigl(\bm{x}_\gamma^{(i)} \bm{\psi}^{(i)}_\gamma  +    \sum_{k=1}^\kappa \mathbb{E}_\mathcal{O}[M_k | \bm{X} = \bm{x}, A = a] \bm{x}^{(ik)}_\xi\bm{\psi}_\xi^{(ik)}\Bigr),\\
       \delta_\mathcal{E}(a; \bm{x}) &= \sum_{k=1}^\kappa (\mathbb{E}_\mathcal{O}[M_k |\bm{X} = \bm{x}, A = a] - \mathbb{E}_\mathcal{O}[M_k| \bm{X} = \bm{x}, A = 0])\bm{x}^{(k)}_\eta \bm{\psi}_\eta^{(k)} \\
       &+  \sum_{i =1 }^l \sum_{k=1}^\kappa f_i(0) (\mathbb{E}_\mathcal{O}[M_k |\bm{X} = \bm{x}, A = a] - \mathbb{E}_\mathcal{O}[M_k| \bm{X} = \bm{x}, A = 0]) \bm{x}^{(ik)}_\xi\bm{\psi}_\xi^{(ik)},\\
         \delta_{\mathcal{E}}^k(a; \bm{x}) &=(\mathbb{E}_\mathcal{O}[M_k |\bm{X} = \bm{x}, A = a] - \mathbb{E}_\mathcal{O}[M_k| \bm{X} = \bm{x}, A = 0])\bm{x}^{(k)}_\eta \bm{\psi}_\eta^{(k)} \\
       &+  \sum_{i =1 }^l f_i(0) (\mathbb{E}_\mathcal{O}[M_k |\bm{X} = \bm{x}, A = a] - \mathbb{E}_\mathcal{O}[M_k| \bm{X} = \bm{x}, A = 0]) \bm{x}^{(ik)}_\xi\bm{\psi}_\xi^{(ik)} .
\end{align*}
Unlike for mediator-mediator interactions and transformations of mediators, the estimation procedure must also be modified when the model contains transformations of the exposure. For basis function index $i = 1, \dots, l$, form the set of estimating equations:
\begin{align*}
    \bm{U}^{(i)}_1(\bm{\psi}) &= \bm{X}_\gamma (f_i(A)- \mathbb{E}_\mathcal{O}[f_i(A)|\bm{X}]) (\Tilde{Y}- \mu_0(\bm{X})),\\
    \bm{U}_2^{(k)}(\bm{\psi}) &= \bm{X}_\eta^{(k)} (M_k - \mathbb{E}_\mathcal{O}[M_k|\bm{X}, A])(\Tilde{Y}- \mu_0(\bm{X})), \; k = 1, \dots, \kappa, \\
    \bm{U}_3^{(ik)}(\bm{\psi}) &= \bm{X}_\xi^{(k)} (f_i(A)- \mathbb{E}_\mathcal{O}[f_i(A)|\bm{X}]) (M_k - \mathbb{E}_\mathcal{O}[M_k|\bm{X}, A])(\Tilde{Y}- \mu_0(\bm{X})), \; k = 1, \dots, \kappa.
\end{align*}
We define the joint estimating function $\bm{U}(\bm{\psi}) = (\bm{U}_1(\bm{\psi}), \bm{U}_2(\bm{\psi}), \bm{U}_3(\bm{\psi}))^\top$, where $\bm{U}_1(\bm{\psi})) = (\bm{U}_1^{(1)}(\bm{\psi}), \dots, \bm{U}_1^{(l)}(\bm{\psi}))^\top$, $\bm{U}_2(\bm{\psi})) = (\bm{U}_2^{(1)}(\bm{\psi}), \dots, \bm{U}_2^{(\kappa)}(\bm{\psi}))^\top$, and  $\bm{U}_3(\bm{\psi})) = (\bm{U}_3^{(11)}(\bm{\psi}), \dots, \bm{U}_3^{(l\kappa)}(\bm{\psi}))^\top$. The full blip parameter vector $\bm{\psi}$ can be estimated by solving the joint set of estimating equations. 
\subsection{Proof of Theorem 1}
We first show that $\bm{U}$ is an unbiased estimating equation provided either the treatment-free component of the outcome model $\mu_0(\bm{X}; \bm{\alpha})$, or the treatment model $\mathbb{E}_\mathcal{O}[A|X;\bm{\alpha}]$ and the mediator models $\mathbb{E}_\mathcal{O}[M_k|\bm{X}, A;\bm{\alpha}], \; k = 1, \dots, \kappa$, are correctly specified.  For $k = 1, \dots, \kappa$,
\begin{align*}
     \bm{U}_1(\bm{\psi}) &= \bm{X}_\gamma (A- \mathbb{E}_\mathcal{O}[A|\bm{X};\bm{\alpha}]) (Y- \Gamma(\bm{X}, A, \bm{M}) - \mu_0(\bm{X};\bm{\alpha})),\\
      \bm{U}_2^{(k)}(\bm{\psi}) &= \bm{X}_\eta^{(k)} (M_k - \mathbb{E}_\mathcal{O}[M_k|\bm{X}, A;\bm{\alpha}])(Y-\Gamma(\bm{X}, A, \bm{M}) -\mu_0(\bm{X};\bm{\alpha})), \\
      \bm{U}_3^{(k)}(\bm{\psi}) &= \bm{X}_\xi^{(k)} (A- \mathbb{E}_\mathcal{O}[A|\bm{X}];\bm{\alpha}) (M_k - \mathbb{E}_\mathcal{O}[M_k|\bm{X}, A;\bm{\alpha}])(Y- \Gamma(\bm{X}, A, \bm{M})-\mu_0(\bm{X};\bm{\alpha})).
\end{align*}
Under correct specification of the blip function, we have that
\begin{align*}
       \mathbb{E}_\mathcal{O}[(Y- \Gamma(\bm{X}, A, \bm{M})-\mu_0(\bm{X};\bm{\alpha}))|\bm{X}, A, \bm{M}] &= \mu_0(\bm{X}) - \mu_0(\bm{X};\bm{\alpha}).
\end{align*}
As a result, for the first equation, 
\begin{align*}
     \mathbb{E}_\mathcal{O}[\bm{U}_1(\bm{\psi})|\bm{X}] &=  \mathbb{E}_\mathcal{O}[\mathbb{E}_\mathcal{O}[\bm{U}_1(\bm{\psi})| \bm{X},A, \bm{M}]|\bm{X}]\\
     &=\bm{X}_\gamma \mathbb{E}_\mathcal{O}\Bigl[(A- \mathbb{E}_\mathcal{O}[A|\bm{X}; \bm{\alpha}]) \Bigr| \bm{X}\Bigr](\mu_0(\bm{X}) - \mu_0(\bm{X};\bm{\alpha})) \\
     &= \bm{X}_\gamma (\mathbb{E}_\mathcal{O}[A|\bm{X}]- \mathbb{E}_\mathcal{O}[A|\bm{X}; \bm{\alpha}])(\mu_0(\bm{X}) - \mu_0(\bm{X};\bm{\alpha})).
\end{align*}
By the tower property, $\bm{U}_1$ is unbiased provided either the treatment model or the outcome model is correctly specified. 

For the second equation, for $k = 1, \dots, \kappa$,
\begin{align*}
     \mathbb{E}_\mathcal{O}[\bm{U}_2^{(k)}(\bm{\psi})|\bm{X}, A] &=   \mathbb{E}_\mathcal{O}[\mathbb{E}_\mathcal{O}[\bm{U}_2^{(k)}(\bm{\psi})|\bm{X}, A, \bm{M}]|\bm{X}, A]\\
      &= \bm{X}_\eta^{(k)} \mathbb{E}_\mathcal{O}\Bigl[(M_k - \mathbb{E}_\mathcal{O}[M_k|\bm{X}, A; \bm{\alpha}])\Bigr| \bm{X}, A\Bigr](\mu_0(\bm{X}) - \mu_0(\bm{X};\bm{\alpha}))\\
      &= \bm{X}_\eta^{(k)} (\mathbb{E}_\mathcal{O}[M_k| \bm{X}, A] - \mathbb{E}_\mathcal{O}[M_k|\bm{X}, A; \bm{\alpha}])(\mu_0(\bm{X}) - \mu_0(\bm{X};\bm{\alpha})).
\end{align*}
By the tower property, $\bm{U}_2^{(k)}$ is unbiased provided either the $k$th mediator model or the outcome model is correctly specified.

For the third equation, for $k = 1, \dots, \kappa$,
\begin{align*}
\mathbb{E}_\mathcal{O}[\bm{U}_3^{(k)}(\bm{\psi})|\bm{X}, A] &=   \mathbb{E}_\mathcal{O}[\mathbb{E}_\mathcal{O}[\bm{U}_3^{(k)}(\bm{\psi})|\bm{X}, A, \bm{M}]|\bm{X}, A]\\
&= \bm{X}_\xi^{(k)} \mathbb{E}_\mathcal{O}\Bigl[(M_k - \mathbb{E}_\mathcal{O}[M_k|\bm{X}, A; \bm{\alpha}])\Bigr| \bm{X}, A\Bigr](\mu_0(\bm{X}) - \mu_0(\bm{X};\bm{\alpha}))\\
&= \bm{X}_\xi^{(k)} (A- \mathbb{E}_\mathcal{O}[A|\bm{X}; \bm{\alpha}]) \mathbb{E}_\mathcal{O}\Bigl[(M_k - \mathbb{E}_\mathcal{O}[M_k|\bm{X}, A; \bm{\alpha}])\Bigr| \bm{X}, A\Bigr](\mu_0(\bm{X}) - \mu_0(\bm{X};\bm{\alpha}))\\
&= \bm{X}_\xi^{(k)} (A- \mathbb{E}_\mathcal{O}[A|\bm{X}; \bm{\alpha}]) (\mathbb{E}_\mathcal{O}[M_k| \bm{X}, A] - \mathbb{E}_\mathcal{O}[M_k|\bm{X}, A; \bm{\alpha}])(\mu_0(\bm{X}) - \mu_0(\bm{X};\bm{\alpha})).
\end{align*}
By the tower property, $\bm{U}_3^{(k)}$ is unbiased provided either the $k$th mediator model or the outcome model is correctly specified.

Consistency of the G-estimator then follows by standard arguments. One possible set of assumptions that could be used to ensure consistency is: (1) $\Psi$ is compact, (2) the equation $\mathbb{E}_\mathcal{O}[\bm{U}(\bm{\psi}, \bm{\alpha}_0)] = \bm{0}$ is uniquely solved at $\bm{\psi}_0$, (3) $\sup_{\psi \in \Psi} \Bigl\|\mathbb{P}_n \bm{U}(\bm{\psi}, \hat{\bm{\alpha}}_n) - \mathbb{E}_\mathcal{O}[\bm{U}(\bm{\psi}, \bm{\alpha}_0)]\Bigr\|\to 0$ \citep{newey1994large, Vaart_1998}.

Uniqueness of the solution to the estimating equation with known nuisance component $\bm{\alpha}_0$,  $\mathbb{E}_\mathcal{O}[\bm{U}(\bm{\psi}, \bm{\alpha}_0)] = \bm{0}$ follows directly by observing that $\bm{U}(\bm{\psi}, \bm{\alpha}_0)$ is equivalent to the estimating equation used by the standard G-estimator for a multivariate exposure, with modified weights \citep{robins1992gest, stephens2015gest}. Consequently, it is also equivalent to a weighted version of the equation used by the standard generalized estimating equations (GEE) estimator \citep{zeger1986}.

Alternatively, when the exposure $A$ is binary, existing semiparametric methods can be employed to construct asymptotically efficient estimators of conditional mediation effects \citep{tchetgen2012semiparametric, tchetgen2014condmed}. 

\subsection{Regularity conditions for Theorem 2}
Suppose $\bm{\psi}_0 \in (\mathbb{R}^p, \|\cdot\|)$, and $\alpha_0$ belong to some metric space $(\mathcal{A}, \rho)$. 
We assume that the class of functions $\{\bm{U}(\bm{\psi}; \bm{\alpha}): \|\bm{\psi} - \bm{\psi}_0\| < \delta, \rho(\bm{\alpha}, \bm{\alpha}_0) < \delta\}$ is Donsker for some $\delta > 0$, and $\mathbb{E}_\mathcal{O} \| \bm{U}(\bm{\psi}, \bm{\alpha}) - \bm{U}(\bm{\psi}_0, \bm{\alpha}_0)\|^2 \to 0$ as $(\bm{\psi}, \bm{\alpha}) \to (\bm{\psi}_0, \bm{\alpha}_0)$. Additionally, assume that the map $\bm{\psi} \to \mathbb{E}_\mathcal{O} [\bm{U}(\bm{\psi}, \bm{\alpha})]$ is differentiable at $\bm{\psi}_0$, uniformly in $\bm{\alpha}$ in a neighbourhood of $\bm{\alpha}_0$, with non-singular derivative matrix $V_{\bm{\psi}_0, \bm{\alpha}}$ such that $V_{\bm{\psi}_0, \bm{\alpha}} \to V_{\bm{\psi}_0, \bm{\alpha}_0}$ as $\bm{\alpha} \to \bm{\alpha}_0$. We also assume the covariate vectors $\bm{X}_\gamma$, $\bm{X}_\eta$, $\bm{X}_\eta$ are bounded almost surely. 

\subsection{Proof of Theorem 2}
  Without loss of generality, we assume that $\text{dim}(\bm{X}_\gamma) = \text{dim}(\bm{X}_\eta) =\text{dim}(\bm{X}_\xi) =1$. For non-trivial design matrices, the following proof can be applied for each coordinate of the resulting estimating equations. Under the regularity conditions stated in the previous section, it suffices to show that the drift term $\mathbb{E}_\mathcal{O} [\bm{U}(\bm{\psi}_0, \hat{\bm{\alpha}}_n)]  = o_p(n^{-1/2})$. Then, the result follows directly from theorem 5.31 of \cite{Vaart_1998}.  

  Let $L > 0$ be some positive constant such that $|X_\gamma| +  |X_\eta|$ +  $|X_\eta| \le L$  almost surely. This constant exists as the covariates are assumed to be bounded with probability 1. Let $\|\cdot\|_q$ denote the $L^q(\mathbb{P}_\mathcal{O})$ norm. Then, by the generalized Holder's inequality, we that the following inequalities hold almost surely
    \begin{align*}
        \mathbb{E}_\mathcal{O}[ U_1(\bm{\psi}_0, \hat{\bm{\alpha}}_n)] &\le L \|A - \mathbb{E}_\mathcal{O}[A| \bm{X}; \hat{\alpha}_n]\|_2\cdot  \|\Tilde{Y} - \mu_0(\bm{X}; \hat{\alpha}_n)\|_2, \\
       \mathbb{E}_\mathcal{O}[ U_2^{(k)}(\bm{\psi}_0, \hat{\bm{\alpha}}_n)]  &\le L \|M_k - \mathbb{E}_\mathcal{O}[M_k| \bm{X},A; \hat{\alpha}_n]\|_2\cdot  \|\Tilde{Y} - \mu_0(\bm{X}; \hat{\alpha}_n)\|_2,  \; k = 1, \dots, \kappa, \\
       \mathbb{E}_\mathcal{O}[ U_3^{(k)}(\bm{\psi}_0, \hat{\bm{\alpha}}_n)]  &\le L \|A - \mathbb{E}_\mathcal{O}[A| \bm{X}; \hat{\alpha}_n]\|_3 \cdot \|M_k - \mathbb{E}_\mathcal{O}[M_k| \bm{X},A; \hat{\alpha}_n]\|_3\cdot  \|\Tilde{Y} - \mu_0(\bm{X}; \hat{\alpha}_n)\|_3, \; k = 1, \dots, \kappa.
    \end{align*}
    Thus, provided the nuisance models converge at the rates described in the statement of theorem 2, the drift term satisfies $\mathbb{E}_\mathcal{O} [\bm{U}(\bm{\psi}_0, \hat{\bm{\alpha}}_n)]  = o_p(n^{-1/2})$.
\subsection{Proof of Theorem 3}
Under the semiparametric partially linear model,  we have the following representation for the direct and indirect effects:
\begin{align*}
    \zeta_\mathcal{E}(a; \bm{x}) &= a \Bigl( \bm{x}_\gamma \bm{\psi}_\gamma + \sum_{k=1}^\kappa \mathbb{E}_\mathcal{O}[M_k | \bm{X} = \bm{x}, A = a] \bm{x}^{(k)}_\xi\bm{\psi}_\xi^{(k)}\Bigr), \\ 
     \delta_\mathcal{E}(a; \bm{x}) &=  \sum_{k=1}^\kappa (\mathbb{E}_\mathcal{O}[M_k |\bm{X} = \bm{x}, A = a] - \mathbb{E}_\mathcal{O}[M_k| \bm{X} = \bm{x}, A = 0])\bm{x}^{(k)}_\eta \bm{\psi}_\eta^{(k)}.
\end{align*} 
Without loss of generality, we assume that the design matrices in the partially linear model are one-dimensional, i.e., $\text{dim}(\mathbf{X}_\gamma) = \text{dim}(\mathbf{X}_\eta) =\text{dim}(\mathbf{X}_\xi) =1$. Then, for any fixed $a\in \mathcal{A}, x \in \mathcal{X}$, using the triangle inequality, we have that
\begin{align*}
    |\zeta_\mathcal{E}(a, x) - \hat{\zeta}_\mathcal{E}(a, x)| &\le |ax_\gamma |\cdot \left|\psi_\gamma - \hat{\psi}_\gamma \right|\\&+\sum_{k = 1}^\kappa \Bigl|\mathbb{E}_\mathcal{O}[M_k|A = a, \bm{X} = \bm{x}] - \mathbb{E}_\mathcal{O}[M_k|A = a, \bm{X} = \bm{x}; \hat{\bm{\alpha}}_n] \Bigr| \cdot |a x^{(k)}_\xi \psi_\xi^{(k)}| \\ &+ \sum_{k = 1}^\kappa \left|\mathbb{E}_\mathcal{O}[M_k|A = a, \bm{X} = \bm{x}; \hat{\bm{\alpha}}_n]a x^{(k)}_\xi \right| \cdot \left| \psi^{(k)}_\xi - \hat{\psi}_\xi^{(k)}\right|\\
    &= O_p(c_n)+O_p(n^{-1/2}) , \\
    |\delta_\mathcal{E}(a, x) - \hat{\delta}_\mathcal{E}(a, x)| &\le |ax_\gamma |\cdot \left|\psi_\gamma - \hat{\psi}_\gamma \right|\\&+\sum_{k = 1}^\kappa \Bigl|\mathbb{E}_\mathcal{O}[M_k|A = a, \bm{X} = \bm{x}] - \mathbb{E}_\mathcal{O}[M_k|A = a, \bm{X} = \bm{x}; \hat{\bm{\alpha}}_n] \Bigr| \cdot  |x^{(k)}_\eta \psi_\eta^{(k)}| \\
    &+\sum_{k = 1}^\kappa \Bigl|\mathbb{E}_\mathcal{O}[M_k|A = 0, \bm{X} = \bm{x}] - \mathbb{E}_\mathcal{O}[M_k|A = 0, \bm{X} = \bm{x}; \hat{\bm{\alpha}}_n] \Bigr| \cdot  |x^{(k)}_\eta \psi_\eta^{(k)}| \\
    &+ \sum_{k = 1}^\kappa \left|\mathbb{E}_\mathcal{O}[M_k|A = a, \bm{X} = \bm{x}; \hat{\bm{\alpha}}_n]x^{(k)}_\eta\right|\cdot \left| \psi^{(k)}_\eta - \hat{\psi}_\eta^{(k)}\right|\\
    &+ \sum_{k = 1}^\kappa \left|\mathbb{E}_\mathcal{O}[M_k|A = 0, \bm{X} = \bm{x}; \hat{\bm{\alpha}}_n]x^{(k)}_\eta\right|\cdot \left| \psi^{(k)}_\eta - \hat{\psi}_\eta^{(k)}\right|\\
    &= O_p(c_n)+O_p(n^{-1/2}) , 
\end{align*}
provided the covariates are bounded in probability. The proof for the cost function follows in a similar manner.
\newpage
\subsection{Sampling Algorithm}
\begin{algorithm}
    \caption{Posterior sampling algorithm for semiparametric model}
    
\begin{enumerate}
    \item Draw sets of weights $\bm{w}^{(1)}, \dots, \bm{w}^{(B)} \sim \text{Dir}(1, \dots, 1)$. 
    \item For $b = 1, \dots, B$,
    \begin{enumerate}
        \item Fit nuisance models using the weights $\bm{\omega}^{(b)}$ to obtain estimate $\hat{\bm{\alpha}}_n^{(b)}$.
        \item Solve the weighted estimating equation in $\bm{\psi}^{(b)}$
        \begin{align*}
            \sum_{i=1}^n \omega^{(b)}_i \bm{U}_i(\bm{\psi}^{(b)}, \hat{\bm{\alpha}}_n^{(b)}) = \bm{0}
        \end{align*}
        to obtain the point estimate $\hat{\bm{\psi}}_n^{(b)}$.
        \item For the chosen $\bm{x} \in \mathcal{X}$, $a \in \mathcal{A}$, estimate direct, indirect, and total effects using the blip parameters and the nuisance model estimates:
        \begin{align*}
            \hat{\zeta}_\mathcal{E}^{(b)}(a, \bm{x}) &= \zeta_\mathcal{E}(a, \bm{x}; \hat{\bm{\psi}}_n^{(b)}, \hat{\bm{\alpha}}_n^{(b)}),\\
            \hat{\delta}_\mathcal{E}^{(b)}(a,  \bm{x}) &= \delta_\mathcal{E}(a, \bm{x}; \hat{\bm{\psi}}_n^{(b)}, \hat{\bm{\alpha}}_n^{(b)}),\\
            \hat{\tau}_\mathcal{E}^{(b)}(a, \bm{x}) &= \hat{\zeta}^{(b)}(a,  \bm{x}) +  \hat{\delta}^{(b)}(a,  \bm{x}).
        \end{align*}
        \item For the desired $\bm{x} \in \mathcal{X}$, estimate the optimal treatment allocation
        \begin{align*}
            \hat{d}^{\text{opt}}_\mathcal{E}(\bm{x}; \hat{\bm{\psi}}_n^{(b)},\hat{\bm{\alpha}}_n^{(b)} ) &= \arg \max_a \hat{\tau}_\mathcal{E}^{(b)}(a, \bm{x}) - f(\bm{x}, a) - \sum_{k=1}^\kappa c_k\mathbb{E}_\mathcal{O}[M_k| \bm{X} = \bm{x}, A = a; \hat{\bm{\alpha}}_n^{(b)}].
        \end{align*}
    \end{enumerate}
\end{enumerate}
\end{algorithm}

\newpage
\section{Extension to survival outcomes}
To each subject, we associate the vector of observable quantities $(\bm{X}, A, \Delta, C, T)$. Our objective is to maximize the survival time $T \in \mathbb{R}_+$. Denote by $\Delta$, the event indicator and by $C$, the right censoring time, where $\Delta = 1$ if an event of any type was observed and $0$ otherwise. The observed data are given by the collection of random variates $\{(\bm{X}_i, A_i, \Delta_i, (1-\Delta_i)C_i, \Delta_i T_i)\}_{i=1}^n$.

For estimation via the semiparametric partially linear model, we require the following additional assumption: 
\begin{itemize}
    \item Independent censoring: $\Delta \perp T| \bm{X}, A, \bm{M}$ \citep{kalbfleisch2002survival}.
\end{itemize}
This condition requires that the censoring indicator is conditionally independent of the survival time conditional on the pre-treatment covariates, the exposure, and the mediators.

If censoring occurs prior to the generation of $\bm{M}$ such that $\bm{M}$ is unobserved when $\Delta = 0$, then we must assume that censoring is conditionally independent of the survival time given only covariates and treatment. The nuisance censoring model used to construct the estimating equations should then be modified accordingly.

We consider an accelerated failure time \citep[AFT;][]{buckley1979linear} model for the survival time; in our case, this is equivalent to the partially linear model with outcome $Y = \log T$. Estimation of $\bm{\psi}$ is based on the following set of unbiased estimating equations:
\begin{align*}
    U_1(\bm{\psi}) &= \bm{X}_\gamma \frac{\Delta}{\mathbb{P}_\mathcal{O}(\Delta = 1| \bm{X}, A,\bm{M}; \bm{\alpha})} (A- \mathbb{E}_\mathcal{O}[A|\bm{X}; \bm{\alpha}]) (\Tilde{Y}- \mu_0(\bm{X}; \bm{\alpha})),\\
    U_2^{(k)}(\bm{\psi}) &= \bm{X}_\eta \frac{\Delta}{\mathbb{P}_\mathcal{O}(\Delta = 1| \bm{X},A, \bm{M}; \bm{\alpha})}(M_k - \mathbb{E}_\mathcal{O}[M_k|\bm{X}, A; \bm{\alpha}])(\Tilde{Y}- \mu_0(\bm{X}; \bm{\alpha})), \\
    U_3^{(k)}(\bm{\psi}) &= \bm{X}_\xi \frac{\Delta}{\mathbb{P}_\mathcal{O}(\Delta = 1| \bm{X},A, \bm{M}; \bm{\alpha})}(A- \mathbb{E}_\mathcal{O}[A|\bm{X}; \bm{\alpha}]) (M_k - \mathbb{E}_\mathcal{O}[M_k|\bm{X}, A; \bm{\alpha}])(\Tilde{Y}- \mu_0(\bm{X; \bm{\alpha}})).
\end{align*}
As before, we define the joint estimating function $U(\bm{\psi}) = (U_1(\bm{\psi}), U_2(\bm{\psi}), U_3(\bm{\psi}))^\top$, where $U_j(\bm{\psi})) = (U_j^{(1)}(\bm{\psi}), \dots, U_j^{(\kappa)}(\bm{\psi}))^\top$ for $j = 2,3$. The only difference relative to the uncensored case is the addition of the first term, which depends on the nuisance model $\mathbb{P}_\mathcal{O}(\Delta =1 | \bm{X}, A, \bm{M}; \bm{\alpha})$. Solving this estimating equation yields a doubly-robust estimator of $\bm{\psi}$ as in the standard case. The properties of the estimator are summarized in the following modified versions of Theorems 1 and 2. Theorem 3 extends directly to the survival setting.

\begin{theorem}
    (Survival) Assuming the semiparametric AFT model holds, the G-estimator $\hat{\bm{\psi}}_n$ is consistent for the true value of the parameter $\bm{\psi}_0$ provided either of the following sets of nuisance models is correctly specified:
    \begin{enumerate}
        \item[(i)] The treatment-mediator-free component $\mu_0(\bm{X})$;  
        \item[(ii)] The treatment model $\mathbb{E}_\mathcal{O}[A|\bm{X}]$, the mediator models $\mathbb{E}_\mathcal{O}[M_k|\bm{X}, A], \; k = 1, \dots, \kappa$, and the censoring model $\mathbb{P}_\mathcal{O}(\Delta = 1| \bm{X},A, \bm{M}; \bm{\alpha})$.
    \end{enumerate}
\end{theorem} 
\begin{proof}
    The proof of the theorem for survival outcomes relies on the same arguments as in the standard setting. Unbiasedness of the estimating equations follows by noting the equality
\begin{align*}
       \mathbb{E}_\mathcal{O}\left[\frac{\Delta}{\mathbb{P}_\mathcal{O}(\Delta = 1| \bm{X},A, \bm{M}; \bm{\alpha})}(Y- \Gamma(\bm{X}, A, \bm{M})-\mu_0(\bm{X};\bm{\alpha}))\Bigr|\bm{X}, A, \bm{M}\right] &=  \\\mathbb{E}_\mathcal{O}\left[\frac{\Delta}{\mathbb{P}_\mathcal{O}(\Delta = 1| \bm{X},A, \bm{M}; \bm{\alpha})}\Bigr|\bm{X}, A, \bm{M}\right] \mathbb{E}_\mathcal{O}\left[(Y- \Gamma(\bm{X}, A, \bm{M})-\mu_0(\bm{X};\bm{\alpha}))|\bm{X}, A, \bm{M}\right] &=\\
    \frac{\mathbb{P}_\mathcal{O}(\Delta = 1| \bm{X},A, \bm{M})}{\mathbb{P}_\mathcal{O}(\Delta = 1| \bm{X},A, \bm{M}; \bm{\alpha})}(\mu_0(\bm{X})-\mu_0(\bm{X};\bm{\alpha})). &
\end{align*}
The second equality follows by the independent censoring assumption.
\end{proof}
\begin{theorem}
(Survival) Suppose that the estimators $(\hat{\bm{\psi}}_n, \hat{\bm{\alpha}}_n)$ are consistent for $(\bm{\psi}_0, \bm{\alpha}_0)$ and the nuisance parameter estimators $\hat{\bm{\alpha}}_n$ converge at a sufficiently fast rate, i.e., for all $\bm{\psi} \in \Psi$, for all $k = 1, \dots, \kappa$, 
\begin{align*}
       \Bigl\| \frac{\mathbb{P}_\mathcal{O}(\Delta = 1| \bm{X},A, \bm{M})}{\mathbb{P}_\mathcal{O}(\Delta = 1| \bm{X},A, \bm{M}; \bm{\alpha})}\Bigr\|_3\cdot \|A - \mathbb{E}_\mathcal{O}[A| \bm{X}; \hat{\bm{\alpha}}_n]\|_3\cdot  \|\Tilde{Y} - \mu_0(\bm{X}; \hat{\bm{\alpha}}_n)\|_3 &= o_p(n^{-1/2}), \\
           \Bigl\| \frac{\mathbb{P}_\mathcal{O}(\Delta = 1| \bm{X},A, \bm{M})}{\mathbb{P}_\mathcal{O}(\Delta = 1| \bm{X},A, \bm{M}; \bm{\alpha})}\Bigr\|_3\cdot\|M_k - \mathbb{E}_\mathcal{O}[M_k| \bm{X},A; \hat{\bm{\alpha}}_n]\|_3\cdot  \|\Tilde{Y} - \mu_0(\bm{X}; \hat{\bm{\alpha}}_n)\|_3 &= o_p(n^{-1/2}), \\
           \Bigl\| \frac{\mathbb{P}_\mathcal{O}(\Delta = 1| \bm{X},A, \bm{M})}{\mathbb{P}_\mathcal{O}(\Delta = 1| \bm{X},A, \bm{M}; \bm{\alpha})}\Bigr\|_4\cdot\|A - \mathbb{E}_\mathcal{O}[A| \bm{X}; \hat{\bm{\alpha}}_n]\|_4& \cdot\\ \|M_k - \mathbb{E}_\mathcal{O}[M_k| \bm{X},A; \hat{\bm{\alpha}}_n]\|_4\cdot  \|\Tilde{Y} - \mu_0(\bm{X}; \hat{\alpha}_n)\|_4 &= o_p(n^{-1/2}), 
    \end{align*}
where $\|\cdot\|_q$ denotes the $L^q(\mathbb{P}_\mathcal{O})$ norm. Then, under suitable regularity conditions,
    \begin{align*}
        \sqrt{n}(\hat{\bm{\psi}}_n - \bm{\psi}_0) = - V_{\bm{\psi}_0, \bm{\alpha}_0}^{-1} \frac{1}{\sqrt{n}}\sum_{i=1}^n  U_i(\bm{\psi}_0; \bm{\alpha}_0) + o_p(1)
    \end{align*}
    for some invertible matrix $V_{\bm{\psi}_0, \bm{\alpha}_0}$.
\end{theorem}

\section{Nonparametric estimator}
\subsection{Regularity conditions for Theorem 4}
Let $\bm{V} = [\bm{X}, A] \in \mathcal{V} \subseteq \mathbb{R}^{d_X + 1}$ and let $\bm{v} \in \mathcal{V}$ be arbitrary. We assume the following regularity conditions:
\begin{enumerate}
    \item The random vector $(\bm{M}, \bm{V})$ admits a joint probability density $f_{\bm{M}, \bm{V}}$ with respect to the Lebesgue measure on $\mathbb{R}^{d_M}\times \mathbb{R}^{d_X + 1}$. If any of the components of $\bm{V}$ are discrete, we can simply apply the theorem within cells of the discrete subvector. The asymptotic convergence rate of the quantization estimator then scales inversely to the dimension of the continuous subvector of $\bm{V}$ rather than the dimension of the full vector. 
    \item  There exists a constant $\chi > 0$ such that \begin{align*}
        \mathbb{P}_\mathcal{O}(\|\bm{V} - \bm{v}\| \le \varepsilon) \ge \chi \varepsilon^{d_X + 1}, \quad \forall \varepsilon > 0.
    \end{align*} 
    \item There exists $\delta > 0$ and an integrable function $r: \mathbb{R}^{d_M} \to \mathbb{R}_+$ such that
    \begin{align*}
        |f^\mathcal{O}_{\bm{M}|\bm{V} = \Tilde{\bm{v}}}(\bm{m}) - f^\mathcal{O}_{\bm{M}|\bm{V} = \Tilde{\bm{v}}}(\bm{m})| \le r(\bm{m}) \|\Tilde{\bm{v}} - \bm{v}\|
    \end{align*}
    for all $\bm{m} \in \mathbb{R}^{d_M}$ and all $\Tilde{\bm{v}}$ such that $\|\Tilde{\bm{v}} - \bm{v}\| < \delta$.
\end{enumerate}
\subsection{Rates for alternative mediation effects}
Suppose the regularity conditions for Theorem 4 are satisfied, and fix $\bm{x} \in \mathcal{X}, a \in \mathcal{A}$. Then, for arbitrary $k \in 1, \dots, \kappa$, if there is some $ c_n \to 0$ such that
\begin{align*}
    \int \Bigl(\int |\hat{\Gamma}_n(\bm{x}, a,\bm{m}_{-k}, m_k) - \Gamma(\bm{x}, a, \bm{m}_{-k}, m_k)|d\mathbb{P}^\mathcal{O}_{\bm{M}_{-k}| \bm{X} = \bm{x}, A = 0}(\bm{m}_{-k})\Bigr)d\mathbb{P}^\mathcal{O}_{M_k| \bm{X} = \bm{x}, A = a}(m_k) = O_p(c_n),
\end{align*}
then the quantization estimator of the $k$th individual mediator effect satisfies:
\begin{align*}
    |\hat{\delta}^k_\mathcal{E}(a,\bm{x}) - \delta^k_\mathcal{E}(a,\bm{x})| &= O_p(a_n),\quad  a_n = \max(b_n, c_n), \quad b_n = \max\left(\left(\sqrt{\frac{\log n}{n}}\right)^{\frac{1}{d_X + 3}}, N^{-1/(d_M-1)}\right).
\end{align*}
The same rate holds for the path-specific direct and indirect effects provided
\begin{align*}
    \int \Bigl(\int |\hat{\Gamma}_n(\bm{x}, a,\bm{m}_{-1}, m_1) - \Gamma(\bm{x}, a, \bm{m}_{-1}, m_1)|d\mathbb{P}^\mathcal{O}_{\bm{M}_{-1}| \bm{X} = \bm{x}, A = a}(\bm{m}_{-1})\Bigr)d\mathbb{P}^\mathcal{O}_{M_1| \bm{X} = \bm{x}, A = 0}(m_1) = O_p(c_n).
\end{align*}
\subsection{Lemmas and auxiliary results}\label{sec:lemmas}

\begin{lemma}
    Let $Z, X_n \in L^1(\Omega, \mathcal{F}, \mathbb{P})$ be real-valued random variables. If $\mathbb{E}[|Z|] = O(a_n)$, then $\mathbb{E}\left[|Z|\Bigr| X_n\right] = O_p(a_n)$. 
    \begin{proof}
        Since $\mathbb{E}[|Z|] = O(a_n)$, $\exists N_0, C$ such that $\frac{\mathbb{E}[|Z|] }{a_n} \le C$ for $n \ge N_0$. Let $\varepsilon > 0$ be arbitrary and let $M = C/\varepsilon$. By Markov's inequality, for $n \ge N_0$,
        \begin{align*}
            \mathbb{P}\left(\frac{\mathbb{E}\left[|Z|\Bigr| X_n\right]}{a_n} > M\right) \le \frac{\mathbb{E}\left[\mathbb{E}\left[|Z|\Bigr| X_n\right]\right]}{a_n \cdot M} = \frac{\mathbb{E}[|Z|]}{a_n \cdot M} \le \varepsilon
        \end{align*}
        The result follows from the definition of boundedness in probability.
    \end{proof}
\end{lemma}

\subsection{Proof of Theorem 4}
We provide the proof for the $k$th individual mediator effect $ \delta^k_\mathcal{E}$. Proofs for the other mediation effects can be easily derived using the same argument. Let $\bm{x}\in \mathcal{X}, a \in \mathcal{A}$ be arbitrary and fixed. 

For any arbitrary outcome $Y$, the mean function $\mu(\bm{X}, A, \bm{M}) = \mathbb{E}_\mathcal{O}[Y| \bm{X}, A, \bm{M}]$ can be written as $ \mu(\bm{X}, A, \bm{M})  = \mu_0(\bm{X}) + \Gamma(\bm{X}, A, \bm{M})$. A similar decomposition holds for the estimate of the mean function $\hat{\mu}_n(\bm{X}, A, \bm{M})$. One such decomposition is the trivial form where $\mu_0 = 0$.  Let $\hat{\mathbb{P}}_{\bm{M}_{-k}| \bm{X} = \bm{x}, A = a}  = \sum_{j=1}^N p_{j, -k}^a \mathds{1}_{\{\bm{g}_{j, -k}^a\}}$,  $\hat{\mathbb{P}}_{M_{k}| \bm{X} = \bm{x}, A = a}  = \sum_{j=1}^N p_{j, k}^a \mathds{1}_{\{g_{j, k}^a\}}$  denote the conditional quantizations of $\bm{M}_{-k}$ and $M_k$ with fixed hyperparameters $N$ and $s$, respectively. Then,  
\begin{align*}
    |\hat{\delta}^k_\mathcal{E}(\bm{x}, a) - \delta^k_\mathcal{E}(\bm{x}, a)| &\le I_1 + I_2,\\
    I_1 &= \sum_{j= 1}^N \sum_{l=1}^N p_{j, -k}^0 p_{l,k}^a\hat{\Gamma}_n(\bm{x}, 0, \bm{g}_{j, -k}^0, g_{l, k}^a) - h(Q_k(0, 0, a,\bm{x}))\\
    &= \int \Bigl(\int \hat{\Gamma}_n(\bm{x}, 0, m_k, \bm{m}_{-k}) d\hat{\mathbb{P}}_{\bm{M}_{-k}| \bm{X} = x, A = 0}(\bm{m}_{-k})\Bigr)d\hat{\mathbb{P}}_{M_k| \bm{X} = x, A = a}(m_k) \\&- \int \Bigl(\int \Gamma(\bm{x}, 0, m_k, \bm{m}_{-k}) d\mathbb{P}_{\bm{M}_{-k}| \bm{X} = x, A = 0}(\bm{m}_{-k})\Bigr)d\mathbb{P}_{M_k| \bm{X} = x, A = a}(m_k)\\
     I_2 &= \sum_{j= 1}^N \sum_{l=1}^N p_{j, -k}^0 p_{l,k}^0\hat{\Gamma}_n(\bm{x}, 0, \bm{g}_{j, -k}^0, g_{l, k}^0) - h(Q_k(0, 0, 0,\bm{x}))\\
    &=\int \Bigl(\int \hat{\Gamma}_n(\bm{x}, 0, m_k, \bm{m}_{-k}) d\hat{\mathbb{P}}_{\bm{M}_{-k}| \bm{X} = x, A = 0}(\bm{m}_{-k})\Bigr)d\hat{\mathbb{P}}_{M_k| \bm{X} = x, A = 0}(m_k) \\&- \int \Bigl(\int \Gamma(\bm{x}, 0, m_k, \bm{m}_{-k}) d\mathbb{P}_{\bm{M}_{-k}| \bm{X} = x, A = 0}(\bm{m}_{-k})\Bigr)d\mathbb{P}_{M_k| \bm{X} = x, A = 0}(m_k).
\end{align*}
Without loss of generality, we focus on the second term $I_2$. The first term can be bounded via the same arguments. The term $I_2$ can be bounded using the triangle inequality:
\begin{align*}
I_2 &\le I_{21} + I_{22}\\
    I_{21} &= \Bigl|\int \Bigl(\int \hat{\Gamma}_n(\bm{x}, 0, m_k, \bm{m}_{-k}) d\hat{\mathbb{P}}_{\bm{M}_{-k}| \bm{X} = x, A = 0}(\bm{m}_{-k})\Bigr)d\hat{\mathbb{P}}_{M_k| \bm{X} = x, A = 0}(m_k) \\&- \int \Bigl(\int \hat{\Gamma}_n(\bm{x}, 0, m_k, \bm{m}_{-k}) d\mathbb{P}_{\bm{M}_{-k}| \bm{X} = x, A = 0}(\bm{m}_{-k})\Bigr)d\mathbb{P}_{M_k| \bm{X} = x, A = 0}(m_k)\Bigr|\\
    I_{22} &= \Bigl|\int \Bigl(\int \hat{\Gamma}_n(\bm{x}, 0, m_k, \bm{m}_{-k}) d\mathbb{P}_{\bm{M}_{-k}| \bm{X} = x, A = 0}(\bm{m}_{-k})\Bigr)d\mathbb{P}_{M_k| \bm{X} = x, A = 0}(m_k) \\&- \int \Bigl(\int \Gamma(\bm{x}, 0, m_k, \bm{m}_{-k}) d\mathbb{P}_{\bm{M}_{-k}| \bm{X} = x, A = 0}(\bm{m}_{-k})\Bigr)d\mathbb{P}_{M_k| \bm{X} = x, A = 0}(m_k)\Bigr|,
\end{align*}
The first term $I_{21}$ can be bounded by noting that $\frac{1}{K_n} \hat{\Gamma}_n$ is 1-Lipschitz. As a result, 
\begin{align*}
    I_{21} &\le K_n \sup \Bigl\{\Bigl|\int \int f(\bm{m}_{-k}, m_k)d\hat{\mathbb{P}}_{\bm{M}_{-k}| \bm{X} = x, A = 0}(\bm{m}_{-k})d\hat{\mathbb{P}}_{M_k| \bm{X} = x, A = 0}(m_k)\\  &- \int \int f(\bm{m}_{-k}, m_k)d\mathbb{P}_{\bm{M}_{-k}| \bm{X} = x, A = 0}(\bm{m}_{-k})d\mathbb{P}_{M_k| \bm{X} = x, A = 0}(m_k) \Bigr|: f \text{ is L-Lipschitz with } L \le 1\Bigr\},\\
    &=K_n \cdot W_1(\hat{\mathbb{P}}_{M_k| \bm{X} = x, A = 0} \times\hat{\mathbb{P}}_{\bm{M}_{-k}| \bm{X} = x, A = 0},  \mathbb{P}_{M_k| \bm{X} = x, A = 0} \times\mathbb{P}_{\bm{M}_{-k}| \bm{X} = x, A = 0}),\\
    &\le K_n \cdot W_2(\hat{\mathbb{P}}_{M_k| \bm{X} = x, A = 0} \times\hat{\mathbb{P}}_{\bm{M}_{-k}| \bm{X} = x, A = 0},  \mathbb{P}_{M_k| \bm{X} = x, A = 0} \times\mathbb{P}_{\bm{M}_{-k}| \bm{X} = x, A = 0}),\\
    &= K_n \sqrt{W_2^2(\hat{\mathbb{P}}_{M_k| \bm{X} = x, A = 0}, \mathbb{P}_{M_k| \bm{X} = x, A = 0} ) + W_2^2(\hat{\mathbb{P}}_{\bm{M}_{-k}| \bm{X} = x, A = 0},\mathbb{P}_{\bm{M}_{-k}| \bm{X} = x, A = 0})},\\
    &\le \sqrt{2}K_n \max \{W_2(\hat{\mathbb{P}}_{M_k| \bm{X} = x, A = 0}, \mathbb{P}_{M_k| \bm{X} = x, A = 0} ) , W_2(\hat{\mathbb{P}}_{\bm{M}_{-k}| \bm{X} = x, A = 0},\mathbb{P}_{\bm{M}_{-k}| \bm{X} = x, A = 0})\}
\end{align*}
where $W_p$ is the $p$-Wasserstein distance \citep{villani2009optimaltransport}. The first equality follows from the alternative definition of the $W_1$ distance in terms of the dual Kantorovich problem \citep[Remark 6.5]{villani2009optimaltransport}. The second inequality is a direct consequence of Hölder's inequality, as $W_p \le W_q$ for $p\le q$. The second equality follows from the additivity of the squared 2-Wasserstein metric over product measures \citep[Section 2]{panaretos2019wassersteinprops}. 

We can now exploit the connection between the Wasserstein metric and the optimal quantization problem. The squared 2-Wasserstein distance between the quantized conditional distribution of $\bm{M}_{-k}$ and the true conditional distribution is equal to the conditional expected squared distortion \citep[Lemma 3.1]{canas2012quantequiv}:
\begin{align*}
     W_2^2(\mathbb{P}_{\bm{M}_{-k}| A= a, \bm{X} = \bm{x}},\hat{\mathbb{P}}_{\bm{M}_{-k}| A= a, \bm{X} = \bm{x}}) &= \mathbb{E}\left[\min_{1\le j \le N} \|\bm{M}_{-k} - \bm{g}^a_j\|^2 \Bigr| \bm{X} = \bm{x}, A = a\right].
\end{align*}
By Corollary  1 of \cite{loubes2017condquant}, under the assumed regularity conditions with $s\asymp n^{\frac{2}{d_X + 3}}$, we have that 
\begin{align*}
     \mathbb{E}\left[\min_{1\le j \le N} \|\bm{M}_{-k} - \bm{g}^a_j\|^2\Bigr| \bm{X} = \bm{x}, A = a\right] - \inf_{G} \mathbb{E}\left[\min_{1\le j \le N} \|\bm{M}_{-k} - \bm{g}\|^2\Bigr| \bm{X} = \bm{x}, A = a\right] = O_p\left(\left(\frac{\log n}{n}\right)^{\frac{1}{d_X + 3}}\right).
\end{align*}
In other words, the expected squared distortion under the estimated optimal quantization grid converges to the expected squared distortion based on the true optimal quantization grid. 

We note however that the rate in probability does not follow immediately from Corollary 1 of \cite{loubes2017condquant}. Their result gives the convergence rate for the expectation of the above difference, integrating out the randomness of the random quantization grid. In order to convert from the rate in expectation to the rate in probability, we can use lemma 1 in section \ref{sec:lemmas}.

Moreover, by theorem 2.1(b) of \cite{pages2015introduction}, the expected distortion for the true optimal quantization grid converges to zero as the number of quantization points increases:
\begin{align*}
    \inf_{G} \mathbb{E}\left[\min_{1\le j \le N} \|\bm{M}_{-k} - \bm{g}\|^2\Bigr| \bm{X} = \bm{x}, A = a\right] = O(N^{-2/d_{M_{-k}}}) =  O(N^{-2/(d_{M}- 1)}) ,
\end{align*}
Combining the two rates, we get the asymptotic convergence rate for the (unsquared) 2-Wasserstein distance
\begin{align*}
       W_2(\mathbb{P}_{\bm{M}_{-k}| A= a, \bm{X} = \bm{x}},\hat{\mathbb{P}}_{\bm{M}_{-k}| A= a, \bm{X} = \bm{x}}) = O_p\left(\left(\sqrt{\frac{\log n}{n}}\right)^{\frac{1}{d_X + 3}}\right) +  O(N^{-1/(d_M- 1)}).
\end{align*}
By the same argument, we also have the rate
\begin{align*}
    W_2(\mathbb{P}_{M_{k}| A= a, \bm{X} = \bm{x}},\hat{\mathbb{P}}_{M_{k}| A= a, \bm{X} = \bm{x}}) = O_p\left(\left(\sqrt{\frac{\log n}{n}}\right)^{\frac{1}{d_X + 3}}\right) +  O(N^{-1}).
\end{align*}
As a result, we have the upper bound
\begin{align*}
    I_{21} = O_p\left(\left(\sqrt{\frac{\log n}{n}}\right)^{\frac{1}{d_X + 3}}\right) +  O_p(N^{-1/(d_M- 1)}).
\end{align*}

The term $I_{22}$ satisfies the following:
\begin{align*}
    I_{22} \le \int \Bigl(\int |\hat{\Gamma}_n(\bm{x}, a,\bm{m}_{-k}, m_k) - \Gamma(\bm{x}, a, \bm{m}_{-k}, m_k)|d\mathbb{P}^\mathcal{O}_{\bm{M}_{-k}| \bm{X} = \bm{x}, A = 0}(\bm{m}_{-k})\Bigr)d\mathbb{P}^\mathcal{O}_{M_k| \bm{X} = \bm{x}, A = a}(m_k)
\end{align*}
By assumption, this term is $O_p(c_n)$. The final result follows by combing the rates for $I_{21}$ and $I_{22}$.

\subsection{Hyperparameter selection}

The choice of hyperparameters $s$ and $N \le s$ is essential to obtaining effective finite-sample approximations to the joint mediator distribution. The number of nearest neighbours $s$ can be chosen as in nearest neighbours regression to minimize the cross-validated empirical squared error \citep{loubes2017condquant}. The choice of $N$ is more straightforward, as a larger quantization grid  leads to a better approximation to the mediator distribution, at the cost of increased computational burden. More specifically, increasing the number of quantization points increases the number of evaluations of the outcome model in the computation of the mediation effects; when computing the conditional expectation of $Y$ is costly, a smaller grid may be preferred. Alternatively, other cluster selection methods such as the silhouette score \citep{rousseeuw1987silhouettes} or stability metrics \citep{ben2001stability} can be used to balance computational efficiency and approximation power. 

\subsection{Identification formulas when using a balancing score}
When performing dimensionality reduction via the propensity score, individual mediator effects are identified as $\delta^k_\mathcal{E}(a,\bm{x}) =Q_k(0, 0, a,\bm{x}) -Q_k(0, 0, 0,\bm{x}) $, where
\begin{align*}
     &Q_k(a,a', a'', \bm{x}) = \\&\int \Bigl(\int \mathbb{E}_\mathcal{O}[Y| \bm{X}=\bm{x}, A= a,m_k, \bm{m}_{-k}] d\mathbb{P}^\mathcal{O}_{\bm{M}_{-k}|\bm{V} = \bm{v}, \pi(\bm{X}) = \pi(\bm{x}), A = a'}(\bm{m}_{-k})\Bigr)d\mathbb{P}^\mathcal{O}_{M_k|\bm{V} = \bm{v}, \pi(\bm{X}) = \pi(\bm{x}), A = a''}(m_k).
\end{align*}
Similarly, path-specific effects are identified as $\zeta_{\mathcal{E}, -1}(a,\bm{x}) =Q(a, 0,\bm{x}) -Q(0, 0,\bm{x}) $, and $\delta_{\mathcal{E}, -1}(a,\bm{x}) =Q(0, a,\bm{x}) -Q(0, 0,\bm{x}) $, where
\begin{align*}
       &Q_{-1}(a,a', \bm{x}) =\\ &\int \Bigl(\int \mathbb{E}_\mathcal{O}[Y| \bm{X}=\bm{x}, A= a,m_1, \bm{m}_{-1}] d\mathbb{P}^\mathcal{O}_{\bm{M}_{-1}|\bm{V} = \bm{v}, \pi(\bm{X}) = \pi(\bm{x}), A = a'}(\bm{m}_{-1})\Bigr)d\mathbb{P}^\mathcal{O}_{M_1|\bm{V} = \bm{v}, \pi(\bm{X}) = \pi(\bm{x}), A = 0}(m_1).
\end{align*}
We provide the identification proof for the conditional direct effect. Proofs for the other mediation effects follow in the same way by assuming equivalent homogeneity assumptions.

Under the second homogeneity assumption, the conditional direct effect is equal to
\begin{align*}
     \zeta_\mathcal{E}(a,\bm{x}) 
    &= \mathbb{E}_\mathcal{E}[Y(a, \bm{M}(a)) | \bm{V} = \bm{v}, \pi(\bm{X}) = \pi(\bm{x})] -\mathbb{E}_\mathcal{E}[Y(0, \bm{M}(a)) | \bm{V} = \bm{v}, \pi(\bm{X}) = \pi(\bm{x})].
\end{align*}
We consider the first term in the difference. By the tower property, we have that
\begin{align*}
    \mathbb{E}_\mathcal{E}[Y(a, \bm{M}(a)) | \bm{V} = \bm{v}, \pi(\bm{X}) = \pi(\bm{x})] &=  \mathbb{E}_\mathcal{E}\Bigl[ \mathbb{E}_\mathcal{E}[Y(a, \bm{M}(a)) | \bm{X}] \Bigr|\bm{V} = \bm{v}, \pi(\bm{X}) = \pi(\bm{x})]\Bigr].
\end{align*}
The inner expectation can be identified in the usual way:
\begin{align*}
    \mathbb{E}_\mathcal{E}[Y(a, \bm{M}(a)) | \bm{X}]&= \int\mathbb{E}_\mathcal{O}[Y|\bm{X}, A= a, \bm{M} = \bm{m}] d\mathbb{P}^\mathcal{O}_{\bm{M}|\bm{X}, A = a}(\bm{m}).
\end{align*}
However, by the first homogeneity assumption, 
\begin{align*}
    \int\mathbb{E}_\mathcal{O}[Y| \bm{X}, A= a, \bm{M}] d\mathbb{P}^\mathcal{O}_{\bm{M}|\bm{X}, A = a} &= \mu_0(\bm{X}) +  \int \Gamma(\bm{V}, a, \bm{m}) d\mathbb{P}^\mathcal{O}_{\bm{M}|\bm{X}, A = a}(\bm{m}),\\
    &= \mu_0(\bm{X}) + \mathbb{E}_\mathcal{O}[\Gamma(\bm{V}, a, \bm{M}) | \bm{X}, A = a],\\
    &= \mu_0(\bm{X}) + \mathbb{E}_\mathcal{O}[\Gamma(\bm{V}, a, \bm{M}(a)) | \bm{X}].
\end{align*}
Then, by the tower property,
\begin{align*}
    &\mathbb{E}_\mathcal{E}\Bigl[ \mathbb{E}_\mathcal{E}[Y(a, \bm{M}(a)) | \bm{X} ] \Bigr|\bm{V} = \bm{v}, \pi(\bm{X}) = \pi(\bm{x})]\Bigr] = \\
    &\mathbb{E}_\mathcal{O}\Bigl[\mu_0(\bm{X}) + \Gamma(\bm{V}, a, \bm{M}(a)) \Bigr|\bm{V} = \bm{v}, \pi(\bm{X}) = \pi(\bm{x}) \Bigr] = \\
    &\mathbb{E}_\mathcal{O}\Bigl[\mu_0(\bm{X})\Bigr|\bm{V} = \bm{v}, \pi(\bm{X}) = \pi(\bm{x}) \Bigr] + \mathbb{E}_\mathcal{O}\Bigl[ \Gamma(\bm{v}, a, \bm{M}(a)) \Bigr|\bm{V} = \bm{v}, \pi(\bm{X}) = \pi(\bm{x}) \Bigr]= \\
    &h(\bm{x}) +  \mathbb{E}_\mathcal{O}\Bigl[ \Gamma(\bm{v}, a, \bm{M}) \Bigr|\bm{V} = \bm{v}, \pi(\bm{X}) = \pi(\bm{x}) , A = a\Bigr] = \\
    &h(\bm{x}) + \int \Gamma(\bm{v}, a, \bm{m}) d\mathbb{P}^\mathcal{O}_{\bm{M}|\bm{V} = \bm{v}, \pi(\bm{X}) = \pi(\bm{x}) , A = a}(\bm{m})
\end{align*}
where the penultimate equality follows from the balancing property of the propensity score, and $h$ is some arbitrary function.

Now, since we're only integrating with respect to the mediator, we have that 
\begin{align*}
    &h(\bm{x}) + \int \Gamma(\bm{v}, a, \bm{m}) d\mathbb{P}^\mathcal{O}_{\bm{M}|\bm{V} = \bm{v}, \pi(\bm{X}) = \pi(\bm{x}) , A = a}(\bm{m}) =\\  &h(\bm{x}) - \mu_0(\bm{x}) +  \int \mu_0(\bm{x}) + \Gamma(\bm{v}, a, \bm{m}) d\mathbb{P}^\mathcal{O}_{\bm{M}|\bm{V} = \bm{v}, \pi(\bm{X}) = \pi(\bm{x}) , A = a}(\bm{m}) = \\
    & h(\bm{x}) - \mu_0(\bm{x}) +  \int \mathbb{E}_\mathcal{O}[Y| \bm{X} = \bm{x}, A = a, \bm{M} = 
    \bm{m}] d\mathbb{P}^\mathcal{O}_{\bm{M}|\bm{V} = \bm{v}, \pi(\bm{X}) = \pi(\bm{x}) , A = a}(\bm{m}). 
\end{align*}
We note that the same argument could not be used above as $\bm{X}$ is a random variable rather than a fixed quantity.

Finally, by applying the exact same argument to the second term in the difference, we have that the terms $h(\bm{x}) - \mu_0(\bm{x})$ cancel out, leaving
\begin{align*}
    \zeta_\mathcal{E}(a,\bm{x})  =&\int \mathbb{E}_\mathcal{O}[Y| \bm{X} = \bm{x}, A = a, \bm{M} = \bm{m}] d\mathbb{P}^\mathcal{O}_{\bm{M}|\bm{V} = \bm{v}, \pi(\bm{X}) = \pi(\bm{x}) , A = a}(\bm{m})  -  \\&\int \mathbb{E}_\mathcal{O}[Y| \bm{X} = \bm{x}, A = 0, \bm{M} = \bm{m}] d\mathbb{P}^\mathcal{O}_{\bm{M}|\bm{V} = \bm{v}, \pi(\bm{X}) = \pi(\bm{x}) , A = a}(\bm{m}),
\end{align*}
which completes the proof.

\subsection{Difficulty of extending dimensionality reduction to continuous exposures}

Let $A$ be an absolutely continuous exposure, and let $\pi_a(\bm{x}) = f_{A|\bm{X}}(a|\bm{x})$ denote the generalized propensity score \citep[GPS;][]{hirano2004gps}.
Under $f_\mathcal{E}$, the GPS is a balancing score for the relationship between $\bm{M}$ and $A$. However, it only satisfies weak (pointwise) unconfoundedness, equivalent to $\bm{M}(a) \perp \mathds{1}(A = a) | \pi_a(\bm{X})$ for $a \in \mathcal{A}$. As a result, identifying counterfactuals at different set values of the exposure requires the use of different propensity scores. 

In order to recover the desired dimensionality reduction in the continuous exposure case, the following homogeneity condition for the indirect effect would have to hold:
\begin{align*}
     \delta_\mathcal{E}(a,\bm{x}) 
    &= \mathbb{E}_\mathcal{E}[Y(0, \bm{M}(a)) | \bm{V} = \bm{v}, \pi_a(\bm{X}) = \pi_a(\bm{x})] -\mathbb{E}_\mathcal{E}[Y(0, \bm{M}(0)) | \bm{V} = \bm{v}, \pi_{0}(\bm{X}) = \pi_{0}(\bm{x})].
\end{align*}
However, unlike the binary exposure case, this assumption does not directly follow from the homogeneity condition 
\begin{align*}
     \delta_\mathcal{E}(a,\bm{x}) 
    &= \mathbb{E}_\mathcal{E}[Y(0, \bm{M}(a)) | \bm{V} = \bm{v}] -\mathbb{E}_\mathcal{E}[Y(0, \bm{M}(0)) | \bm{V} = \bm{v}]
\end{align*}
due to the differing conditioning sets. Consequently, the argument provided in the previous section can not be used to obtain similar results for continuous exposures.

\subsection{Algorithm}
Algorithm \ref{alg:np} details estimation of conditional direct and indirect effects using the nonparametric quantization procedure and subsequent statistical inference using the Bayesian bootstrap. The algorithm also includes the extension to mixed mediator types and dimensionality reduction using the propensity score. Estimation of other conditional mediation effects can be performed in an analogous way.

\begin{algorithm}
    \caption{Posterior sampling algorithm for nonparametric model}\label{alg:np}
    
\begin{enumerate}
    \item For $b = 1, \dots, B$, draw weights $\bm{w}^{(1)}, \dots, \bm{w}^{(B)} \sim \text{Dir}(1, \dots, 1)$.
    \begin{enumerate}
        \item Fit outcome model $\hat{\mu}^{(b)}_n(\bm{X}, A, \bm{M}) = \hat{\mu}^{(b)}_{0n}(\bm{X}) + \hat{\Gamma}_n^{(b)}(\bm{V}, A, \bm{M})$ using weights $\bm{\omega}^{(b)}$.
        \item (optional) Fit propensity score model $\hat{\pi}^{(b)}(\bm{X})$ using weights $\bm{\omega}^{(b)}$.
        \item If applicable, estimate joint conditional distribution of binary mediator subvector $f_{\bm{M}^b| \bm{X}, A}$ via regression by fitting the following mean models using weights $\bm{\omega}^{(b)}$
        \begin{align*}
            \mathbb{E}[M^b_i | \bm{X}, A, \bar{\bm{M}}_{i-1}], \quad i = 1, \dots, \dim(\bm{M}^b).
        \end{align*}
        \item Fix $\bm{x} \in \mathcal{X}$, $a \in \mathcal{A}$. Then, for $l = 1, \dots, N_{\text{mc}}$
        \begin{enumerate}
            \item Sample binary mediators $\bm{m}^b_l$ from $\hat{f}^{(b)}_{\bm{M}^b| \bm{X} = x, A = a}$.
            \item Perform conditional quantization with weights $W_{n,i}^{(b)} = W_{n,i} \cdot \omega_i^{(b)}$ to obtain the quantized measure 
            \begin{align*}
                \hat{\mathbb{P}}_{\bm{M}^c | \bm{V} = \bm{v}, A = a, \hat{\pi}_a(\bm{X}) = \hat{\pi}(\bm{x}), \bm{M}^b = \bm{m}^b_l}= \sum_{j = 1}^N p_{jl}^a \mathds{1}_{\{\bm{g}^{jl}_a\}}. 
            \end{align*}
            \item Compute the following quantities
            \begin{align*}
    \hat{\zeta}^{(b, l)}_\mathcal{E}(\bm{x}, a) &= \sum_{j= 1}^N p_{jl}^a\hat{\mu}^{(b)}_n(\bm{x}, a, \bm{g}_{jl}^a)  - \sum_{j= 1}^N p_{jl}^a\hat{\mu}^{(b)}_n(\bm{x}, 0, \bm{g}_{jl}^a),\\
    \hat{\delta}^{(b, l)}_\mathcal{E}(\bm{x}, a) &= \sum_{j= 1}^N p_{jl}^a\hat{\mu}^{(b)}_n(\bm{x}, 0, \bm{g}_{jl}^a)  - \sum_{j= 1}^N p_{jl}^0\hat{\mu}^{(b)}_n(\bm{x}, 0, \bm{g}_{jl}^0),\\
    \hat{\mathbb{E}}^{(b, l)}_\mathcal{O}[C(a, \bm{x}, \bm{M})| \bm{X} = \bm{x}, A = a] &= \sum_{j= 1}^N p_{jl}^a c(\bm{x}, a, \bm{g}_{jl}^a). 
\end{align*}
        \item Estimate mediation effects by averaging over the Monte Carlo samples
        \begin{align*}
              \hat{\zeta}^{(b)}_\mathcal{E}(\bm{x}, a) = \sum_{l = 1}^{N_{\text{mc}}}\hat{\zeta}^{(b)}_\mathcal{E}(\bm{x}, a), \quad  \hat{\delta}^{(b)}_\mathcal{E}(\bm{x}, a) &= \sum_{l = 1}^{N_{\text{mc}}}\hat{\delta}^{(b)}_\mathcal{E}(\bm{x}, a), \quad \tau_\mathcal{E}^{(b)}(a, \bm{x}) =   \hat{\zeta}^{(b)}_\mathcal{E}(\bm{x}, a) +   \hat{\delta}^{(b)}_\mathcal{E}(\bm{x}, a),\\
                \hat{\mathbb{E}}^{(b)}_\mathcal{O}[C(a, \bm{x}, \bm{M})| \bm{X} = \bm{x}, A = a] &=   \sum_{l = 1}^{N_{\text{mc}}} \hat{\mathbb{E}}^{(b, l)}_\mathcal{O}[C(a, \bm{x}, \bm{M})| \bm{X} = \bm{x}, A = a].
        \end{align*}
        \end{enumerate}
        \item For the desired $\bm{x} \in \mathcal{X}$, estimate the optimal treatment allocation
        \begin{align*}
            \hat{d}^{\text{opt}}_\mathcal{E}(\bm{x}) &= \arg \max_a \hat{\tau}_\mathcal{E}^{(b)}(a, \bm{x}) -  \hat{\mathbb{E}}^{(b)}_\mathcal{O}[C(a, \bm{x}, \bm{M})| \bm{X} = \bm{x}, A = a].
        \end{align*}
    \end{enumerate}
\end{enumerate}
\end{algorithm}

\newpage
\bibliography{refsAppendix}


\title{\bf Supplementary materials: Appendix B}
\date{}
\maketitle

\def\spacingset#1{\renewcommand{\baselinestretch}%
{#1}\small\normalsize} \spacingset{1}
\spacingset{1.8}

\section{Simulations: ITR evaluation}
In this section, we describe the data generating processes and fitted models for all five simulation settings. For each setting, we also provide a barplot depicting mediation effects for five randomly sampled individuals in the test set, allowing for easier interpretation of simulation results. 
\subsection{Simulation settings}
\subsubsection{Setting 1}

Let $i = 1,\dots, n$ denote each subject. The data generating mechanism is given by the following:
\begin{align*}
    X_{i1} &\sim \mathcal{U}(0,1), \; X_{i2} \sim \mathcal{U}(-1,1), X_{i3} \sim \mathcal{N}(0, 4),\\
    A_i &\sim \text{Bernoulli}(\text{expit}(-0.1 + 0.3X_{i1} + 0.3X_{i2})),\\
   M_i &= -1 + 0.3X_{i1} -0.2X_{i2}^2 + 0.2X_{i3} + A_i(0.5 + 0.5X_{i2}) + \varepsilon_i,\\
   Z_i &\sim \text{Bernoulli}(0.5),\; \varepsilon_i \sim 0.3(1-Z_i)(\mathcal{G}(5,5)-2) + 0.3Z_i\mathcal{G}(5,5),\\ 
    p_{i} &= \text{expit}(-1 + 0.5X_{i1} -0.5X_{i2} + 0.2X_{i1}X_{i2} + 0.5A_i + 0.5A_iX_{i3} -0.5M_i + 0.2A_iM_i),\\
    Y_i &\sim \text{Bernoulli}(p_i^Y).
\end{align*} 
We also assume a cost function which depends on the mediator: 
\begin{align*}
    C(a, \bm{M}) = 0.1 M_i. 
\end{align*}
The following models were fitted for the nonparametric estimator using weighted maximum likelihood estimation:
\begin{align*}
    \text{logit}\; \mathbb{P}(Y=1|\bm{X}, A, M) &= \beta_0 + \beta_1X_{1} + \beta_2X_{2} + \beta_3X_{1}X_{2} + \psi_1 A + \psi_2AX_{3} + \psi_3M + \psi_4AM.
\end{align*}
\begin{figure}
    \centering
    \includegraphics[]{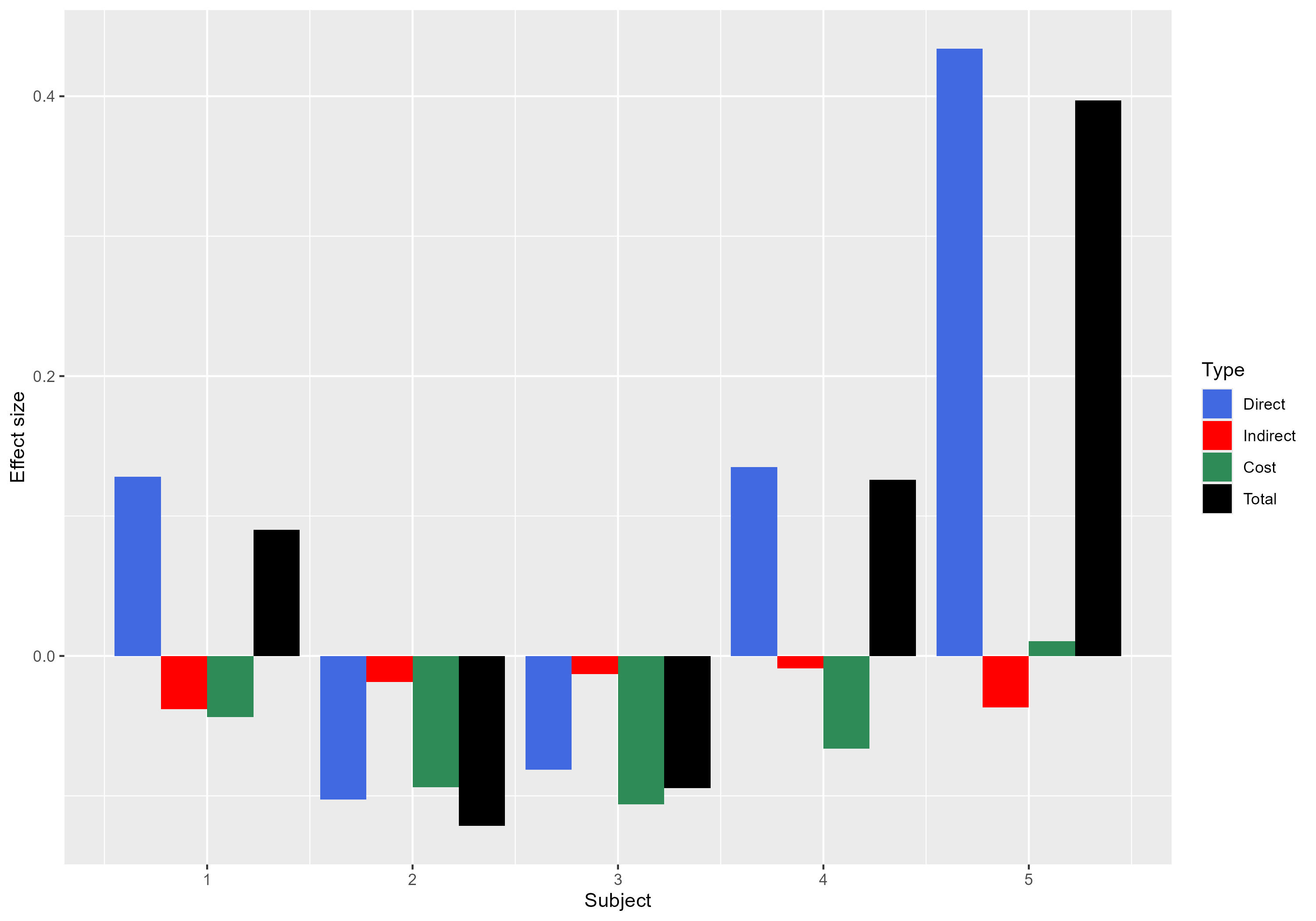}
    \caption{Mediation effect decomposition for 5 randomly sampled individuals in the test set under simulation setting 1.}
    \label{fig:sim1}
\end{figure}

\subsubsection{Setting 2}

Let $i = 1,\dots, n$ denote each subject. The data generating mechanism is given by the following:
\begin{align*}
    \mathbf{X}_i&\sim \bm{\mathcal{N}}(\bm{0}, \mathbf{I}_5),\; V_{i1} \sim \mathcal{U}(-0.5, 0.5), \; V_{i2} \sim \text{Bernoulli}(0.5),\\
    A_i &\sim \text{Bernoulli}(p^A_i), \quad p^A_i = \text{expit}(\mathbf{X}_i \beta_A + V_{i1} + V_{i2}), \; \beta_A = [1, 1, 1, 1, 1]^\top, \\
    \varepsilon_1, \varepsilon_2 &\sim \mathcal{N}(0, 0.1), \; \varepsilon_{M_1} \sim \mathcal{N}(0, 0.25), \; \varepsilon_{M_2} \sim \mathcal{G}(5,5) -1, \; \varepsilon_{M_3} \sim \mathcal{U}(-0.5, 0.5), \; \varepsilon_Y \sim \mathcal{N}(0, 0.5),\\
    M_{i1} &= 1 - \mathbf{X}_i \beta_1 + V_{i1} + V_{i2} + A_i(-2 + 0.3V_{i1}) + \varepsilon_{M_1} + \varepsilon_1, \\
    M_{i2} &= 2 - \mathbf{X}_i \beta_1 - V_{i1} - V_{i2} + A_i(2 - 3V_{i2})  + \varepsilon_{M_2} +\varepsilon_1 - \varepsilon_2,\\
     M_{i3} &= \mathbf{X}_i \beta_1 + V_{i1} + V_{i2} + A_i(0.5 + V_{i1} + V_{i2}) + \varepsilon_{M_3} +\varepsilon_2,\\
     Y &= 3 + \mathbf{X}_i \beta_Y - 3V_{i1} + 4V_{i2} + A_i(1+ V_{i1} - 2V_{i2}) + M_{i1} + M_{i2} + M_{i3} + \varepsilon_Y,\\
     \beta_Y &= [1, -1, 1, -0.5, 2]^\top.
\end{align*} 
We also assume a cost function, which depends on the exposure and the covariates: 
\begin{align*}
    C(a, \bm{x}) = a(-0.2 -0.2v_2).
\end{align*}
For the nonparametric estimator, we employ the dimensionality reduction scheme based on the propensity score, using the fact that the conditional mediation effects of interest only vary within levels of $\bm{V} = [V_1, V_2]$. We fit the following models:
\begin{align*}
    \mathbb{E}[Y|\bm{X}, A, \bm{M}] &= \beta_0 + \mathbf{X} \bm{\beta}_1 + \beta_2V_{1} + \beta_3V_{2} + \psi_1A + \psi_2AV_{1} + \psi_3AV_{2} + \psi_4M_{1} + \psi_5M_{2} + \psi_6M_{3}, \\
    \text{logit}\; \mathbb{P}(A = 1 | \bm{X}) &= \bm{X} \bm{\alpha}_1 + \alpha_2 V_1 + \alpha_3 V_2.
\end{align*}
For the semiparametric estimator, we fit the following models:
\begin{align*}
    \Gamma(\bm{X}, A, \bm{M}) &=  \psi_1A + \psi_2AV_{1} + \psi_3AV_{2} + \psi_4M_{1} + \psi_5M_{2} + \psi_6M_{3},\\
    \mu_0(\bm{X}) &= \beta_0 + \mathbf{X} \bm{\beta}_1 + \beta_2V_{1} + \beta_3V_{2},\\
    \text{logit}\; \mathbb{P}(A = 1 | \bm{X}) &= \bm{X} \bm{\alpha}_1 + \alpha_2 V_1 + \alpha_3 V_2,\\
    \mathbb{E}[M_1 | \bm{X}, A] &= \theta_0 + \bm{X}\theta_1 + \theta_2V_1 + \theta_3V_2 + \theta_4 A +  \theta_5 AV_1,\\
    \mathbb{E}[M_2 | \bm{X}, A] &= \theta_0 + \bm{X}\theta_1 + \theta_2V_1 + \theta_3V_2 + \theta_4 A +  \theta_5 AV_2,\\
    \mathbb{E}[M_2 | \bm{X}, A] &= \theta_0 + \bm{X}\theta_1 + \theta_2V_1 + \theta_3V_2 + \theta_4 A +  \theta_5 AV_1 + \theta_6 AV_2.
\end{align*}
The parameters of the nuisance component $\mu_0$ were estimated via joint estimating equations along with the blip parameters. The rest of the nuisance models were fitted via weighted maximum likelihood.

\begin{figure}
    \centering
    \includegraphics[]{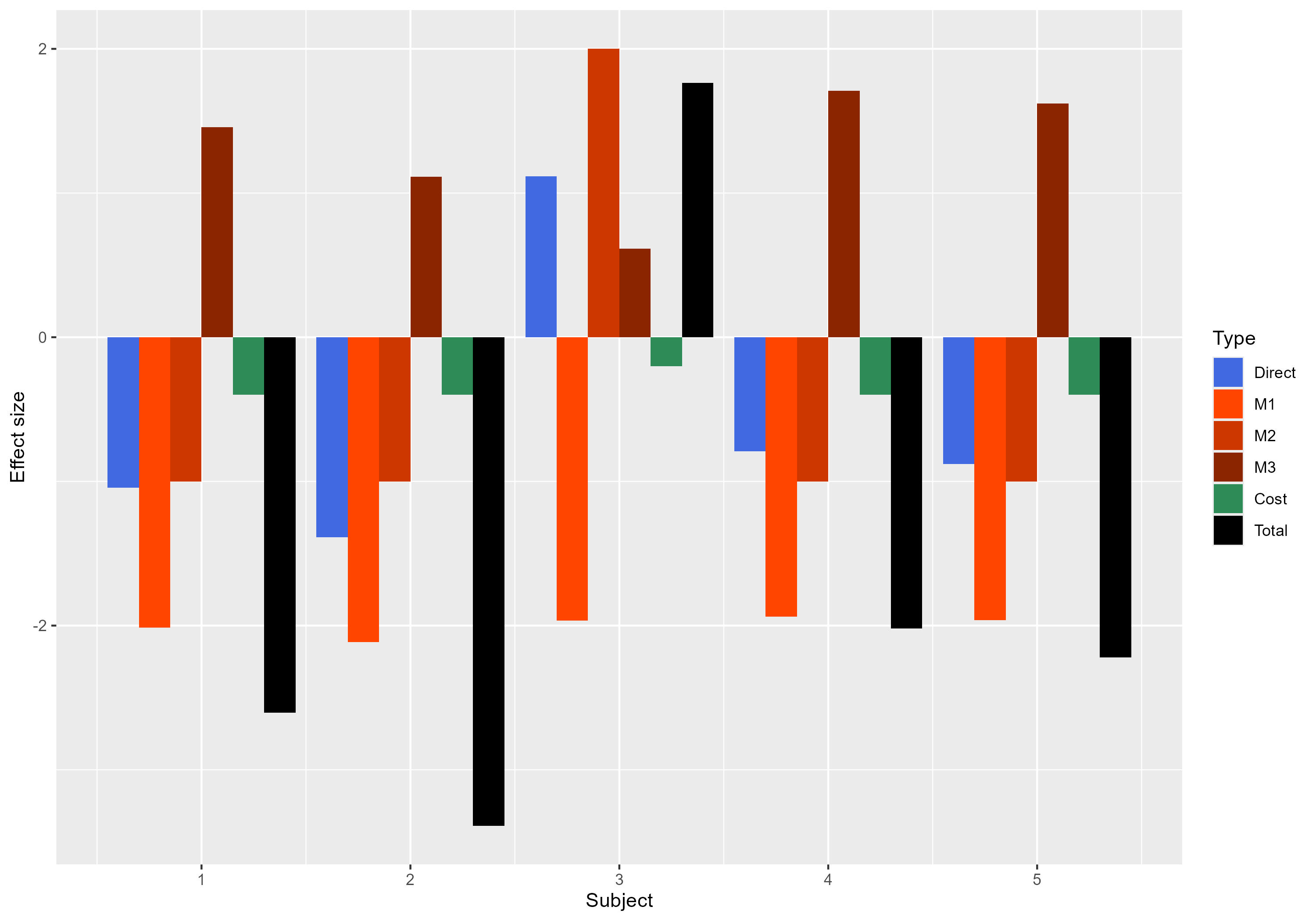}
    \caption{Mediation effect decomposition for 5 randomly sampled individuals in the test set under simulation setting 2.}
    \label{fig:sim2}
\end{figure}

\subsubsection{Setting 3}
Let $i = 1,\dots, n$ denote each subject. The data generating mechanism is given by the following:
\begin{align*}
    \bm{X}_i &\sim \mathcal{N}_5(0, 0.2 I),\\ 
    V_{i1} &\sim \mathcal{U}(0,1), \; V_{i2} \sim \text{Bernoulli}(0.5),\\
    A_i &\sim \text{Bernoulli}(\text{expit}(\bm{X}_iI - 0.5V_{i1} -3V_{i2})),\\
    M_i &\sim \text{Bernoulli}(\text{expit}(\bm{X}_iI - V_{i1} - V_{i2} + A_i(1 + V_{i1} +  V_{i2})))\\
    \log T &= \bm{X}_iI + 0.1 V_{i1} - 0.1 V_{i2} + 1.5 M_{i} + 0.5M_{i} V_{i1} + A_i (-1.5 + 0.5 V_{i2} + 0.5 M_{i} + 0.5 M_{i}V_{i2}) + \varepsilon_i,\\
    \varepsilon_i &\sim \mathcal{G}(5,5) - 1,\\
    \Delta_i &\sim \text{Bernoulli}(\text{expit}(-1 + V_{i1} + V_{i2} + V_{i1}V_{i2} + A_i + A_iV_{i1} )).
\end{align*}
We assume a multiplicative cost given by 
\begin{align*}
    \log C(M_i) = \log(1/0.5) M_i.
\end{align*}
For the nonparametric estimator, we fit the following models:
\begin{align*}
     \mathbb{E}[\log T| \bm{X}, A, M] &= \bm{X}\beta_1 + \beta_2 V_{1} + \beta_3 V_{2} + \psi_1 M + \psi_2M V_{1} + \psi_3A+ \psi_4AV_{2} + \psi_5 AM + \psi_6 AMV_{2},\\
     \text{logit}\;\mathbb{P}(M = 1|\bm{X}, A) &= \bm{X}\theta_1 + \theta_2 V_1 + \theta_3 V_2 + \theta_4 A + \theta_5 AV_1 + \theta_6AV_2,\\
     \text{logit}\; \mathbb{P}(\Delta = 1| \bm{X}, A) &= \alpha^*_0 + \alpha^*_1 V_1 + \alpha^*_2 V_2 + \alpha^*_3 V_1V_2 + \alpha^*_4 A + \alpha^*_5 AV_1.
\end{align*}
For the semiparametric estimator, we fit the following models:
\begin{align*}
     \mu_0(\bm{X}) &= \bm{X}\beta_1 + \beta_2 V_{1} + \beta_3 V_{2},\\
     \Gamma(\bm{X}, A, M) &= \psi_1 M + \psi_2M V_{1} + \psi_3A+ \psi_4AV_{2} + \psi_5 AM + \psi_6 AMV_{2},\\
     \text{logit}\; \mathbb{P}(A = 1| \bm{X}, A) &= \bm{X}\alpha_1 + \alpha_2V_1 + \alpha_3 V_2,\\
     \text{logit}\;\mathbb{P}(M = 1|\bm{X}, A) &= \bm{X}\theta_1 + \theta_2 V_1 + \theta_3 V_2 + \theta_4 A + \theta_5 AV_1 + \theta_6AV_2,\\
     \text{logit}\; \mathbb{P}(\Delta = 1| \bm{X}, A) &= \alpha^*_0 + \alpha^*_1 V_1 + \alpha^*_2 V_2 + \alpha^*_3 V_1V_2 + \alpha^*_4 A + \alpha_5^* AV_1.
\end{align*}
Parameters of the nuisance component $\mu_0$ were estimated via joint estimating equations along with the blip parameters. The remaining nuisance models were fitted via weighted maximum likelihood.
\begin{figure}
    \centering
    \includegraphics[]{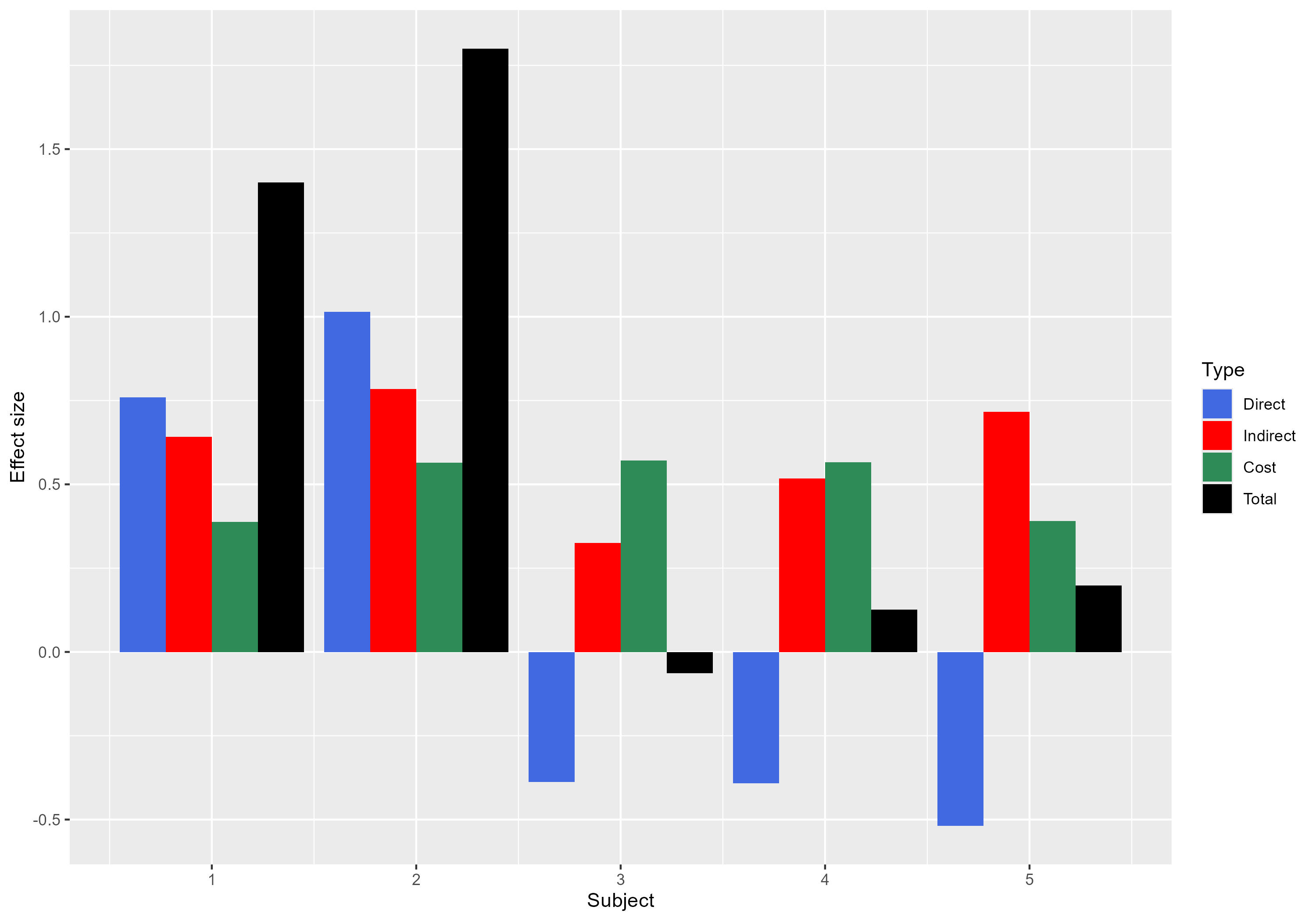}
    \caption{Mediation effect decomposition for 5 randomly sampled individuals in the test set under simulation setting 3.}
    \label{fig:sim3}
\end{figure}

\subsubsection{Setting 4}
Let $i = 1,\dots, n$ denote each subject. The data generating mechanism is given by the following:
\begin{align*}
    X_{i1} &\sim 0.5\times \mathcal{U}(-1,0), \; X_{i2} \sim 0.5\times \mathcal{U}(0.6,1.5),\\
    A_i &= 0.3X_{i1} + 0.3X_{i2} + u_i, \quad u_i \sim \mathcal{U}(-0.4, 0.4),\\
    M_{i1} &\sim \text{Bernoulli}(p_i^{M_1}), \; p_i^{M_1} = \text{expit}(-0.5 + 0.5A_i + 0.3X_{i1} +0.8X_{i2} + 0.5X_{i1}X_{i2}),\\
    M_{i2} &= -1 + 0.5X_{i1} + A_i^2 + 2A_i X_{i2} -X^2_{i2} + 0.5M_{i1} + g_i, \quad g_i \sim 0.3\times (\mathcal{G}(5,5) -1),\\
    \mu_{0i} &= 4 + 0.5X_{i1} -0.5 X_{i2} + 0.5X_{i1}X_{i2},\\
    Y_i &= \mu_{0i}- A_i + 0.7AX_{i2} + M_{i1} + M_{i2} + M_{i1}M_{i2} + 0.3A_i M_{i1} + 0.3A_i M_{i2} + 0.3A_i M_{i1}M_{i2} + \varepsilon_i,\\
    \varepsilon_i &\sim \mathcal{N}(0, 0.5).
\end{align*} 
We assume a trivial cost function, i.e., $C = 0$. Since the exposure is continuous, we estimate the optimal treatment allocation via a grid search with grid $(0, 0.1, 0.2, 0.3, 0.4, 0.5)$. This grid is also used to compute the AMSE for the mediation effects of interest.

For the nonparametric estimator, we consider the nested sampling scheme for mixed mediator types and fit an additional model for the binary mediator from which to sample from. We fit the following models:
\begin{align*}
    \mathbb{E}[Y|\bm{X}, A, \bm{M}] &= \beta_0 + \beta_1 X_1 + \beta_2 X_2 + \beta_3 X_1 X_2 + \psi_1 A + \psi_2 M_1 + \psi_3 M_2 + \psi_4 M_1 M_2\\
    &+ \psi_5 AM_1 + \psi_6 AM_2 + \psi_7 AM_1M_2,\\
    \text{logit}\; \mathbb{P}(M_1 = 1| \bm{X}, A) &= \theta_0 + \theta_1 A + \theta_2 X_1 + \theta_3 X_2 + \theta_4 X_1 X_2.
\end{align*}
For the semiparametric estimator, we fit the following models:
\begin{align*}
    \mu_0(\bm{X}) &=  \beta_0 + \beta_1 X_1 + \beta_2 X_2 + \beta_3 X_1 X_2,\\
    \Gamma(\bm{X}, A, \bm{M}) &= \psi_1 A + \psi_2 M_1 + \psi_3 M_2 + \psi_4 M_1 M_2+ \psi_5 AM_1 + \psi_6 AM_2 + \psi_7 AM_1M_2,\\
    \mathbb{E}[A| \bm{X}] &= \alpha_0 + \alpha_1 X_1 + \alpha_2 X_2,\\
     \text{logit}\; \mathbb{P}(M_1 = 1| \bm{X}, A) &= \theta_0 + \theta_1 A + \theta_2 X_1 + \theta_3 X_2 + \theta_4 X_1 X_2,\\
     \mathbb{E}[M_2| \bm{X}, A] &= \theta_0 + \theta_1 X_1 + \theta_2 A^2 + \theta_3 AX_2 - \theta_4 X_2^2 + \theta_5 M_1,\\
     \mathbb{E}[M_1 M_2| \bm{X}, A] &= m(\bm{X}, A).
\end{align*}
Parameters of the nuisance component $\mu_0$ were estimated via joint estimating equations along with the blip parameters. The rest of the parametric nuisance models were fitted via weighted maximum likelihood. The exception is the mean model $m(\bm{X}, A)$ for the interaction term $M_1M_2$ which was estimated using generalized random forests \citep{athey2019generalized}. The generalized regression forests were fitted using $100$ total trees. The rest of the random forest hyperparameters were set to their default value. The implementation was provided by the R package \texttt{grf} \citep{grfpackage}.

\begin{figure}[H]
    \centering
    \includegraphics[]{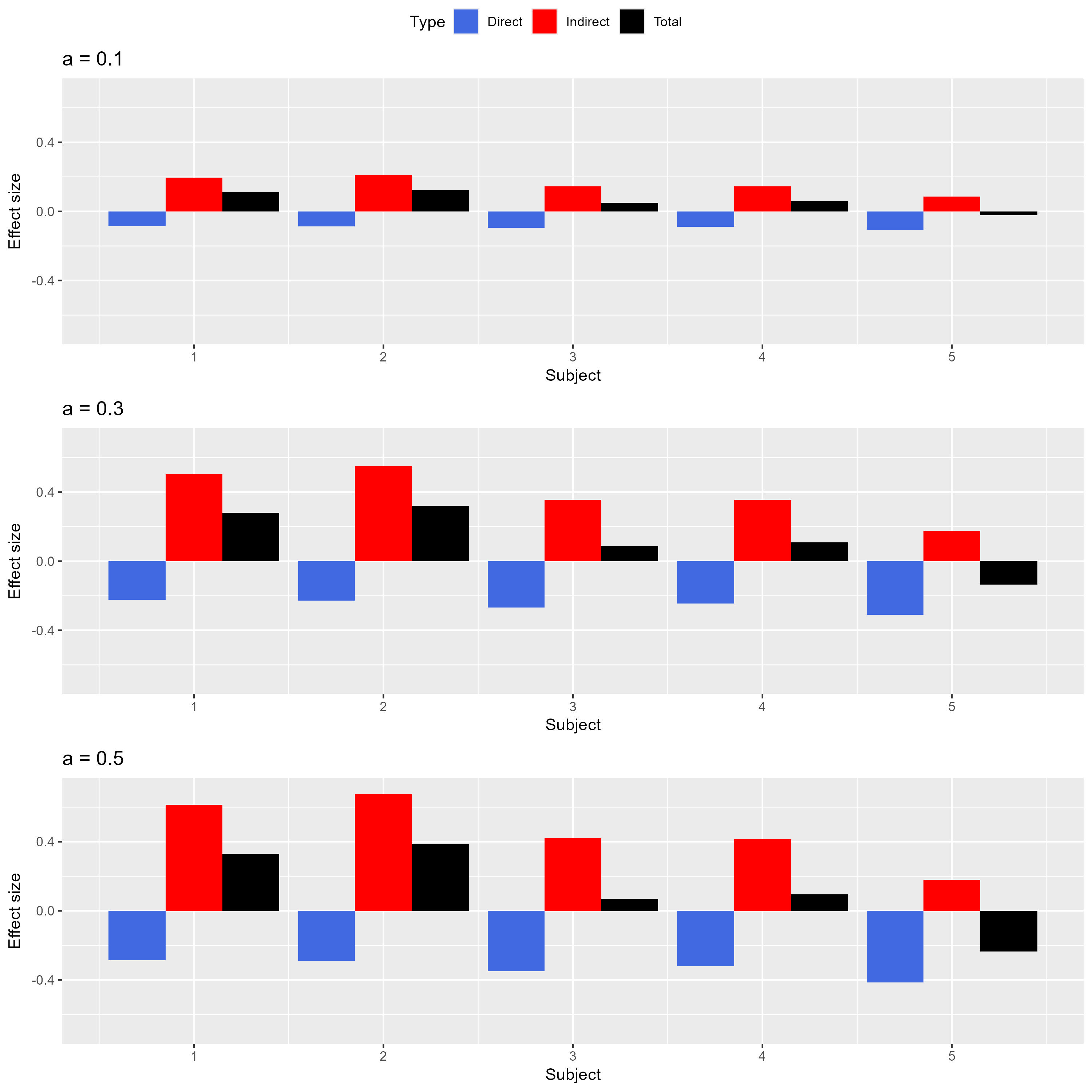}
    \caption{Mediation effect decomposition for 5 randomly sampled individuals in the test set under simulation setting 4.}
    \label{fig:sim4}
\end{figure}

\subsubsection{Setting 5}
This setting is a modification of setting 2. In this altered setting, $M_1$ is made conditionally independent of the other two mediators, allowing for identification of path-specific effects. Here, we compute the path-specific direct, indirect and individual mediator effects (through $M_2$ and $M_3$).

Let $i = 1,\dots, n$ denote each subject. The data generating mechanism is given by the following:
\begin{align*}
    \mathbf{X}_i&\sim \bm{\mathcal{N}}(\bm{0}, \mathbf{I}_5),\; V_{i1} \sim \mathcal{U}(-0.5, 0.5), \; V_{i2} \sim \text{Bernoulli}(0.5),\\
    A_i &\sim \text{Bernoulli}(p^A_i), \quad p^A_i = \text{expit}(\mathbf{X}_i \beta_A + V_{i1} + V_{i2}), \; \beta_1 = [1, 1, 1, 1, 1]^\top,\\
    \varepsilon_1, \varepsilon_2, \varepsilon_3 &\sim \mathcal{N}(0, 0.1), \; \varepsilon_{M_1} \sim \mathcal{N}(0, 0.25), \; \varepsilon_{M_2} \sim \mathcal{G}(5,5) -1, \; \varepsilon_{M_3} \sim \mathcal{U}(-0.5, 0.5), \; \varepsilon_Y \sim \mathcal{N}(0, 0.5),\\
    M_{i1} &= 1 - \mathbf{X}_i \beta_1 + V_{i1} + V_{i2} + A_i(-2 + 0.3V_{i1}) + \varepsilon_{M_1} + \varepsilon_1,\\
    M_{i2} &= 2 - \mathbf{X}_i \beta_1 - V_{i1} - V_{i2} + A_i(2 - 3V_{i2})  + \varepsilon_{M_2} +\varepsilon_3 - \varepsilon_2,\\
     M_{i3} &= \mathbf{X}_i \beta_1 + V_{i1} + V_{i2} + A_i(0.5 + V_{i1} + V_{i2}) + \varepsilon_{M_3} +\varepsilon_2,\\
     Y &= 3 + \mathbf{X}_i \beta_Y - 3V_{i1} + 4V_{i2} + A_i(1+ V_{i1} - 2V_{i2}) + M_{i1} + M_{i2} + M_{i3} + \varepsilon_Y,\\
     \beta_Y &= [1, -1, 1, -0.5, 2]^\top.
\end{align*} 
We assume a cost function, which depends on the exposure and the covariates: 
\begin{align*}
    C(a, \bm{x}) = a(-0.2 -0.2v_2).
\end{align*}
For the nonparametric estimator, we employ the dimensionality reduction scheme based on the propensity score, using the fact that the conditional mediation effects of interest only vary within levels of $\bm{V} = [V_1, V_2]$. We fit the following models:
\begin{align*}
    \mathbb{E}[Y|\bm{X}, A, \bm{M}] &= \beta_0 + \mathbf{X} \bm{\beta}_1 + \beta_2V_{1} + \beta_3V_{2} + \psi_1A + \psi_2AV_{1} + \psi_3AV_{2} + \psi_4M_{1} + \psi_5M_{2} + \psi_6M_{3}, \\
    \text{logit}\; \mathbb{P}(A = 1 | \bm{X}) &= \bm{X} \bm{\alpha}_1 + \alpha_2 V_1 + \alpha_3 V_2.
\end{align*}
For the semiparametric estimator, we fit the following models:
\begin{align*}
    \Gamma(\bm{X}, A, \bm{M}) &=  \psi_1A + \psi_2AV_{1} + \psi_3AV_{2} + \psi_4M_{1} + \psi_5M_{2} + \psi_6M_{3},\\
    \mu_0(\bm{X}) &= \beta_0 + \mathbf{X} \bm{\beta}_1 + \beta_2V_{1} + \beta_3V_{2},\\
    \text{logit}\; \mathbb{P}(A = 1 | \bm{X}) &= \bm{X} \bm{\alpha}_1 + \alpha_2 V_1 + \alpha_3 V_2,\\
    \mathbb{E}[M_1 | \bm{X}, A] &= \theta_0 + \bm{X}\theta_1 + \theta_2V_1 + \theta_3V_2 + \theta_4 A +  \theta_5 AV_1,\\
    \mathbb{E}[M_2 | \bm{X}, A] &= \theta_0 + \bm{X}\theta_1 + \theta_2V_1 + \theta_3V_2 + \theta_4 A +  \theta_5 AV_2,\\
    \mathbb{E}[M_2 | \bm{X}, A] &= \theta_0 + \bm{X}\theta_1 + \theta_2V_1 + \theta_3V_2 + \theta_4 A +  \theta_5 AV_1 + \theta_6 AV_2.
\end{align*}
The parameters of the nuisance component $\mu_0$ were estimated via joint estimating equations along with the blip parameters. The rest of the nuisance models were fitted via weighted maximum likelihood.

\newpage
\section{Simulations: Doubly robust estimation}
In this section, we detail additional simulations which explore the double robustness of the G-estimator for the semiparametric partially linear model. For both the linear and survival settings, we consider four model specifications: (a) all models correct, (b) outcome model correct, treatment and mediator models (and censoring model, if applicable) incorrect, (c) outcome model incorrect, treatment and mediator models (and censoring model, if applicable) correct, and (d) all models incorrect. Blip parameters were estimated over $500$ replicate training sets $(n_\text{train} = 1000)$ and posterior inference was then carried out using $B = 100$ Bayesian bootstrap draws. Posterior means of blip parameters for each replicate training set were used to compute estimates of the scaled bias and variance.

\subsection{Simulation settings}
\subsubsection{Linear setting}
 The data generating mechanism is given by the following:
\begin{align*}
    X_{i1} &\sim \mathcal{U}(0,1), \; X_{i2} \sim \mathcal{U}(-1,0), X_{i3} \sim \mathcal{N}(0, 1),\\
    A_i &\sim \text{Bernoulli}(\text{expit}(0.5 + X_{i1} + X_{i2} -0.5X_{i3})),\\
   M_i &= -0.5 + A_i + X_{i1}  + X_{i2} + X_{i3}  + \varepsilon_{i1},\quad \varepsilon_{i1} \sim \mathcal{N}(0, 1),\\ 
   \mu_{0i} &= 4 + 5X_{i1} -5X_{i2} + X_{i3} + 2X_{i1}X_{i2} + X_{i1}X_{i3},\\
   \Gamma_i &= \psi_1A_{i} + \psi_2A_{i1}X_{i1} + \psi_3M_i + \psi_4M_iX_{i2} + \psi_5A_i M_{i} + \psi_6A_i M_i X_{i3},\\
   \bm{\psi} &= (-1, 0.7, 1, 2, 0.3, 0.5)^\top,\\
   Y_i &= \mu_{0i} + \Gamma_i + \varepsilon_{i2}, \quad \varepsilon_{i2} \sim \mathcal{N}(0,1). 
\end{align*}
The following sets of models were fitted:
\begin{enumerate}
    \item[(a)] All models correct:
    \begin{align*}
        \mu_0(\bm{X}) &= \beta_0 + \beta_1X_1 + \beta_2 X_2 + \beta_3X_3 + \beta_4X_1X_2 + \beta_5X_1X_3,\\
        \mathbb{P}(A = 1| \bm{X}) &= \text{expit}(\alpha_0 + \alpha_1X_1 + \alpha_2X_2 + \alpha_3X_3),\\
        \mathbb{E}[M|\bm{X}, A] &= \theta_0 + \theta_1X_1 +  \theta_2X_2 +  \theta_3X_3 + \theta_4A.
    \end{align*}
    \item[(b)] Outcome correct, treatment and mediator incorrect
    \begin{align*}
        \mu_0(\bm{X}) &= \beta_0 + \beta_1X_1 + \beta_2 X_2 + \beta_3X_3 + \beta_4X_1X_2 + \beta_5X_1X_3,\\
        \mathbb{P}(A = 1| \bm{X}) &= \text{expit}(\alpha_0),\\
        \mathbb{E}[M|\bm{X}, A] &= \theta_0.
    \end{align*}
    \item[(c)] Outcome incorrect, treatment and mediator correct
    \begin{align*}
        \mu_0(\bm{X}) &= \beta_0, \\
        \mathbb{P}(A = 1| \bm{X}) &= \text{expit}(\alpha_0 + \alpha_1X_1 + \alpha_2X_2 + \alpha_3X_3),\\
        \mathbb{E}[M|\bm{X}, A] &= \theta_0 + \theta_1X_1 +  \theta_2X_2 +  \theta_3X_3 + \theta_4A.
    \end{align*}
    \item[(d)] All models incorrect
    \begin{align*}
        \mu_0(\bm{X}) &= \beta_0,\\
        \mathbb{P}(A = 1| \bm{X}) &= \text{expit}(\alpha_0),\\
        \mathbb{E}[M|\bm{X}, A] &= \theta_0. 
    \end{align*}
\end{enumerate}
The treatment-mediator-free component $\mu_0$ was estimated along with the blip parameters via joint estimating equations. The treatment model and the mediator model were estimated via weighted maximum likelihood.

\subsubsection{Survival setting}
Let $i = 1,\dots, n$ denote each subject. The data generating mechanism is given by the following:
\begin{align*}
    \bm{X}_i &\sim \mathcal{N}_5(0, 0.2 I),\\ 
    V_{i1} &\sim \mathcal{U}(0,1), \; V_{i2} \sim \text{Bernoulli}(0.5),\\
    A_i &\sim \text{Bernoulli}(\text{expit}(\bm{X}_i \mathds{1} - 0.5V_{i1} -V_{i2})),\\
    M_i &\sim \text{Bernoulli}(\text{expit}(0.5\bm{X}_i\mathds{1} - V_{i1} - V_{i2} + A_i(1 + V_{i1} +  V_{i2}))),\\
    \mu_{0i} &= 0.5\bm{X}_i\mathds{1} + V_{i1} - V_{i2},\\
    \Gamma_i &= \psi_1 A_i + \psi_2 A V_{i2} + \psi_3 M_{i} + \psi_4M_{i} V_{i1} + \psi_5AM_{i} + \psi_6 AM_{i}V_{i2}, \\
    \bm{\psi} &= (-1.5, 0.5, 1.5, 0.5, 0.5, 0.5)^\top,\\
    \log T &= \mu_{0i} + \Gamma_i + 0.5\varepsilon_i\\
    \varepsilon_i &\sim \mathcal{G}(5,5) - 1,\\
    \Delta_i &\sim \text{Bernoulli}(\text{expit}(-0.5 + V_{i1} + V_{i2} + V_{i1}V_{i2} + A_i + A_iV_{i1} + M_i)).
\end{align*}
The following sets of models were fitted:
\begin{enumerate}
    \item[(a)] All models correct:
    \begin{align*}
        \mu_0(\bm{X}) &= \beta_0 + \beta_1X_1 + \beta_2 X_2 + \beta_3X_3 + \beta_4X_4 + \beta_5X_5 + \beta_6V_1 + \beta_7V_2,\\
        \mathbb{P}(A = 1| \bm{X}) &= \text{expit}(\alpha_0 + \alpha_1X_1 + \alpha_2X_2 + \alpha_3X_3 + \alpha_4X_4 + \alpha_5X_5 + \alpha_6V_1 + \alpha_7V_2),\\
        \mathbb{P}(\Delta = 1|\bm{X}, A, M) &= \text{expit}(\alpha_0^* + \alpha_1^*V_1 + \alpha_2^*V_2 + \alpha_3^*A + \alpha_4^*AV_1 + \alpha_5^*M),\\ 
        \mathbb{E}[M|\bm{X}, A] &= \theta_0 + \theta_1X_1 +  \theta_2X_2 +  \theta_3X_3 + \theta_4A.
    \end{align*}
    \item[(b)] Outcome correct, treatment, mediator and censoring models incorrect
    \begin{align*}
        \mu_0(\bm{X}) &= \beta_0 + \beta_1X_1 + \beta_2 X_2 + \beta_3X_3 + \beta_4X_4 + \beta_5X_5 + \beta_6V_1 + \beta_7V_2,\\
        \mathbb{P}(A = 1| \bm{X}) &= \text{expit}(\alpha_0 ),\\
        \mathbb{P}(\Delta = 1|\bm{X}, A, M) &= \text{expit}(\alpha_0^*),\\ 
        \mathbb{E}[M|\bm{X}, A] &= \theta_0.
    \end{align*}
    \item[(c)] Outcome incorrect, treatment, mediator and censoring models correct
    \begin{align*}
         \mu_0(\bm{X}) &= \beta_0, \\
        \mathbb{P}(A = 1| \bm{X}) &= \text{expit}(\alpha_0 + \alpha_1X_1 + \alpha_2X_2 + \alpha_3X_3 + \alpha_4X_4 + \alpha_5X_5 + \alpha_6V_1 + \alpha_7V_2),\\
        \mathbb{P}(\Delta = 1|\bm{X}, A, M) &= \text{expit}(\alpha_0^* + \alpha_1^*V_1 + \alpha_2^*V_2 + \alpha_3^*A + \alpha_4^*AV_1 + \alpha_5^*M),\\ 
        \mathbb{E}[M|\bm{X}, A] &= \theta_0 + \theta_1X_1 +  \theta_2X_2 +  \theta_3X_3 + \theta_4A.
    \end{align*}
    \item[(d)] All models incorrect
    \begin{align*}
         \mu_0(\bm{X}) &= \beta_0, \\
        \mathbb{P}(A = 1| \bm{X}) &= \text{expit}(\alpha_0 ),\\
        \mathbb{P}(\Delta = 1|\bm{X}, A, M) &= \text{expit}(\alpha_0^*),\\ 
        \mathbb{E}[M|\bm{X}, A] &= \theta_0. 
    \end{align*}
\end{enumerate}
The treatment-mediator-free component $\mu_0$ was estimated along with the blip parameters via joint estimating equations. The treatment model, the censoring model and the mediator model were estimated via weighted maximum likelihood.

\subsection{Results}
\begin{figure}
    \centering
    \includegraphics[]{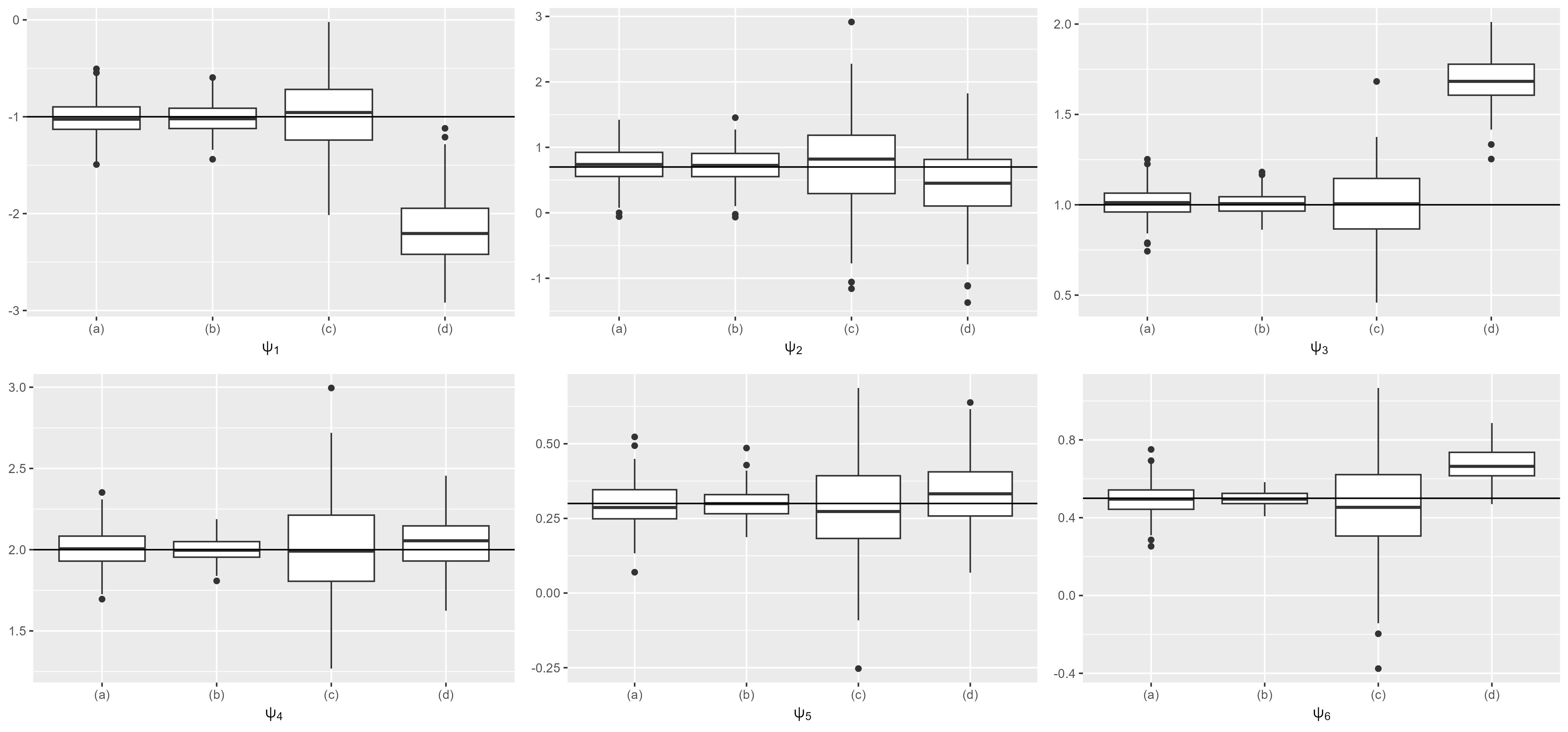}
    \caption{Distribution of blip parameter estimates in the linear setting across $4$ model specifications: (a) all models correct, (b) outcome model correct, treatment and mediator models incorrect, (c) outcome model incorrect, treatment and mediator models correct, and (d) all models incorrect. }
    \label{fig:linear}
\end{figure}

\begin{figure}
    \centering
    \includegraphics[]{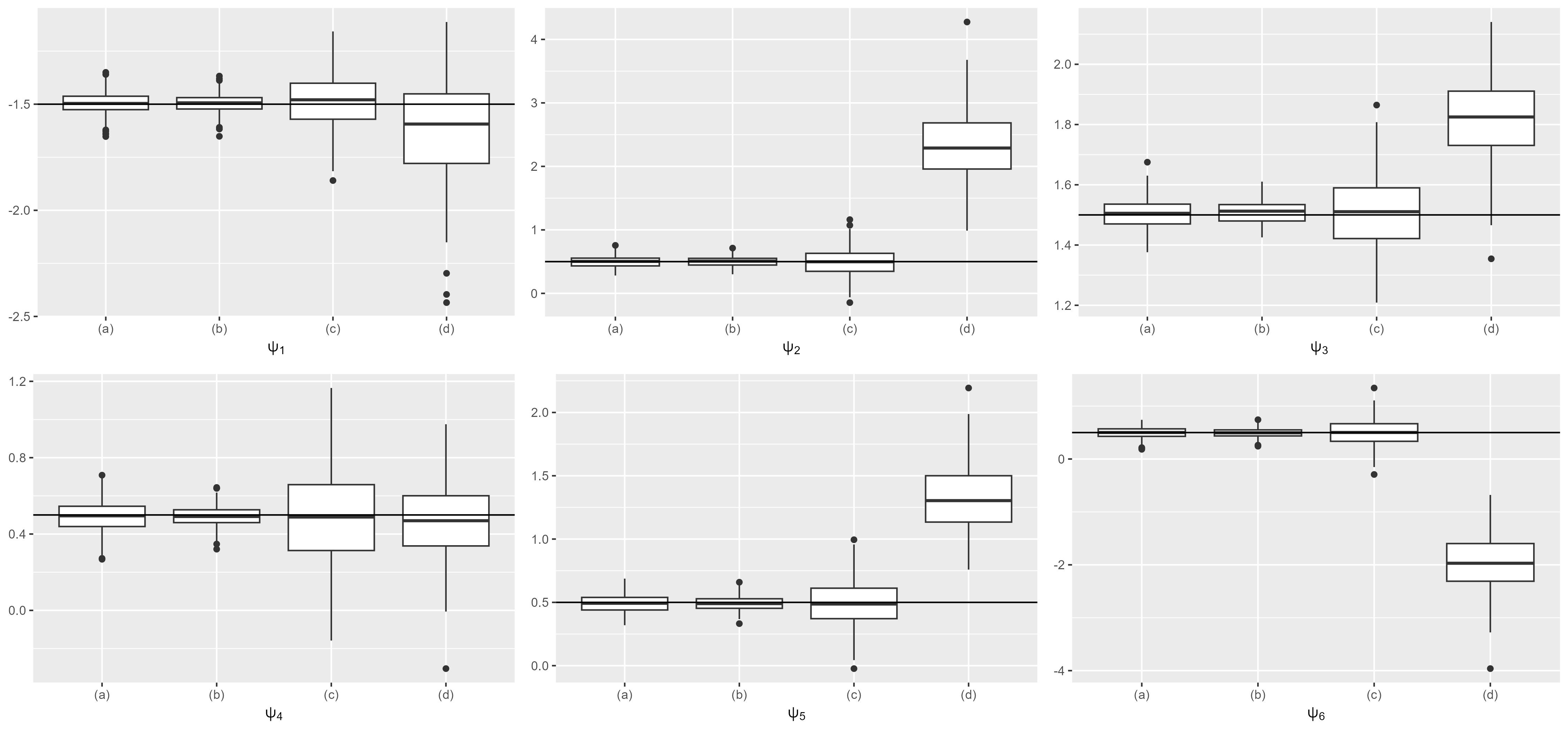}
    \caption{Distribution of blip parameter estimates in the survival (AFT) setting across $4$ model specifications: (a) all models correct, (b) outcome model correct, treatment, mediator and censoring models incorrect, (c) outcome model incorrect, treatment, mediator and censoring models correct, and (d) all models incorrect. }
    \label{fig:survival}
\end{figure}

Table \ref{tab:dr} presents root-$n$ adjusted biases and variances for the blip parameter estimates in the linear and survival settings. Figures \ref{fig:linear} and \ref{fig:survival} present the distributions of the blip parameter estimates in both settings along with horizontal lines indicating the true blip parameter values. In both settings, blip parameter estimators are unbiased provided either the outcome model, or the treatment and mediator models (and censoring, if applicable) are correctly specified.
\begin{table}[H]
\centering
\caption{Root-$n$ adjusted bias and variance for blip parameter estimators in the linear and survival settings. Estimates were computed over $500$ replicate datasets, with posterior inference carried out using $B = 100$ Bayesian bootstrap draws. Posterior means were used as Bayesian point estimates for each replicate dataset. The following model specifications were considered: (a) all models correct, (b) outcome model correct, treatment and mediator models (and censoring model, if applicable) incorrect, (c) outcome model incorrect, treatment and mediator models (and censoring model, if applicable) correct, and (d) all models incorrect. }
\begin{tabular}{lccccccccc}
  \hline
    & & \multicolumn{4}{c}{Linear} & \multicolumn{4}{c}{Survival (AFT)} \\
 & & (a) & (b) & (c) & (d) & (a) & (b) & (c) & (d) \\ 
  \hline
 $\sqrt{n}\; \times$ Bias&$\psi_1$ & -0.55 & -0.46 & 0.80 & -37.22 & 0.16 & 0.15 & 0.41 & -3.85 \\ 
 & $\psi_2$ & 0.81 & 0.66 & 1.43 & -8.40 & 0.00 & 0.07 & -0.34 & 57.45\\ 
  &$\psi_3$  & 0.39 & 0.22 & 0.09 & 21.66 & 0.21 & 0.28 & 0.28 & 10.23 \\ 
  &$\psi_4$ & 0.27 & 0.07 & 0.37 & 1.28 & -0.19 & -0.25 & -0.16 & -1.01  \\ 
  &$\psi_5$ & -0.17 & -0.06 & -0.39 & 1.03 & -0.19 & -0.24 & -0.28 & 26.25\\ 
  &$\psi_6$ &-0.11 & -0.08 & -1.34 & 5.50 & -0.03 & -0.09 & 0.14 & -78.66 \\ 
  \hline
   $n \; \times$ Var&$\psi_1$ & 31.23 & 23.21 & 162.02 & 121.51 & 2.96 & 2.26 & 14.68 & 57.05 \\ 
  &$\psi_2$ & 78.98 & 68.81 & 477.62 & 321.49 & 7.90 & 6.22 & 45.90 & 260.07 \\ 
  &$\psi_3$ & 7.35 & 3.51 & 40.94 & 16.77 & 2.64 & 1.69 & 16.57 & 18.56\\ 
  &$\psi_4$ & 12.64 & 5.18 & 97.58 & 28.67 & 6.60 & 3.11 & 67.17 & 41.43 \\ 
  &$\psi_5$ & 4.95 & 2.49 & 24.54 & 11.68 & 4.98 & 3.07 & 29.34 & 67.77 \\ 
  &$\psi_6$ & 7.02 & 1.31 & 54.61 & 8.36 & 10.95 & 8.29 & 66.55 & 277.12 \\ 
   \hline
   \label{tab:dr}
\end{tabular}
\end{table}

\bibliography{refsAppendix}


\title{\bf Supplementary materials: Appendix C}
\date{}
\maketitle

 \section{Additional details for data analysis}
 The following models were used for the data analysis. Donor and recipient variables are prefixed by Don and Rec, respectively. The treatment variable $A$ is the donor's HCV status (DonHCV), and the mediator $M$ is the binary indicator of graft rejection (GraftRejection).
 \begin{align*}
     \mathbb{P}(A = 1 | \bm{X}) = \text{expit}\Bigl(&\text{RecAge} + \text{RecGender} + \text{RecAge} + \text{RecEthnicity} + \text{RecDiabetes} +\text{RecHCV} + \\ &\text{RecImmunoGroup} \Bigr)\\
      P(\Delta= 1 | \bm{X}, A)= \text{expit}\Bigl(&\text{RecAge} + \text{RecGender} + \text{RecAge} + \text{RecEthnicity} + \text{RecDiabetes} +\text{RecHCV} + \\ &\text{RecImmunoGroup} + \text{DonorHCV} + \text{DonorType} + \text{DonorAge} + \\&
      \text{DonorEthnicity} +  \text{DonorGender} +  \text{DonorCigUse} +  \text{DonorStroke} +  \\&
      \text{KidneyIschemicTime} + \text{OrganAllocationType}\Bigr)\\
      \mathbb{E}[\log T | \bm{X}, A, M] = & \;\mu_0(\bm{X}) + \Gamma(\bm{X},A, M)\\
      \mu_0(\bm{X}) = &\text{RecAge} + \text{RecGender} + \text{RecAge} + \text{RecEthnicity} + \text{RecDiabetes} +\text{RecHCV} + \\ &\text{RecImmunoGroup} +  \text{DonorHCV} + \text{DonorType} + \text{DonorAge} + \\&
      \text{DonorEthnicity} +  \text{DonorGender} +  \text{DonorCigUse} +  \text{DonorStroke} +  \\&
      \text{KidneyIschemicTime} + \text{OrganAllocationType}\\
      \Gamma(\bm{X}, A, M) = & \text{DonorHCV} \times \Bigl(\text{RecHCV} + \text{DonorType} + \text{RecAge}\Bigr) + \text{GraftRejection} + \\
      & \text{DonorHCV} \times \text{GraftRejection} \times\Bigl(\text{RecHCV} + \text{DonorType} + \text{RecAge}\Bigr)\\
      \mathbb{P}(M = 1 | \bm{X}, A) =\text{expit}\Bigl(&\text{RecAge} + \text{RecGender} + \text{RecAge} + \text{RecEthnicity} + \text{RecDiabetes} +\text{RecHCV} + \\ &\text{RecImmunoGroup} + \text{DonorHCV} + \text{DonorType} + \text{DonorAge} + \\&
      \text{DonorEthnicity} +  \text{DonorGender} +  \text{DonorCigUse} +  \text{DonorStroke} +  \\&
      \text{KidneyIschemicTime} + \text{OrganAllocationType}\Bigr).
 \end{align*}

  \begin{table}[h]
  \caption*{Table 1: OPTN kidney transplant characteristics (January 1, 2001-December 31, 2022) stratified by donor HCV status along with standardized mean differences (SMD).}
\centering
\begin{tabular}{lllll}
  \hline
  Characteristics & HCV-  & HCV+ & Overall & SMD \\ 
  \hline
 & ($n$=300,161) & ($n$=11,313) & ($n$=311,474) &  \\ 
  Recipient gender &  &  &  & 0.249 \\ 
    \quad Female & 119857 (39.9\%) & 3190 (28.2\%) & 123047 (39.5\%) &  \\ 
    \quad Male & 180304 (60.1\%) & 8123 (71.8\%) & 188427 (60.5\%) &  \\ 
  Recipient ethnicity &  &  &  & 0.399 \\ 
   \quad Non-African American & 220256 (73.4\%) & 6179 (54.6\%) & 226435 (72.7\%) &  \\ 
    \quad African American & 79905 (26.6\%) & 5134 (45.4\%) & 85039 (27.3\%) &  \\ 
  Immuno-group &  &  &  & 0.099 \\ 
    \quad 1 & 186856 (62.3\%) & 7518 (66.5\%) & 194374 (62.4\%) &  \\ 
    \quad 2 & 63193 (21.1\%) & 2012 (17.8\%) & 65205 (20.9\%) &  \\ 
    \quad 3 & 9908 (3.3\%) & 410 (3.6\%) & 10318 (3.3\%) &  \\ 
   \quad  4 & 40204 (13.4\%) & 1373 (12.1\%) & 41577 (13.3\%) &  \\ 
  Recipient HCV status &  &  &  & 0.922 \\ 
    \quad HCV- & 290861 (96.9\%) & 7191 (63.6\%) & 298052 (95.7\%) &  \\ 
    \quad HCV+ & 9300 (3.1\%) & 4122 (36.4\%) & 13422 (4.3\%) &  \\ 
  Donor type &  &  &  & 0.726 \\ 
    \quad Deceased & 214903 (71.6\%) & 10925 (96.6\%) & 225828 (72.5\%) &  \\ 
    \quad Living & 85258 (28.4\%) & 388 (3.4\%) & 85646 (27.5\%) &  \\ 
  Recipient age (years) &  &  &  & 0.547 \\ 
    \quad Mean (SD) & 49.8 (15.6) & 57.1 (10.9) & 50.0 (15.5) &  \\ 
  Donor age (years) &  &  &  & 0.467 \\ 
    \quad 0-18 & 22956 (7.6\%) & 44 (0.4\%) & 23000 (7.4\%) &  \\ 
    \quad 19-49 & 197611 (65.8\%) & 9314 (82.3\%) & 206925 (66.4\%) &  \\ 
    \quad $>$ 49 & 79594 (26.5\%) & 1955 (17.3\%) & 81549 (26.2\%) &  \\ 
  Donor ethnicity &  &  &  & 0.176 \\ 
    \quad Non-African American & 261997 (87.3\%) & 10473 (92.6\%) & 272470 (87.5\%) &  \\ 
    \quad African American & 38164 (12.7\%) & 840 (7.4\%) & 39004 (12.5\%) &  \\ 
  Donor gender &  &  &  & 0.151 \\ 
   \quad  Female & 137046 (45.7\%) & 4324 (38.2\%) & 141370 (45.4\%) &  \\ 
    \quad Male & 163115 (54.3\%) & 6989 (61.8\%) & 170104 (54.6\%) &  \\ 
  Donor history of cigarette use &  &  &  & 0.338 \\ 
    \quad No & 232353 (77.4\%) & 7027 (62.1\%) & 239380 (76.9\%) &  \\ 
    \quad Yes & 67808 (22.6\%) & 4286 (37.9\%) & 72094 (23.1\%) &  \\ 
  Donor cerebrovascular/stroke &  &  &  & 0.113 \\ 
   \quad  No & 237035 (79.0\%) & 9431 (83.4\%) & 246466 (79.1\%) &  \\ 
   \quad  Yes & 63126 (21.0\%) & 1882 (16.6\%) & 65008 (20.9\%) &  \\ 
  Kidney cold ischemic time (hours) &  &  &  & 0.507 \\ 
   \quad  Mean (SD) & 13.6 (10.5) & 18.5 (8.49) & 13.8 (10.5) &  \\ 
  Organ allocation type &  &  &  & 0.835 \\ 
    \quad Local & 38210 (12.7\%) & 3585 (31.7\%) & 41795 (13.4\%) &  \\ 
    \quad Regional & 0 (0\%) & 0 (0\%) & 0 (0\%) &  \\ 
    \quad National & 236221 (78.7\%) & 4642 (41.0\%) & 240863 (77.3\%) &  \\ 
    \quad Foreign & 25730 (8.6\%) & 3086 (27.3\%) & 28816 (9.3\%) &  \\ 
  Follow-up time (days) &  &  &  & 0.614 \\ 
    \quad Median [IQR] & 1526 [695, 2900] & 739 [355, 1469] & 1480 [676, 2875] &  \\ 
    Graft rejection & 48227 (16.1\%) & 1281 (11.3\%) & 49508 (15.9\%) & \\ 
   \hline
\end{tabular}
\end{table}